\documentclass[12pt,twoside,english]{article}
\usepackage[T1]{fontenc} 			
\usepackage{amssymb}
\usepackage{a4wide}
\usepackage{graphicx}
\usepackage{fancyhdr}
\usepackage{amsmath, amssymb}
\usepackage{subfigure}
\usepackage{setspace}
\usepackage{verbatim} 
\usepackage[english]{babel}
\usepackage{caption}
\usepackage{float}
\usepackage{booktabs}

\usepackage{siunitx}
\usepackage{xfrac}

\usepackage{multirow}
\usepackage{tabularx}
\usepackage{array}

\makeatletter
\usepackage{hyperref}

\hypersetup{
	colorlinks = false,
	linktocpage = true,
	allbordercolors = {0.8 0.8 0.8}
}

\addto\captionsenglish{}
\addto\captionsenglish{}
\addto\extrasenglish{}
\addto\extrasenglish{}
\addto\extrasenglish{}
\addto\extrasenglish{}
\addto\extrasenglish{}
\addto\extrasenglish{}
\addto\extrasenglish{}

\newcommand{\lyxaddress}[1]{
\par {\raggedright #1
\vspace{1.4em}
\noindent\par}
}

\numberwithin{equation}{section}

\newcolumntype{L}[1]{>{\raggedright\arraybackslash}m{#1}}
\newcolumntype{M}[1]{>{\centering\arraybackslash}m{#1}}
\newcolumntype{R}[1]{>{\raggedleft\arraybackslash}m{#1}}

\newcommand{\vek}[1]{\mathchoice{\displaystyle\boldsymbol#1}
{\textstyle\boldsymbol#1}{\scriptstyle\boldsymbol#1}
{\scriptscriptstyle\boldsymbol#1}}
\newcommand{\mat}[1]{\mathchoice{\displaystyle\mathbf#1}
{\textstyle\mathbf#1}{\scriptstyle\mathbf#1}
{\scriptscriptstyle\mathbf#1}}

\newcommand{\ScalProd}{\boldsymbol{\cdot}}
\newcommand{\FrobProd}{\boldsymbol{:}}
\newcommand{\jumpl}{[\![}
\newcommand{\jumpr}{]\!]}

\makeatother

\graphicspath{{FiguresNew/}}

\begin{document}

\title{Simultaneous analysis of curved\\Kirchhoff beams and Kirchhoff--Love\\shells embedded in bulk domains}

\author{Jonas Neumeyer, Michael Wolfgang Kaiser, Thomas-Peter Fries}
\maketitle

\lyxaddress{\begin{center}
Institute of Structural Analysis\\
Graz University of Technology\\
Lessingstr. 25/II, 8010 Graz, Austria\\
\texttt{www.ifb.tugraz.at}\\
\texttt{jonas.neumeyer@tugraz.at}
\end{center}}

\vspace{-0.2cm}
\begin{abstract}
A set of curved beams and shells is geometrically implied by level sets of a scalar function over some bulk domain. The mechanical model for each structure is based on the Kirchhoff--Love theory, that is, small displacements without shear deformations are considered. These models for individual geometries are extended to bulk models, simultaneously modeling the whole set of beams/shells on all level sets. A major focus is on the numerical analysis of such models. A mixed-hybrid and higher-order accurate Bulk Trace FEM is proposed that enables the use of standard $C^0$-continuous Lagrange elements with dimensionality of the bulk domain. That is, the higher-order continuity requirements of displacement-based formulations in context of the Kirchhoff--Love theory are successfully alleviated. Several numerical tests confirm the accuracy and higher-order convergence of the proposed methodology, also qualifying as benchmark test cases in future studies.

Keywords: Kirchhoff beam, Kirchhoff--Love shell, tangential differential calculus, Bulk Trace FEM, mixed-hybrid formulation, higher-order convergence

\end{abstract}
\newpage{}\tableofcontents{}\newpage{}

\section{Introduction}\label{Sec:Introduction}
Models in structural mechanics can be classified based on their dimensionality. Starting from three-dimensional continua, geometrical assumptions often allow for reducing the dimensionality of the resulting models. That is, when only one dimension is dominant over the other two, it is often possible to formulate mechanical models in one dimension such as for tension bars and beams. When the thickness is much smaller than the other two dimensions, two-dimensional plate and shell models result. Such reductions of dimensionality are also feasible for \emph{curved} structures, e.g., curved beams \cite{Gimena_2008a, Hansbo_2014b, Kaiser_2023a} and shells \cite{Basar_1985a, Bischoff_2017a, Chapelle_1998a, Zingoni_2017a}, then, these structures are embedded in a higher-dimensional space. From a geometrical perspective, curved beams are idealized by their centerline, and shells by their middle surface. By exploiting the slenderness of these structures, the reduced models are further capable of capturing the dominant stress and strain quantities with much less computational effort.

A crucial aspect of the mechanical modeling of curved structures is the definition of their geometry as manifolds and the introduction of advanced (surface) differential operators. This can be accomplished either for \emph{explicit} geometry definitions through parametrizations or \emph{implicit} descriptions through level-set functions. A suitable and versatile approach is provided by the Tangential Differential Calculus (TDC) \cite{Delfour_2011a, Dziuk_2013a}, being a modern interpretation of differential geometry applicable to both explicit and implicit geometry descriptions. Moreover, formulations in the frame of the TDC do not require (local) curvilinear coordinates, and all quantities are expressed with respect to a global Cartesian coordinate system into which the curved structures are immersed.

For the analysis, a straightforward approach based on the Finite Element Method (FEM) is to discretize some explicit geometry via a surface mesh. This approach is often referred to as Surface FEM and has been established in terms of the TDC for membranes \cite{Fries_2020a, Hansbo_2014a}, curved beams \cite{Hansbo_2014b, Kaiser_2023a}, and shells \cite{Neumeyer_2025a, Neunteufel_2019a, Neunteufel_2024a, Schoellhammer_2019a, Schoellhammer_2019b}. Alternatively, implicitly defined curved structures are rather related to fictitious domain methods such as the Cut FEM or Trace FEM, where a background mesh is employed that is neither aligned with the manifold nor its boundary. This single manifold is usually defined by the zero-isoline or zero-isosurface of a level-set function. Examples for this approach have been applied to transport problems \cite{Burman_2015a, Burman_2018a, Olshanskii_2009b, Olshanskii_2014a, Olshanskii_2017a}, membranes \cite{Cenanovic_2016a, Fries_2020a}, and shells \cite{Gfrerer_2021a, Schoellhammer_2020a}.

Extending the implicit approach beyond a single level set enables the simultaneous analysis of multiple curves or surfaces defined over a bulk domain. In this setting, the bulk domain is explicitly discretized using finite elements. These bulk elements conform to the boundary of the bulk domain but are generally not aligned to the level sets. This approach, later labeled the \emph{Bulk Trace FEM} \cite{Fries_2023a}, was introduced in early works on transport problems \cite{Burger_2009a, Dziuk_2008a, Dziuk_2008b, Dziuk_2013a}. Recent developments extend the method to fluid dynamics \cite{Kaiser_2025a} and to structural mechanics with ropes and membranes \cite{Fries_2023a}, as well as shear-flexible curved beams \cite{Kaiser_2024b} and shells \cite{Kaiser_2024a}. However, shear-rigid structures have not yet been addressed within the Bulk Trace FEM framework, due to the stricter continuity requirements of their governing equations which is the main focus of this work.

Major distinctions for thin structures subjected to bending arise from the consideration of transverse shear deformations. Neglecting these leads to shear-rigid models, such as Kirchhoff beams and Kirchhoff--Love shells \cite{Kirchhoff_1850a, Love_1888a}. In contrast, considering transverse shear deformations results in shear-flexible models, i.e., curved Timoshenko--Ehrenfest beams \cite{Timoshenko_1921a} and Reissner--Mindlin shells \cite{Reissner_1945a, Mindlin_1951a}. \emph{Shear-rigid models} are suitable for thin beams and shells, where shear deformations are negligible. Their governing equations typically consist of a single vector-valued fourth-order partial differential equation (PDE), with only the displacement of the centerline or middle surface as the primary unknown; see \cite{Echter_2013a, Hansbo_2014b, Kaiser_2023a, Kiendl_2009a, Schoellhammer_2019a}. In contrast, \emph{shear-flexible models} are applicable for thin to moderately thick shells. Their governing equations typically comprise two vector-valued second order PDEs with both displacements and rotations as unknowns; see \cite{Echter_2013a, Hansbo_2014b, Schoellhammer_2019b}.

Looking from a numerical perspective, while shear-flexible models incur more degrees of freedom (DOFs), their second-order PDEs permit straightforward FE discretizations with standard $C^0$-continuous Lagrange elements. Shear-rigid models, however, usually require at least $C^1$-continuous shape functions in their displacement-based form due to the fourth-order derivatives in their strong form, which prohibits the use of standard Lagrange elements. This challenge can be overcome using a mixed or mixed-hybrid formulation. Such approaches are well established for Kirchhoff--Love plates \cite{Arnold_1985a, Boffi_2013a, Brezzi_1974a, Comodi_1989a, Johnson_1973a, Rafetseder_2018a} and have recently been extended to Kirchhoff--Love shells \cite{Neumeyer_2025a, Neunteufel_2019a, Neunteufel_2024a, Rafetseder_2019a}. Here, the moment tensor is introduced as an additional primary unknown and a second governing equation is established leading to a higher number of DOFs. This reduces the continuity requirement to $C^0$-continuity and makes Lagrange elements admissible. In order to reduce the number of DOFs, a hybridization and static condensation is performed, where the moment tensor field is eliminated element-wise after introducing Lagrange multipliers that weakly enforce the continuity of moments in tangent direction on the element interfaces.

The objective of this work is to derive a mechanical model for a family of geometrically linear Kirchhoff beams and Kirchhoff--Love shells living in a two- and three-dimensional bulk domain, respectively. Subsequently, we derive a suitable numerical method in form of the Bulk Trace FEM. The main contributions are summarized as follows:
\begin{itemize}
    \item We present the first method capable of solving a family of continuously implied Kirchhoff beams or Kirchhoff--Love shells \emph{simultaneously} within the framework of the Bulk Trace FEM. Compared with shear-flexible structures, these shear-rigid formulations require fewer degrees of freedom, which in turn reduces the overall computational costs.
    \item Difficulties with respect to the continuity requirements are solved by using a mixed-hybrid method, extending the approach of \cite{Neumeyer_2025a} to the simultaneous analysis in this work. The resulting framework enables the use of standard $C^0$-continuous Lagrange elements in the bulk mesh.
    \item A variety of numerical test cases are presented, featuring smooth solutions and achieving higher-order convergence rates, thus also serving as benchmark tests for verifications and future studies.
\end{itemize}

This paper is structured as follows. Section~\ref{Sec:Preliminaries} discusses preliminaries regarding the geometrical description and surface differential operators. Section~\ref{Sec:MechanicalModel} summarizes briefly the mechanical model of the Kirchhoff beam and the Kirchhoff--Love shell, leading to the mixed strong form of the models. A mixed-hybrid weak form of both models in the frame of the Bulk Trace FEM is derived in Section~\ref{Sec:BTFForShearRigidBeams&Shells}. Numerical results demonstrating higher-order convergence are presented in Section~\ref{Sec:NumericalResults}. Finally, Section~\ref{Sec:Conclusions&Outlook} ends the paper with some concluding remarks and an outlook to future research.

\section{Preliminaries and geometric definitions}\label{Sec:Preliminaries}
Starting point is the \emph{geometric} definition of families of curved beams and shells over bulk domains that are later simultaneously considered in the \emph{mechanical} modeling. These geometry definitions are based on level sets of scalar functions over some bulk domains following the level-set method \cite{Osher_2003a,Osher_2001a,Sethian_1999b}. The situation is described for curved beams embedded in two dimensions ($d=2$) and shells embedded in three dimensions ($d=3$).

\subsection{Level sets and bulk domains}\label{Sec:LevelSets&BulkDomains}
\begin{figure}[ht!]
 \centering
 
 \subfigure[$\Omega$ and $\phi$ in $\mathbb{R}^2$]
 {\includegraphics[width=0.36\textwidth]{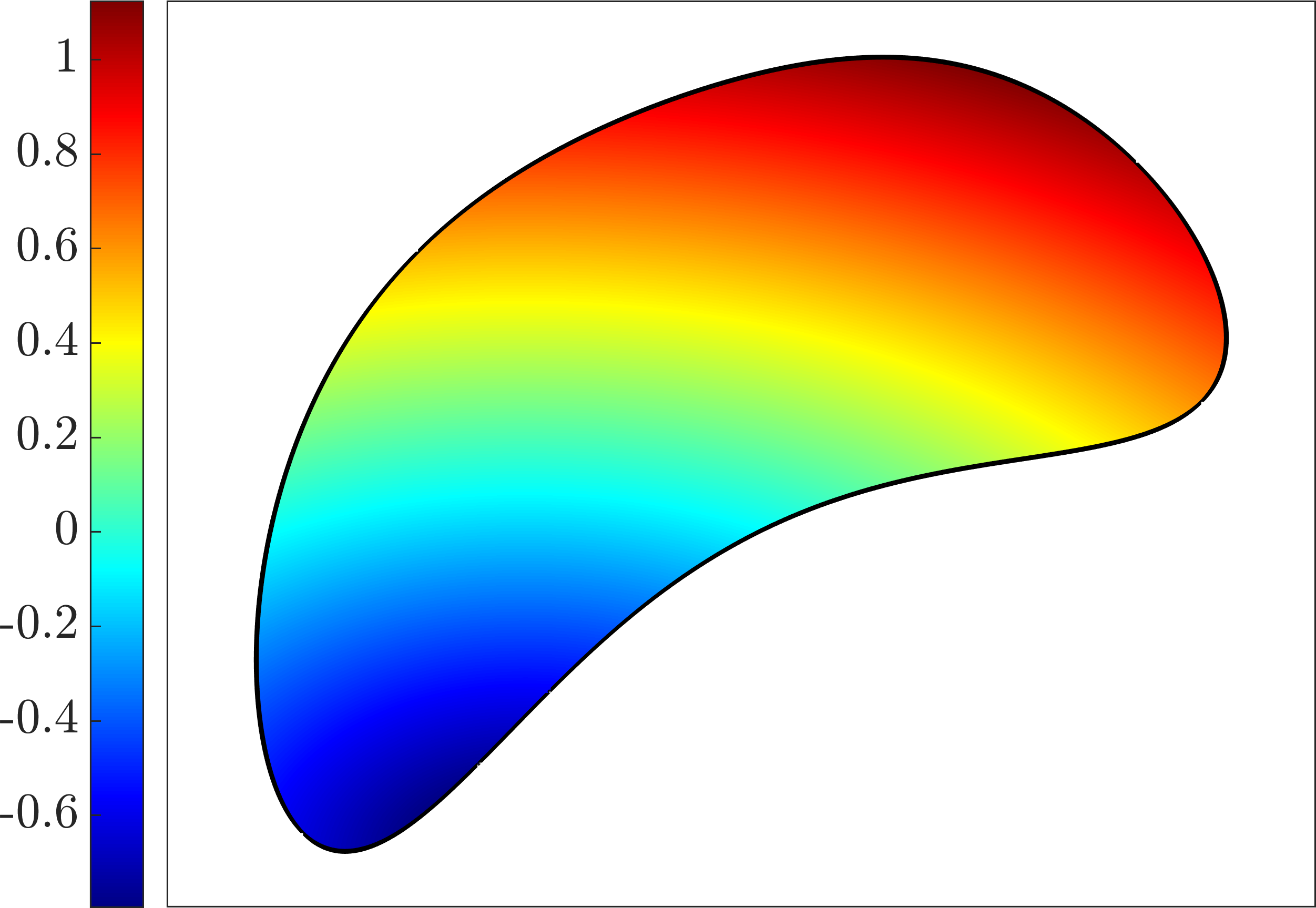}\label{Fig:BulkDomain2d}}\hfill
 \subfigure[$\Omega$ and level sets $\Gamma^c$ in $\mathbb{R}^2$]
 {\includegraphics[width=0.315\textwidth]{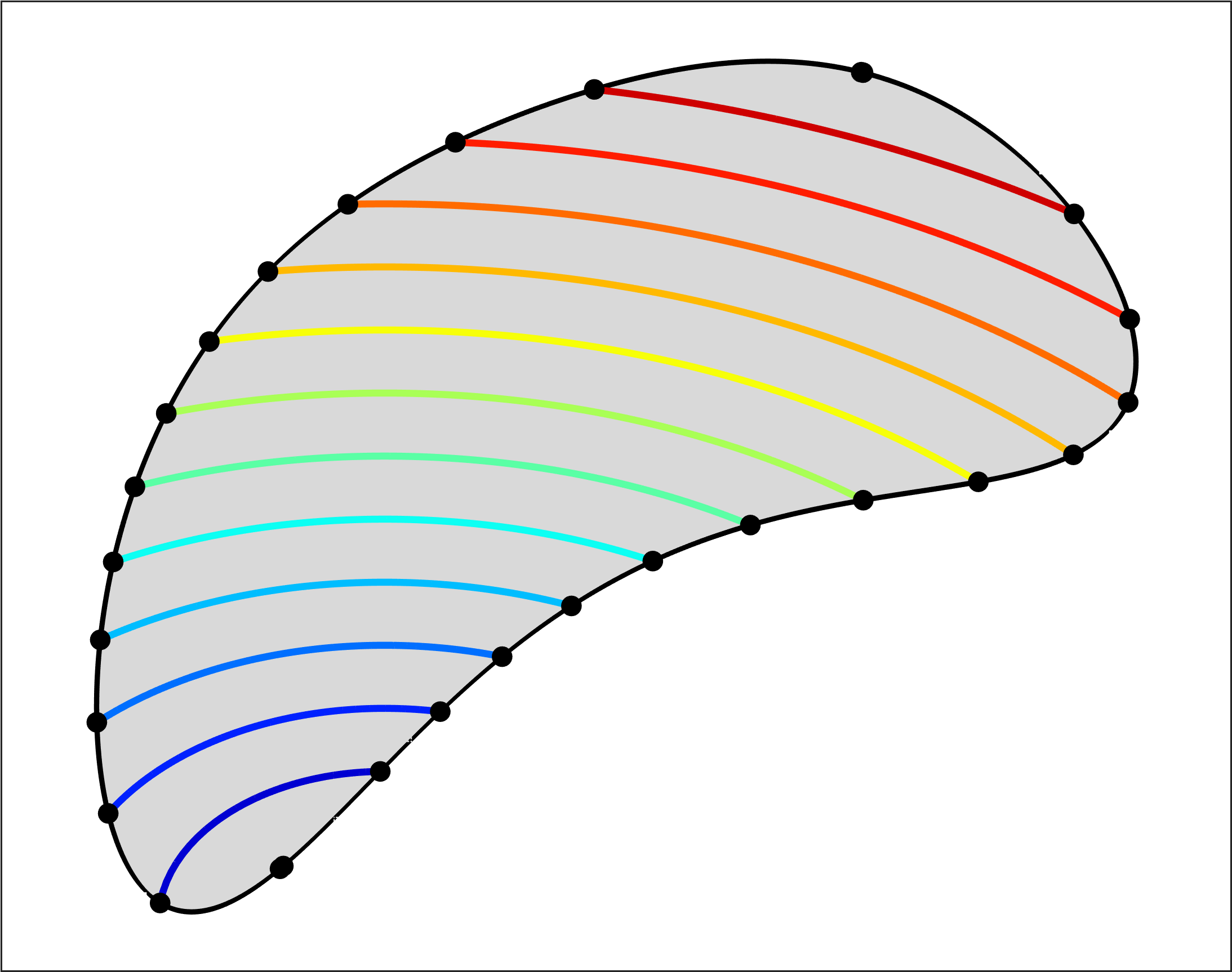}\label{Fig:BulkDomain+LevelSets2d}}\hfill
    \subfigure[$\Omega$ and level sets $\Gamma^c$ in $\mathbb{R}^2$]
 {\includegraphics[width=0.315\textwidth]{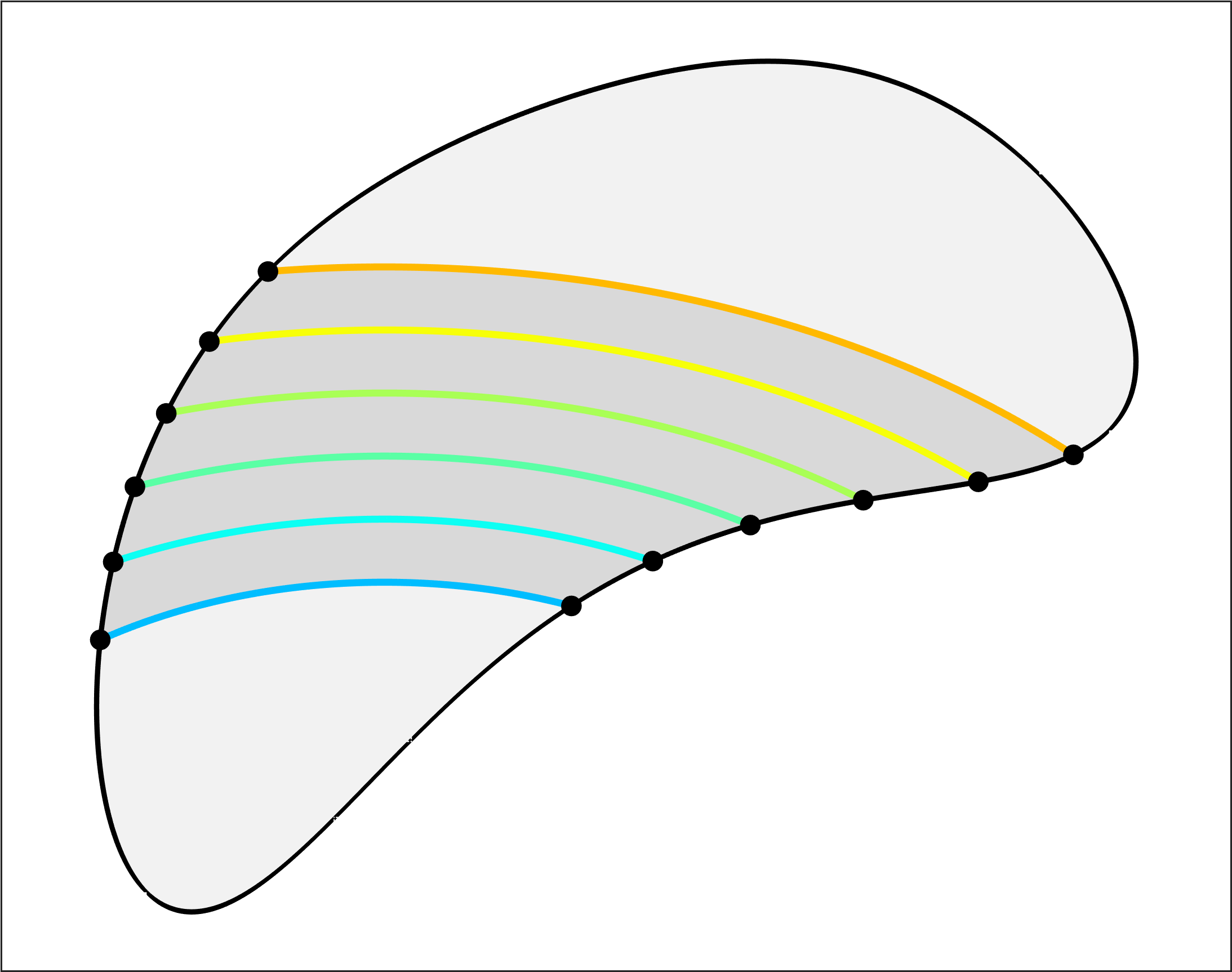}\label{Fig:BulkDomain+LevelSetInterval2d}}\\
    \subfigure[$\Omega$ and $\phi$ in $\mathbb{R}^3$]
 {\includegraphics[width=0.365\textwidth]{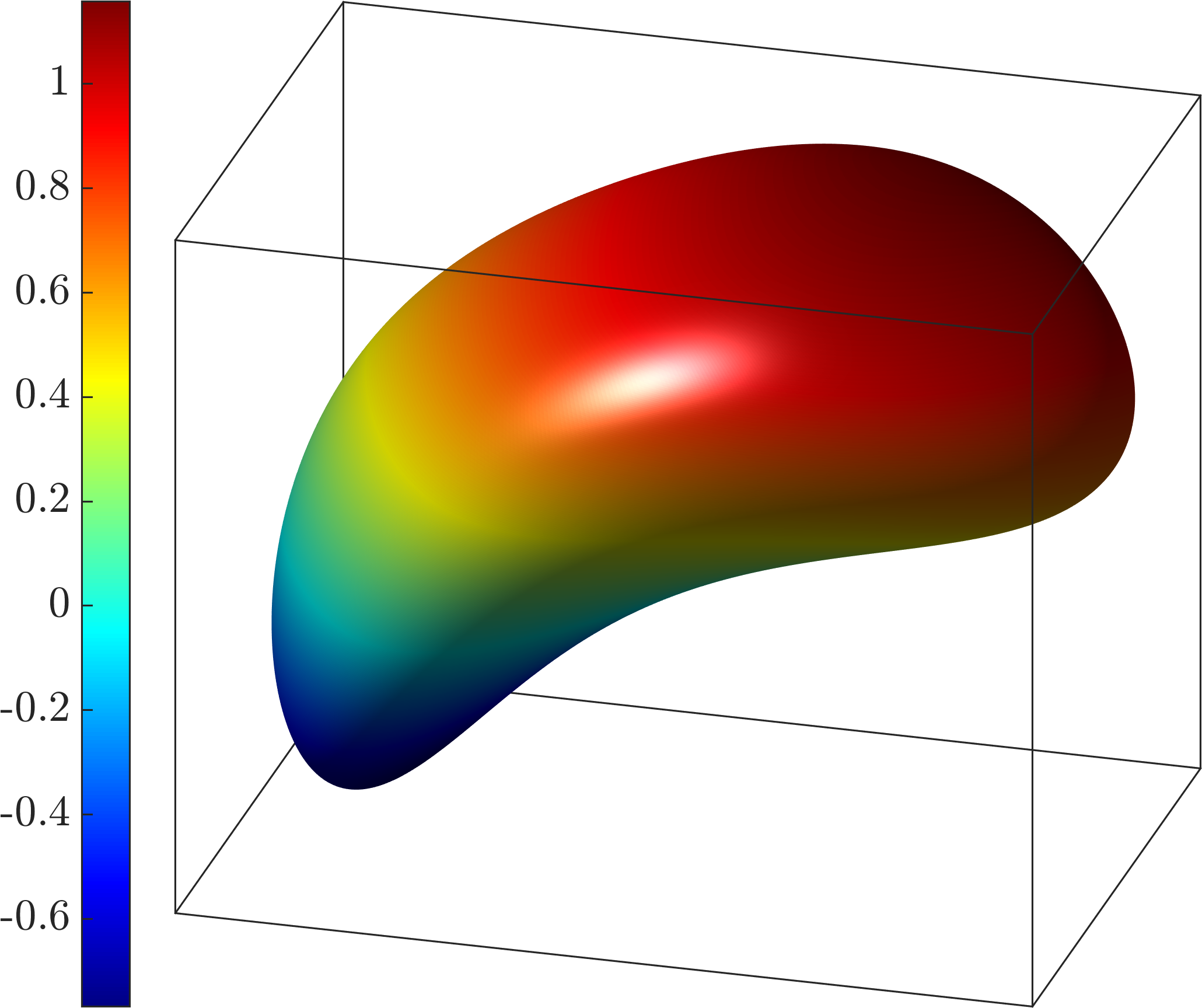}\label{Fig:BulkDomain3d}}\hfill
 \subfigure[$\Omega$ and level sets $\Gamma^c$ in $\mathbb{R}^3$]
 {\includegraphics[width=0.3125\textwidth]{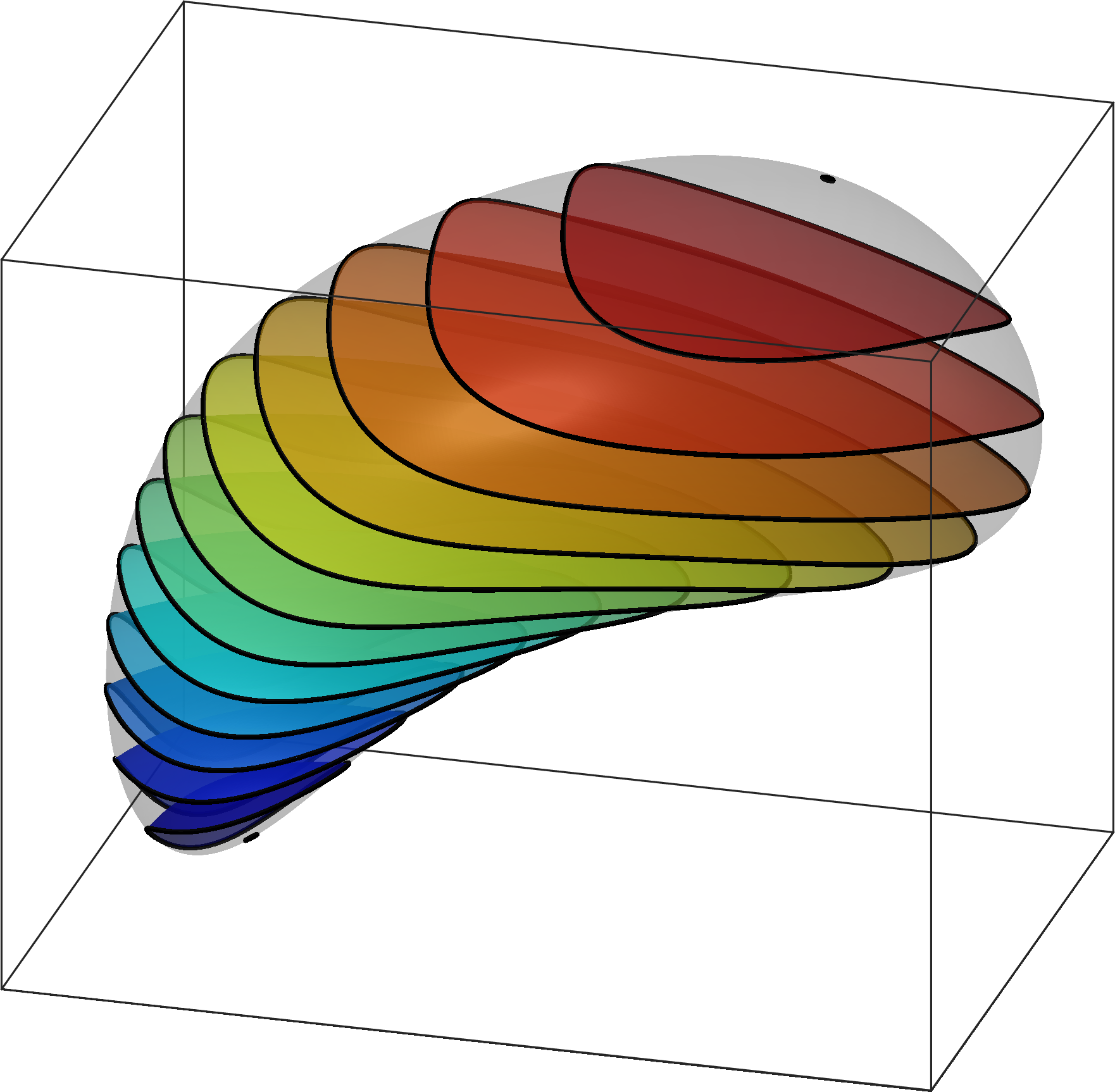}\label{Fig:BulkDomain+LevelSets3d}}\hfill
    \subfigure[$\Omega$ and level sets $\Gamma^c$ in $\mathbb{R}^3$]
 {\includegraphics[width=0.3125\textwidth]{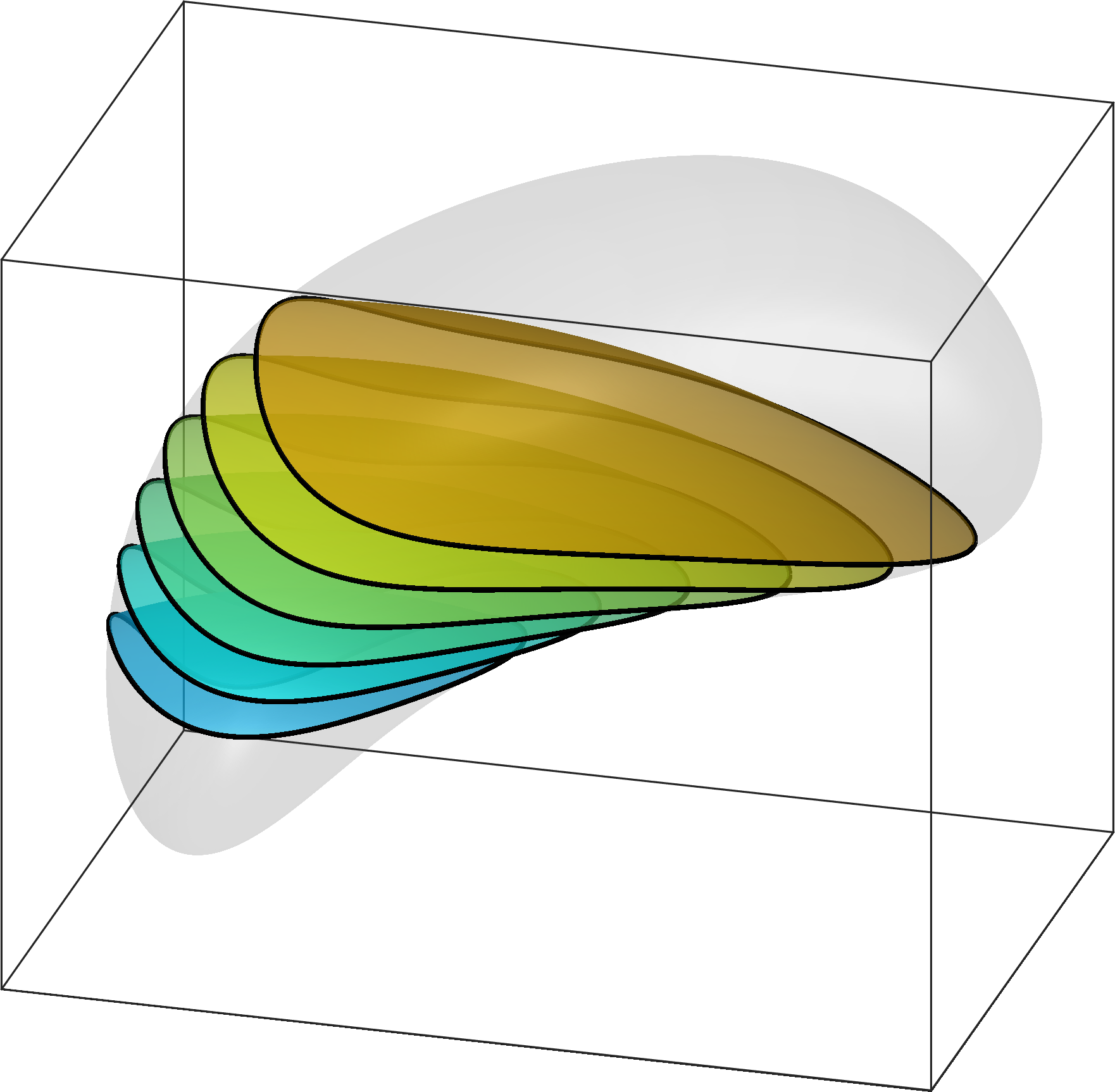}\label{Fig:BulkDomain+LevelSetInterval3d}}
 
 \caption{\label{Fig:BulkDomain}For an arbitrary bulk domain $\Omega$ and level-set function $\phi(\vek{x})$, (a) and (d) show the level-set functions $\phi(\vek{x})$, (b) and (e) some selected level sets $\Gamma^c$ in $\Omega$. When restricting bulk domains to prescribed level-set intervals, see Eq.~(\ref{Eq:BulkDomainInterval}), examples are seen in (c) and (f), including selected levels sets $\Gamma^c$.}
\end{figure}

Let there be a $d$-dimensional bulk domain $\Omega \subset \mathbb{R}^d$ and a level-set function $\phi(\vek{x}): \Omega \rightarrow \mathbb{R}$, where the minimum and maximum value of $\phi$ inside the bulk domain result as $\phi^{\text{min}} = \text{inf} \, \phi(\vek{x})$ and $\phi^{\text{max}} = \text{sup} \, \phi(\vek{x})$, see Figs.~\ref{Fig:BulkDomain2d} and \ref{Fig:BulkDomain3d}. The level sets of $\phi$, i.e., the isolines and isosurfaces, are curved manifolds that imply the geometry of the beams and shells, respectively, as
\begin{equation}\label{Eq:IndividualManifold}
    \Gamma^c =
    \begin{Bmatrix}
        \vek{x} \in \Omega: \phi(\vek{x}) = c \in \mathbb{R}
    \end{Bmatrix}
    \quad \text{with} \quad \phi^{\text{min}} < c < \phi^{\text{max}}.
\end{equation}
In this work, we focus on structures with codimension one, i.e., ($d-1$)-dimensional manifolds. Boundaries of a single manifold $\partial\Gamma^c$ are the intersection points or curves of $\Gamma^c$ and the boundary of the bulk domain $\partial\Omega$. This situation is depicted in Figs.~\ref{Fig:BulkDomain+LevelSets2d} and \ref{Fig:BulkDomain+LevelSets3d}.

In the $d$-dimensional bulk domain $\Omega$, the manifolds may locally degenerate to single points, i.e., where $\phi(\vek{x}) = \phi^{\text{min}}$ and $\phi(\vek{x}) = \phi^{\text{max}}$. As a consequence, the mechanical BVPs on the level sets may be ill-posed there, resulting in local steep gradients in the simultaneous solutions over the bulk domain, leading to suboptimal convergence rates in numerical results as already shown in \cite{Fries_2023a}. To avoid such effects, $\Omega$ may rather be defined taking the level-set function into account, for example, by restricting some super-set $\Omega^{\text{sup}} \subset \mathbb{R}^d$ based on prescribed level-set values,
\begin{equation}\label{Eq:BulkDomainInterval}
    \Omega =
    \begin{Bmatrix}
        \vek{x} \in \Omega^{\text{sup}}: \phi^{\text{min}} \leq \phi(\vek{x}) \leq \phi^{\text{max}}
    \end{Bmatrix},
\end{equation}
as shown in Figs.~\ref{Fig:BulkDomain+LevelSetInterval2d} and \ref{Fig:BulkDomain+LevelSetInterval3d}. In this case, $\partial\Omega$ is defined by the boundary of the superset $\partial\Omega^{\text{sup}} \subset \mathbb{R}^d$ as
\begin{equation}\label{Eq:BoundaryBulkDomainInterval}
    \partial\Omega =
    \begin{Bmatrix}
        \vek{x} \in \partial\Omega^{\text{sup}}: \phi^{\text{min}} \leq \phi(\vek{x}) \leq \phi^{\text{max}}
    \end{Bmatrix}.
\end{equation}

It is also noted that certain combinations of level-set functions and bulk domains may result in invalid situations, typically traced back to rapid topological changes. A detailed discussion on this topic is found in \cite{Dziuk_2008a, Fries_2023a}. In the following, only valid combinations of $\phi$ and $\Omega$ are considered that ensure smooth geometries without topological changes in the level-set interval. \\ \\

\subsection{Geometric quantities and differential operators}\label{Sec:GeometricQuantities&DifferentialOperators}
In order to define the governing equations of a boundary value problem on a manifold, several geometric quantities (e.g., normal and tangential vector fields) and surface differential operators (e.g., gradients and divergence w.r.t.~level sets) are required. This has already been discussed in various publications, see \cite{Delfour_2011a, Dziuk_2013a, Jankuhn_2017a, Fries_2023a, Kaiser_2024a} for further details. For the sake of brevity, we summarize these terms in Tab.~\ref{Tab:GeometricQuantities&DifferentialOperators}. As the geometries are implicitly defined by level sets and, therefore, parametrizations and curvilinear coordinates are not applicable, a coordinate-free description in terms of the TDC is required.\\ \\ \\

\begin{table}[ht!]
    \centering
    \begin{tabularx}{\textwidth}{L{0.22\textwidth}XX}
        \toprule
         &
        Curved beams in $\mathbb{R}^2$ &
        Shells in $\mathbb{R}^3$\\
        \midrule
        Unit normal vector on $\partial\Omega$ w.r.t.~the bulk domain (Fig.~\ref{Fig:VectorFields}) &
        \multicolumn{2}{c}{
            \begin{tabular}{c}
            $\vek{m}$ \\
            (results from boundary\\elements in the bulk mesh)
            \end{tabular}
        }\\
        \midrule
        Unit normal vector in $\Omega$ w.r.t.~level sets (Fig.~\ref{Fig:VectorFields}) &
        \multicolumn{2}{c}{$\vek{n} = \frac{\vek{n}^\star}{\| \vek{n}^\star \|},\, \vek{n}^\star = \nabla\phi$}\\
        \midrule
        Unit tangent vector on $\partial\Omega$ (Fig.~\ref{Fig:VectorFields}) &
        \multicolumn{1}{c}{-} &
        \multicolumn{1}{c}{$\vek{t} = \frac{\vek{t}^\star}{\| \vek{t}^\star \|},\, \vek{t}^\star = \vek{n} \times \vek{m}$}\\
        \midrule
        Unit conormal vector on $\partial\Omega$ (Fig.~\ref{Fig:VectorFields}) &
        \multicolumn{1}{c}{$\vek{q} =
        \begin{cases}
            [-n_y, n_x]^\text{T} \, \text{if} \, n_x m_y \geq n_y m_x\\
            [n_y, -n_x]^\text{T} \, \text{if} \, n_x m_y < n_y m_x
        \end{cases}$\!\!\!\!\!\!}&
        \multicolumn{1}{c}{$\vek{q} = \vek{t} \times \vek{n}$}\\
        \midrule
        Projector field &
        \multicolumn{2}{c}{$\mat{P} = \mat{I} - \vek{n} \otimes \vek{n}$}\\
        \midrule
        Surface gradient of a scalar function &
        \multicolumn{2}{c}{$\nabla_\Gamma f = \mat{P} \cdot \nabla f$}\\
        \midrule
        Dir.~surf.~gradient of a vector function &
        \multicolumn{2}{c}{$\nabla_\Gamma^{\text{dir}} \vek{v} = \nabla \vek{v} \cdot \mat{P}$}\\
        \midrule
        Cov.~surf.~gradient of a vector function &
        \multicolumn{2}{c}{$\nabla_\Gamma^{\text{cov}} \vek{v} = \mat{P} \cdot \nabla_\Gamma^{\text{dir}} \vek{v} = \mat{P} \cdot \nabla \vek{v} \cdot \mat{P}$}\\
        \midrule
        Surface divergence of a vector function &
        \multicolumn{2}{c}{$\text{div}_\Gamma \vek{v} = \text{tr}(\nabla_\Gamma^{\text{dir}} \vek{v}) = \text{tr}(\nabla_\Gamma^{\text{cov}} \vek{v})$}\\
        \midrule
        Surface divergence of a tensor function &
        \multicolumn{1}{c}{$\text{div}_\Gamma \mat{T} = 
     \begin{bmatrix}
      \text{div}_\Gamma [T_{11},T_{12}]\\
      \text{div}_\Gamma [T_{21},T_{22}]
     \end{bmatrix}$} &
        \multicolumn{1}{c}{$\text{div}_\Gamma \mat{T} = 
     \begin{bmatrix}
      \text{div}_\Gamma [T_{11},T_{12},T_{13}]\\
      \text{div}_\Gamma [T_{21},T_{22},T_{23}]\\
      \text{div}_\Gamma [T_{31},T_{32},T_{33}]
     \end{bmatrix}$}\\
        \midrule
        Weingarten map &
        \multicolumn{2}{c}{$\mat{H} = \nabla_\Gamma^{\text{dir}} \vek{n} = \nabla_\Gamma^{\text{cov}} \vek{n}$}\\
        \midrule
        Mean curvature &
        \multicolumn{2}{c}{$\varkappa = \text{tr}(\mat{H})$}\\
        \bottomrule
    \end{tabularx}
    \caption{Geometric quantities and surface differential operators for implicitly defined beams and shells.}
    \label{Tab:GeometricQuantities&DifferentialOperators}
\end{table}

\begin{figure}[ht!]
    \centering
    \raisebox{-0.5\height}{%
      \subfigure[vector fields $\vek{m}$, $\vek{n}$, and $\vek{q}$ in $\mathbb{R}^2$]{%
        \includegraphics[width=0.45\textwidth]{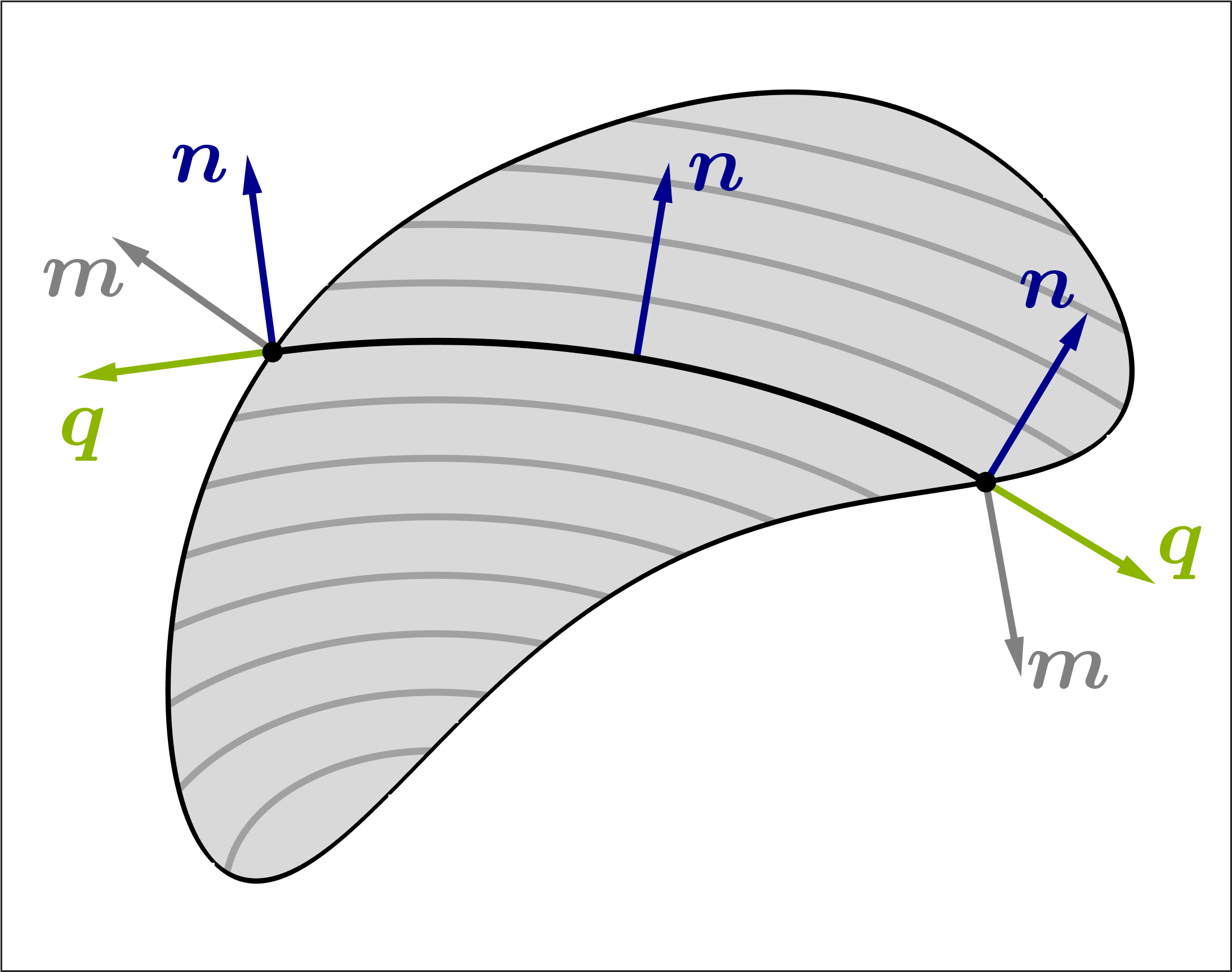}%
        \label{Fig:VectorFields2d}
      }%
    }
    \hspace{0.03\textwidth}
    \raisebox{-0.5\height}{%
      \subfigure[vector fields $\vek{m}$, $\vek{n}$, $\vek{q}$, and $\vek{t}$ in $\mathbb{R}^3$]{%
        \includegraphics[width=0.45\textwidth]{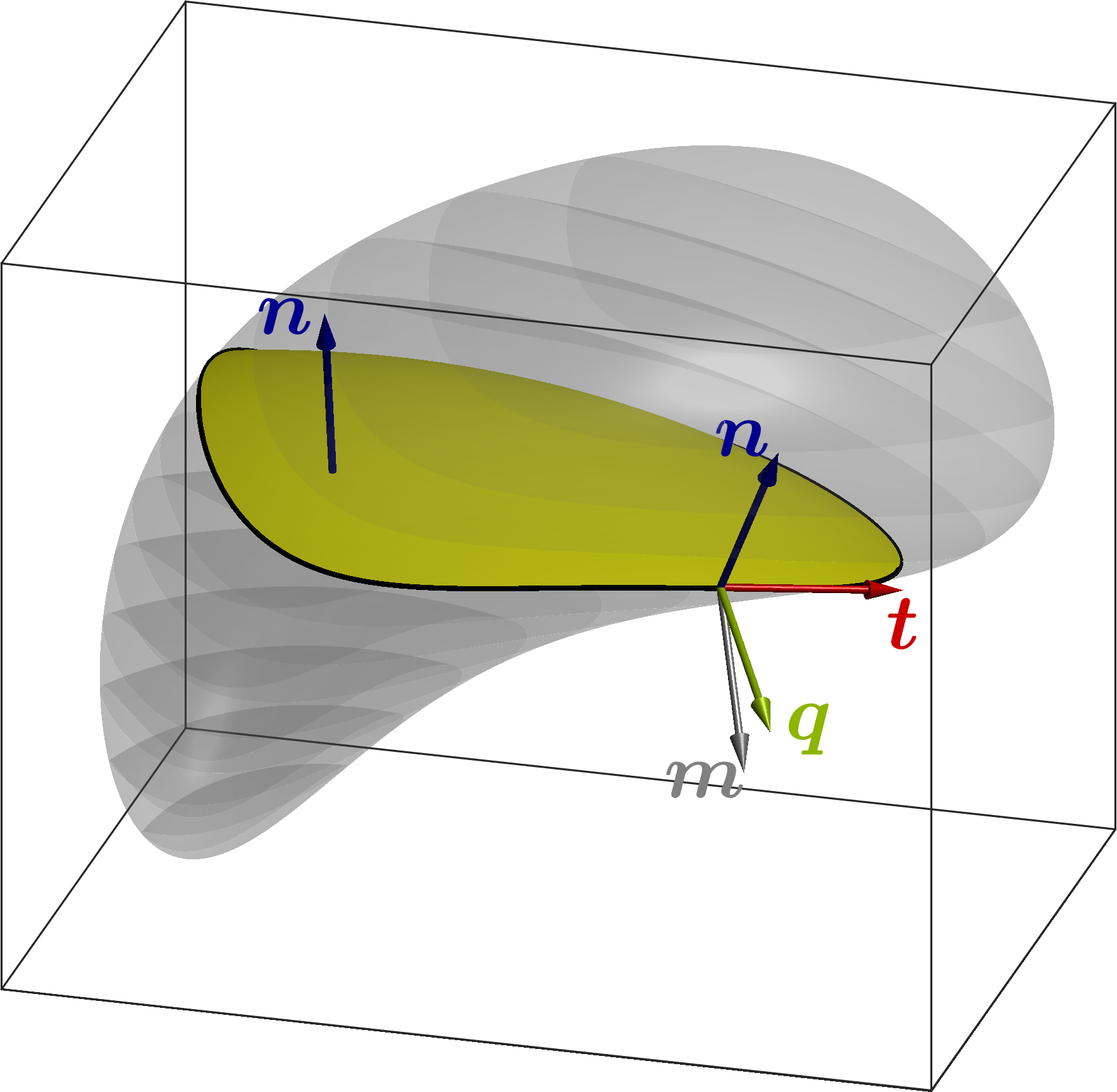}%
        \label{Fig:VectorFields3d}
      }%
    }
    \caption{\label{Fig:VectorFields}Vector fields on a selected manifold $\Gamma^c$ and its boundary $\partial\Gamma^c$. The normal vector $\vek{m}$ with respect to $\partial\Omega$ is shown in gray, the normal vector $\vek{n}$ with respect to $\Gamma^c$ in blue, the conormal vector $\vek{q}$ in green, and the tangent vector $\vek{t}$ in red.}
\end{figure}

\subsection{Integral theorems}
Integral theorems are an essential part of deriving weak formulations of (mechanical) BVPs as needed in any FEM-procedure, starting from a strong formulation based on classical PDEs. In the context of the Bulk Trace FEM, the required integral theorems are a combination of \emph{coarea formulas} formulated over bulk domains and classical integral theorems relevant for \emph{individual} manifolds. First, the formulation of a family of manifolds must be converted from a double integral over each $\Gamma^c$ or $\partial\Gamma^c$ and the interval $[\phi^{\text{min}}, \phi^{\text{max}}]$ to a single integral over $\Omega$ or $\partial\Omega$. This operation can be performed using the \emph{coarea formula} \cite{Federer_1969a, Morgan_1988a, Delfour_1995a, Dziuk_2008a, Dziuk_2013a, Fries_2023a, Kaiser_2024a}, defined as
\begin{align}
\label{Eq:CoareaFormula}
    \int_{\phi^{\text{min}}}^{\phi^{\text{max}}} \int_{\Gamma^c} f \, \text{d}\Gamma  &= \int_\Omega f \, \|\nabla\phi\| \, \text{d}\Omega,\\
\label{Eq:CoareaFormulaBound}
    \int_{\phi^{\text{min}}}^{\phi^{\text{max}}} \int_{\partial\Gamma^c} g \, \vek{q}  \,\, \text{d}\partial\Gamma \, \text{d}c &= \int_{\partial\Omega} g \, \vek{q} \, \big(\vek{q} \ScalProd \vek{m}\big) \, \|\nabla\phi\| \, \text{d}\Omega.
\end{align}
Second, the coarea formula is combined with integral theorems \emph{for a single manifold} \cite{Delfour_1996a, Fries_2018a, Schoellhammer_2019a}, resulting in a divergence theorem \emph{for all level sets over a bulk domain}, as laid out in \cite{Burger_2009a, Fries_2023a, Kaiser_2024a}. Consequently, the integral theorem for a set of manifolds $\Gamma^c$ in a bulk domain $\Omega$, considering some scalar function $f:\Omega\rightarrow\mathbb{R}$ and vector function $\vek{v}:\Omega\rightarrow\mathbb{R}^{d}$, becomes
\begin{equation}\label{Eq:IntegralTheorem1}
    \begin{split}
        \int_\Omega f \, \text{div}_\Gamma \vek{v} \, \|\nabla\phi\| \, \text{d}\Omega =& - \int_\Omega \nabla_\Gamma f \ScalProd \vek{v} \, \|\nabla\phi\| \, \text{d}\Omega + \int_\Omega \varkappa \, f \, \big(\vek{v} \ScalProd \vek{n}\big) \, \|\nabla\phi\| \, \text{d}\Omega \\
        &+ \int_{\partial\Omega} f \, \big(\vek{v} \ScalProd \vek{q}\big) \, \big(\vek{q} \ScalProd \vek{m}\big) \, \|\nabla\phi\| \, \text{d}\partial\Omega.
    \end{split}
\end{equation}
Adapting Eq.~(\ref{Eq:IntegralTheorem1}), the integral theorem for some vector function $\vek{v}$ and tensor function $\mat{T}:\Omega\rightarrow\mathbb{R}^{d\times d}$ is
\begin{equation}\label{Eq:IntegralTheorem2}
    \begin{split}
       \int_\Omega \vek{v} \ScalProd \text{div}_\Gamma \mat{T} \, \|\nabla\phi\| \, \text{d}\Omega =& - \int_\Omega \nabla_\Gamma^{\text{dir}} \vek{v} \FrobProd \mat{T} \, \|\nabla\phi\| \, \text{d}\Omega + \int_\Omega \varkappa \, \vek{v} \ScalProd \big(\mat{T} \cdot \vek{n}\big) \, \|\nabla\phi\| \, \text{d}\Omega\\
       &+ \int_{\partial\Omega} \vek{v} \ScalProd \big(\mat{T} \cdot \vek{q}\big) \, \big(\vek{q} \ScalProd \vek{m}\big) \, \|\nabla\phi\| \, \text{d}\partial\Omega.
    \end{split}
\end{equation}
Note that $\vek{a} \ScalProd \vek{b} = \vek{a}^{\text{T}} \cdot \vek{b}$ refers to the inner product of two vectors and $\mat{A} \FrobProd \mat{B} = \text{tr}\big(\mat{A} \cdot \mat{B}^{\text{T}}\big)$ refers to the Frobenius inner product of two tensors.

\section{Mechanical model of shear-rigid beams and shells}\label{Sec:MechanicalModel}
Next, the mechanical models for curved Kirchhoff beams and Kirchhoff-Love shells are presented. Both models neglect shear deformations. Small deformations are considered, that is, the geometrically linear case. The governing equations are given in a coordinate-free formulation following our previous works in \cite{Schoellhammer_2019a, Kaiser_2023a} where the displacement-based formulations of curved beams and shells are derived. It is then well-known that the resulting BVP involves forth-order derivatives in the strong form and second-order derivatives in the weak form, thus, requiring $C^1$-continuous shape functions in the numerical analyses. Those are naturally provided in isogeometric analysis, however, not in the context of classical FEM which relies on $C^0$-continuous shape functions as, e.g., provided by Lagrange elements. In order to still enable a classical FEM-framework for the analysis, in \cite{Neumeyer_2025a}, we proposed a mixed-hybrid approach for single shells where not only the displacement vector $\vek{u}$, but also the moment tensor $\mat{m}_\Gamma$ serves as a primary unknown. Additional degrees of freedom are statically condensed on the element-level resulting in displacements and tangential rotations as unknowns. This model is summarized below and adapted to the proposed context of the simultaneous solution of beams and shells on all level sets within a bulk domain.

\subsection{Governing equations}\label{Sec:FieldEquations}
The field equations of the strong forms are derived based on kinematics, constitutive relations, and equilibrium. For a purely displacement-based approach, the strong form of a shear-rigid beam or shell results in a single field equation with forth-order derivatives as seen in \cite{Kaiser_2023a, Schoellhammer_2019a}. However, the \emph{mixed} strong form with the two unknown fields $\vek{u}$ and $\mat{m}_\Gamma$ is given by \emph{two} field equations,
\begin{gather}
    \label{Eq:StrongForm1}
 -\vek{\varepsilon}_{\Gamma,\text{Bend}}(\mat{m}_\Gamma) + \vek{\varepsilon}_{\Gamma,\text{Bend}}(\vek{u}) = \mat{0},\\
    \label{Eq:StrongForm2}
    \vek{n} \, \text{div}_\Gamma \big(\mat{P} \cdot \text{div}_\Gamma \mat{m}_\Gamma\big)
    + \mat{H} \cdot \text{div}_\Gamma \mat{m}_\Gamma
    + \text{div}_\Gamma \big(\mat{H} \cdot \mat{m}_\Gamma\big)
    + \text{div}_\Gamma \tilde{\mat{n}}_\Gamma(\vek{u})
    = -\vek{f}.
\end{gather}
It is important to note that these two equations only feature second-order derivatives.

\begin{table}[ht!]
    \centering
    \begin{tabularx}{\textwidth}{L{0.18\textwidth}XX}
        \toprule
         &
        Kirchhoff beams in $\mathbb{R}^2$ &
        Kirchhoff--Love shells in $\mathbb{R}^3$\\
        \midrule
        Linear strain tensor &
        \multicolumn{2}{c}{$\vek{\varepsilon}_\Gamma = \vek{\varepsilon}_{\Gamma,\text{Memb}} + \zeta \, \vek{\varepsilon}_{\Gamma,\text{Bend}}$}\\
        \midrule
        Membrane strain tensor &
        \multicolumn{2}{c}{$\vek{\varepsilon}_{\Gamma,\text{Memb}}(\vek{u}) = \tfrac{1}{2} \, \big[\nabla^\text{cov}_\Gamma\vek{u} + (\nabla^\text{cov}_\Gamma\vek{u})^\text{T}\big]$}\\
        \midrule
        Bending strain tensor &
        \multicolumn{2}{c}{$\vek{\varepsilon}_{\Gamma,\text{Bend}}(\vek{u}) = - \sum_{i=1}^d \nabla^\text{cov}_\Gamma (\nabla_\Gamma u_i) \, n_i$}\\
        \midrule
        Eff. normal force tensor &
        \multicolumn{1}{c}{$\tilde{\mat{n}}_\Gamma(\vek{\varepsilon}_{\Gamma,\text{Memb}}) = E\,A\,\vek{\varepsilon}_{\Gamma,\text{Memb}}$} &
        \multicolumn{1}{c}{$\begin{aligned}
            &\tilde{\mat{n}}_\Gamma(\vek{\varepsilon}_{\Gamma,\text{Memb}}) = t \, \mat{C}_\Gamma : \vek{\varepsilon}_{\Gamma,\text{Memb}}\\
            =& t \, \big[2 \, \mu \, \vek{\varepsilon}_{\Gamma,\text{Memb}} + \lambda \, \text{tr}(\vek{\varepsilon}_{\Gamma,\text{Memb}}) \, \mat{P}\big]
        \end{aligned}$}\\
        \midrule
        Moment tensor &
        \multicolumn{1}{c}{$\mat{m}_\Gamma(\vek{\varepsilon}_{\Gamma,\text{Bend}}) = E\,I\,\vek{\varepsilon}_{\Gamma,\text{Bend}}$} &
        \multicolumn{1}{c}{$\begin{aligned}
            &\mat{m}_\Gamma(\vek{\varepsilon}_{\Gamma,\text{Bend}}) = \tfrac{t^3}{12} \, \mat{C}_\Gamma : \vek{\varepsilon}_{\Gamma,\text{Bend}}\\
            =& \tfrac{t^3}{12} \, \big[2 \, \mu \, \vek{\varepsilon}_{\Gamma,\text{Bend}} + \lambda \, \text{tr}(\vek{\varepsilon}_{\Gamma,\text{Bend}}) \, \mat{P}\big]
        \end{aligned}$}\\
        \midrule
        Physical normal force tensor &
        \multicolumn{2}{c}{$\mat{n}_\Gamma^\text{real} = \tilde{\mat{n}}_\Gamma + \mat{H} \cdot \mat{m}_\Gamma$}\\
        \midrule
        Inverse material law &
        \multicolumn{1}{c}{$\vek{\varepsilon}_{\Gamma,\text{Bend}}(\mat{m}_\Gamma) = \tfrac{1}{E\,I}\,\mat{m}_\Gamma$} &
        \multicolumn{1}{c}{$\begin{aligned}
            &\vek{\varepsilon}_{\Gamma,\text{Bend}}(\mat{m}_\Gamma) = \tfrac{12}{t^3} \, \mat{C}_\Gamma^{-1} : \mat{m}_\Gamma\\
            =& \tfrac{12}{E \, t^3} \, \big[\big(1+\nu\big) \, \mat{m}_\Gamma - \nu \, \text{tr}(\mat{m}_\Gamma) \, \mat{P}\big]
        \end{aligned}$}\\
        \bottomrule
    \end{tabularx}
    \caption{Mechanical quantities of Kirchhoff beams and Kirchhoff--Love shells.}
    \label{Tab:MechanicalQuantities}
\end{table}

For the sake of brevity, all relevant mechanical quantities are summarized in Tab.~\ref{Tab:MechanicalQuantities}. The \emph{kinematics} is defined by a linear strain tensor $\vek{\varepsilon}_\Gamma$ depending on $\vek{u}$, which can be split into a membrane and bending part, $\vek{\varepsilon}_{\Gamma,\text{Memb}}$ and $\vek{\varepsilon}_{\Gamma,\text{Bend}}$, respectively. All three tensors are symmetric and live in the tangent space of the beam or shell.

The \emph{constitutive relations} establish a connection between the strain and stress quantities. Here, the effective normal force tensor $\tilde{\mat{n}}_\Gamma$ and the moment tensor $\mat{m}_\Gamma$ depend on $\vek{\varepsilon}_{\Gamma,\text{Memb}}$ and $\vek{\varepsilon}_{\Gamma,\text{Bend}}$, respectively. Relevant material parameters are Young's modulus $E$, Poisson's ratio $\nu$, and the Lam\'e constants $\mu = \frac{E}{2(1+\nu)}$ and $\lambda = \frac{E\cdot\nu}{1-\nu^2}$. Geometric quantities of the cross section are the area $A$, the area moment of inertia $I$, and the thickness $t$. $\tilde{\mat{n}}_\Gamma$ shall not be mistaken for the physical normal force tensor $\mat{n}_\Gamma^\text{real}$. All three tensors are in-plane, but only $\tilde{\mat{n}}_\Gamma$ and $\mat{m}_\Gamma$ are symmetric in general.

Inverting the material law of the bending part leads to a definition of $\vek{\varepsilon}_{\Gamma,\text{Bend}}$ depending on $\mat{m}_\Gamma$. Combining this relation with the definition of $\vek{\varepsilon}_{\Gamma,\text{Bend}}$ from the kinematics leads to the first equation (\ref{Eq:StrongForm1}) in alignment with the Hellinger--Reissner principle. The second equation (\ref{Eq:StrongForm2}) represents the equilibrium of forces with $\vek{f}$ being the body load vector.

\subsection{Boundary conditions}\label{Sec:BoundaryConditions}
\begin{figure}[ht!]
    \centering
    \raisebox{-0.5\height}{%
      \subfigure[boundary conditions for a curved beam in $\mathbb{R}^2$]{%
        \includegraphics[width=0.45\textwidth]{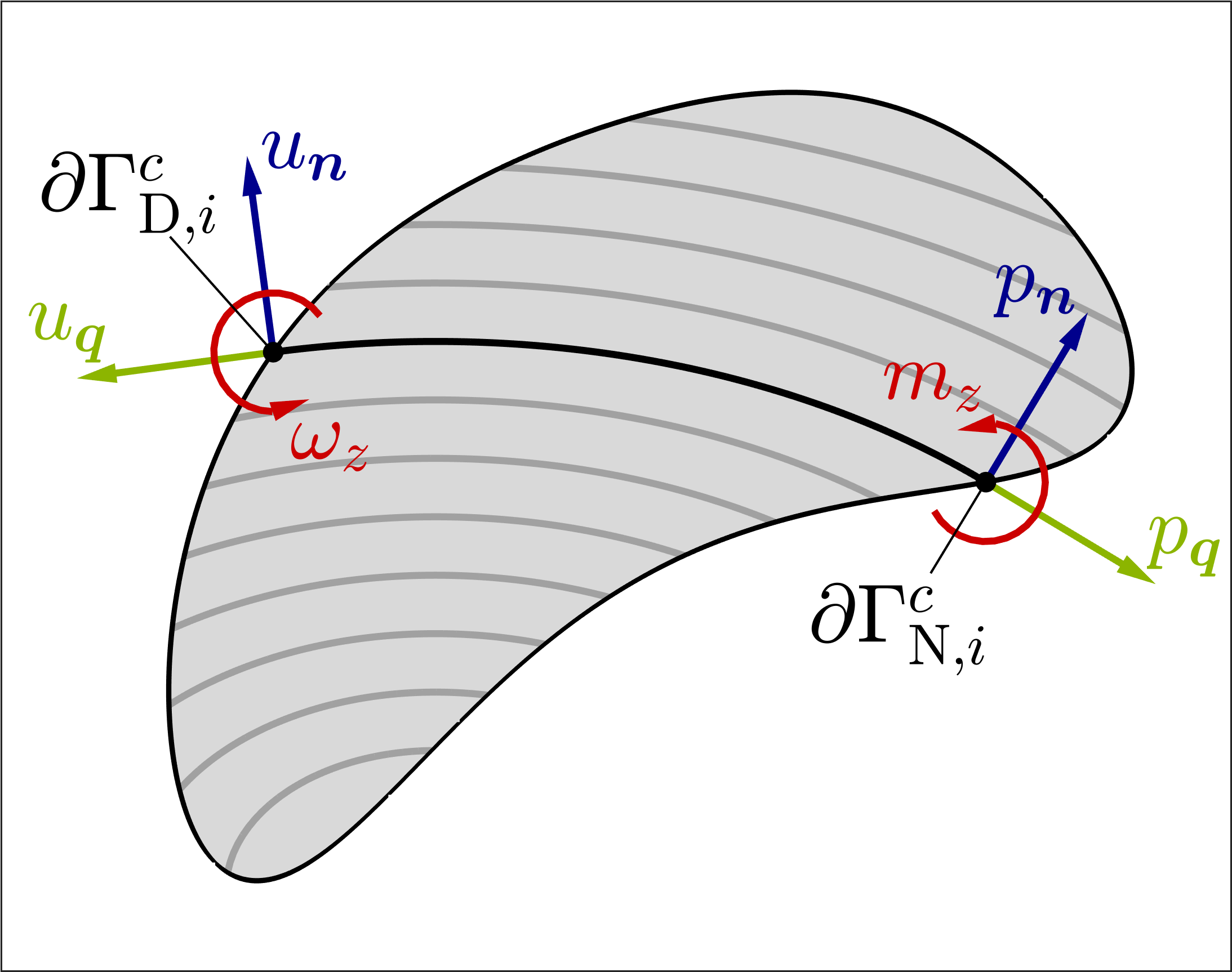}%
        \label{Fig:BoundaryConditions2d}
      }%
    }
    \hspace{0.03\textwidth}
    \raisebox{-0.5\height}{%
      \subfigure[boundary conditions for a shell in $\mathbb{R}^3$]{%
        \includegraphics[width=0.45\textwidth]{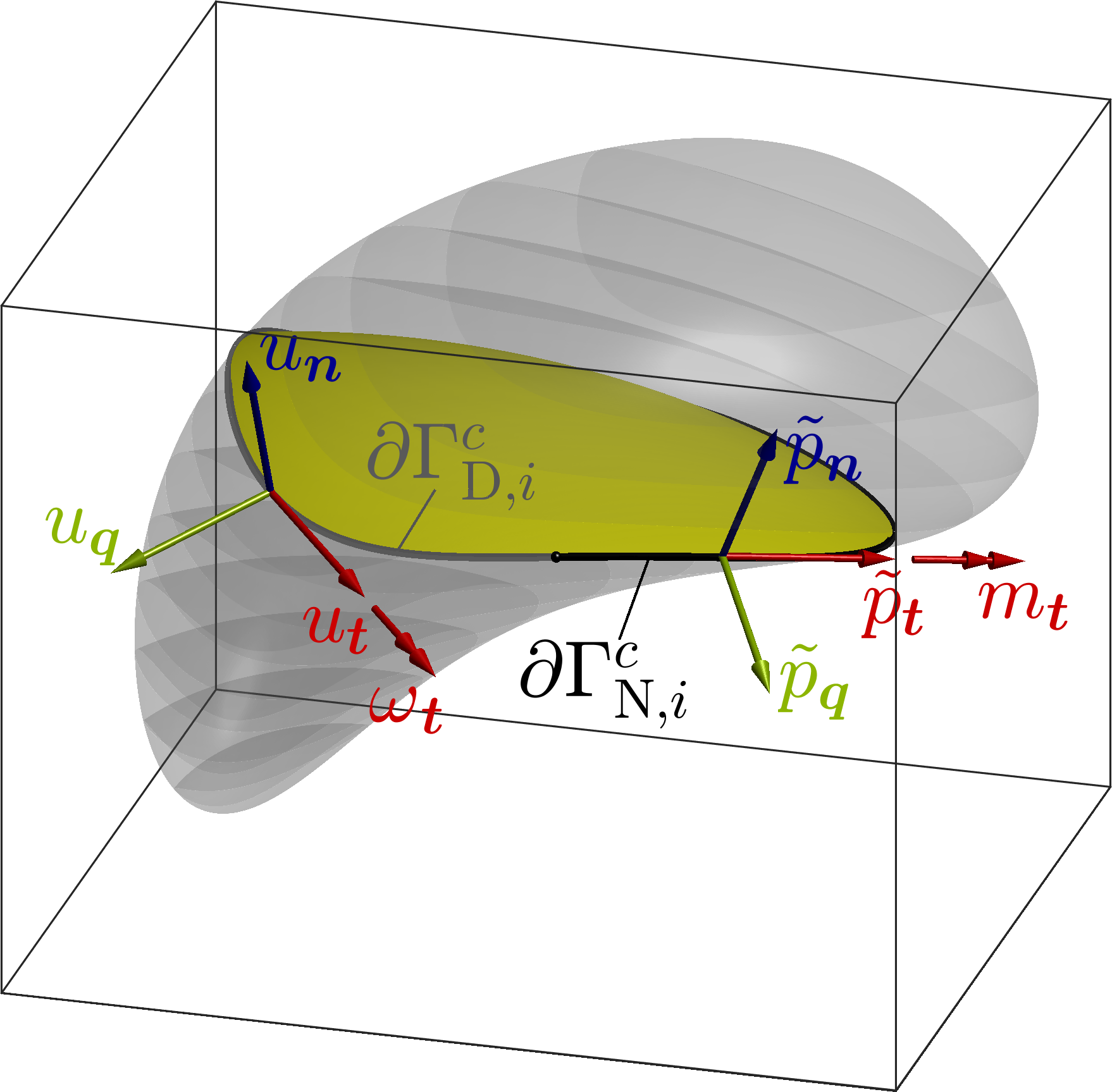}%
        \label{Fig:BoundaryConditions3d}
      }%
    }
    \caption{\label{Fig:BoundaryConditions}Boundary quantities on a selected boundary $\partial\Gamma^c$. Displacements and force components are split in terms of (a) $\vek{n}$ and $\vek{q}$ for beams  or (b) $\vek{n}$, $\vek{q}$, and $\vek{t}$ for shells. Rotations and moments act around the $z$-axis for beams or around $\vek{t}$ for shells. A distinction is made between the Dirichlet boundaries $\partial\Gamma_{\text{D}, i}^c$ and Neumann boundaries $\partial\Gamma_{\text{N}, i}^c$.}
\end{figure}

Boundary conditions are an essential part of a BVP and require special attention in the case of shear-rigid structures. For the displacement fields, there exist two non-overlapping sections of the boundary $\partial\Gamma^c$ of a single shell or beam, i.e., $\partial\Gamma_{\text{D}, i}^c \cup \partial\Gamma_{\text{N}, i}^c = \partial\Gamma^c$ and $\partial\Gamma_{\text{D}, i}^c \cap \partial\Gamma_{\text{N}, i}^c = \emptyset$. Those fields are the translational displacements $\vek{u}$ and, in the case of a shell, the rotation around the tangential direction $\omega_{\vek{t}}$. For beams in $\mathbb{R}^2$, the rotational field is the rotation $\omega_z$ around the (virtual) $z$-axis being perpendicular to the plane of the actual two-dimensional space considered. The boundaries of those fields are either the Dirichlet boundaries $\partial\Gamma_{\text{D},\vek{u}}^c$ and $\partial\Gamma_{\text{D},\omega}^c$ or the corresponding Neumann boundaries $\partial\Gamma_{\text{N},\vek{u}}^c$ and $\partial\Gamma_{\text{N},\omega}^c$. The respective boundary conditions and their fundamental components are summarized in Tab.~\ref{Tab:BoundaryConditions}, and a corresponding visualization is seen in Fig.~\ref{Fig:BoundaryConditions}.

\begin{figure}[ht!]
    \centering
        \includegraphics[width=0.45\textwidth]{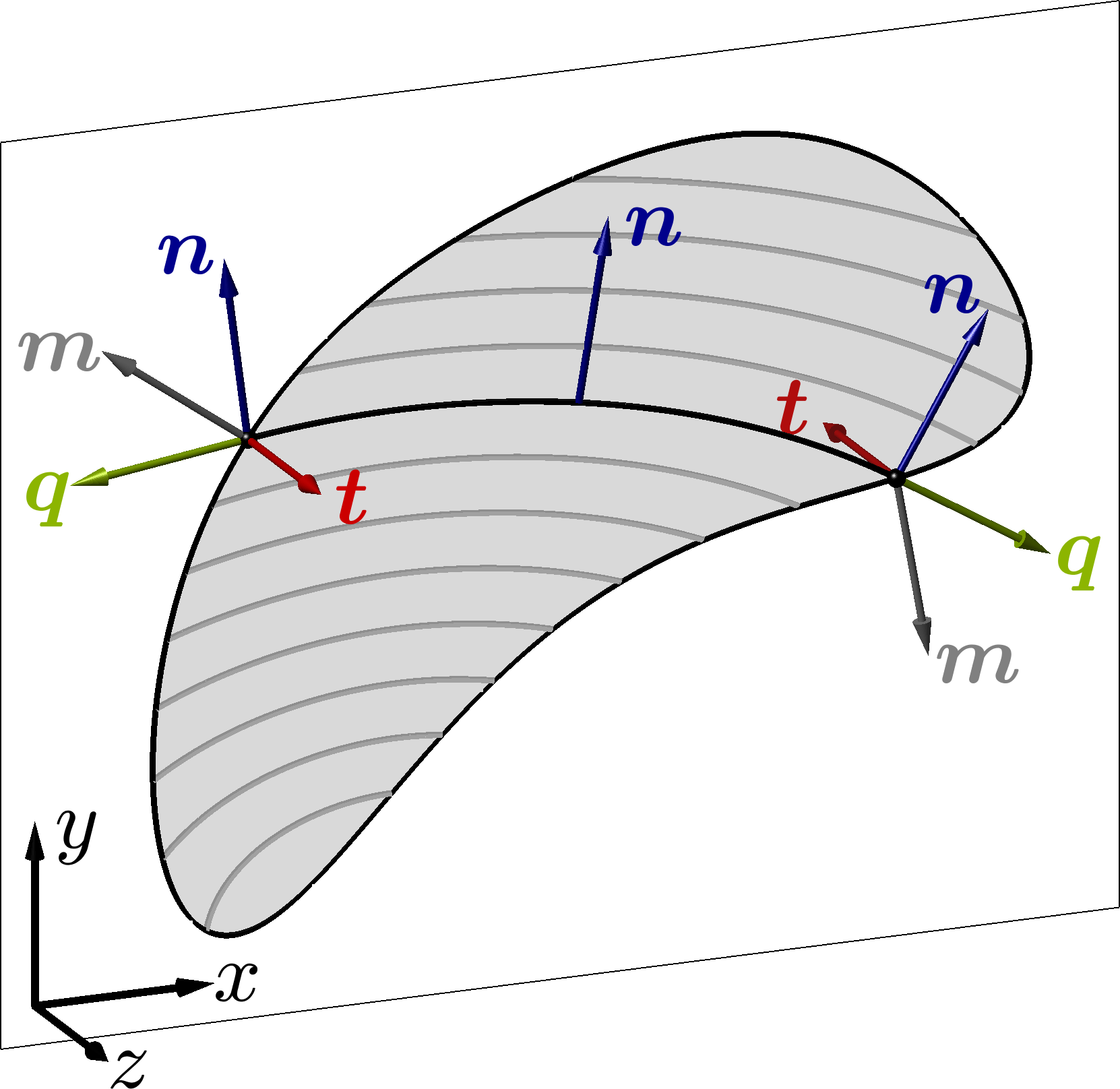}%

    \caption{\label{Fig:VectorFields2dIn3d}A three-dimensional illustration of the two-dimensional plane from Fig.~\ref{Fig:VectorFields2d}. It can be seen that, on both depicted boundaries $\partial\Gamma^c$, $\vek{t}$ is aligned with the $z$-axis, pointing in positive $z$-direction on the left boundary and in negative $z$-direction on the right boundary.}
\end{figure}

\begin{table}[ht!]
    \centering
    \begin{tabularx}{\textwidth}{L{0.26\textwidth}XX}
        \toprule
         &
        Kirchhoff beams in $\mathbb{R}^2$ &
        Kirchhoff--Love shells in $\mathbb{R}^3$\\
        \midrule
        \multirow{2}{\hsize}{Dirichlet BCs} &
        $\vek{u} = \hat{\vek{u}} \:\: \text{on} \:\: \partial\Gamma_{\text{D},\vek{u}}^c$ &
        $\vek{u} = \hat{\vek{u}} \:\: \text{on} \:\: \partial\Gamma_{\text{D},\vek{u}}^c$\\
        &
        $\omega_z = \hat{\omega}_z \:\: \text{on} \:\: \partial\Gamma_{\text{D},\omega}^c$ &
        $\omega_{\vek{t}} = \hat{\omega}_{\vek{t}} \:\: \text{on} \:\: \partial\Gamma_{\text{D},\omega}^c$\\
        \midrule
        \multirow{2}{\hsize}{Neumann BCs} &
        $\vek{p} = \hat{\vek{p}} \:\: \text{on} \:\: \partial\Gamma_{\text{N},\vek{u}}^c$ &
        $\tilde{\vek{p}} = \hat{\tilde{\vek{p}}} \:\: \text{on} \:\: \partial\Gamma_{\text{N},\vek{u}}^c$\\
        &
        $m_z = \hat{m}_z \:\: \text{on} \:\: \partial\Gamma_{\text{N},\omega}^c$ &
        $m_{\vek{t}} = \hat{m}_{\vek{t}} \:\: \text{on} \:\: \partial\Gamma_{\text{N},\omega}^c$\\
        \midrule
        Rotation around $\vek{t}$ &
        \multicolumn{2}{c}{$\omega_{\vek{t}} = -\big[(\nabla^\text{dir}_\Gamma \vek{u})^\text{T} \cdot \vek{n}\big] \ScalProd \vek{q}$}\\
        \midrule
        Rotation around $\vek{q}$ &
        - &
        $\omega_{\vek{q}} = -\big[(\nabla^\text{dir}_\Gamma \vek{u})^\text{T} \cdot \vek{n}\big] \ScalProd \vek{t}$\\
        \midrule
        Rotation around $z$-axis &
        $\omega_z = \omega_{\vek{t}} \, t_z$ &
        -\\
        \midrule
        Bound. force in $\vek{n}$ &
        \multicolumn{2}{c}{$p_{\vek{n}} = \big(\mat{P} \cdot \text{div}_\Gamma \mat{m}_\Gamma\big) \ScalProd \vek{q}$}\\
        \midrule
        Bound. force in $\vek{q}$ &
        \multicolumn{2}{c}{$p_{\vek{q}} = \big(\mat{n}_\Gamma^\text{real} \cdot \vek{q}\big) \ScalProd \vek{q}$}\\
        \midrule
        Bound. force in $\vek{t}$ &
        - &
        $p_{\vek{t}} = \big(\mat{n}_\Gamma^\text{real} \cdot \vek{q}\big) \ScalProd \vek{t}$\\
        \midrule
        Res. bound. force &
        $\vek{p} = p_{\vek{n}} \, \vek{n} + p_{\vek{q}} \, \vek{q}$ &
        $\vek{p} = p_{\vek{n}} \, \vek{n} + p_{\vek{q}} \, \vek{q} + p_{\vek{t}} \, \vek{t}$\\
        \midrule
        Eff. bound. force in $\vek{n}$ &
        - &
        $\tilde{p}_{\vek{n}} = p_{\vek{n}} + \nabla_\Gamma m_{\vek{q}} \ScalProd \vek{t}$\\
        \midrule
        Eff. bound. force in $\vek{q}$ &
        - &
        $\tilde{p}_{\vek{q}} = p_{\vek{q}} + \big(\mat{H} \cdot \vek{t}\big) \ScalProd \vek{q} \, m_{\vek{q}}$\\
        \midrule
        Eff. bound. force in $\vek{t}$ &
        - &
        $\tilde{p}_{\vek{t}} = p_{\vek{t}} + \big(\mat{H} \cdot \vek{t}\big) \ScalProd \vek{t} \, m_{\vek{q}}$\\
        \midrule
        Res. eff. bound. force &
        - &
        $\tilde{\vek{p}} = \tilde{p}_{\vek{n}} \, \vek{n} + \tilde{p}_{\vek{q}} \, \vek{q} + \tilde{p}_{\vek{t}} \, \vek{t}$\\
        \midrule
        Moment around $\vek{t}$ &
        \multicolumn{2}{c}{$m_{\vek{t}} = \big(\mat{m}_\Gamma \cdot \vek{q}\big) \ScalProd \vek{q}$}\\
        \midrule
        Moment around $\vek{q}$ &
        - &
        $m_{\vek{q}} = \big(\mat{m}_\Gamma \cdot \vek{q}\big) \ScalProd \vek{t}$\\
        \midrule
        Moment around $z$-axis &
        $m_z = m_{\vek{t}} \, t_z$ &
        -\\
        \bottomrule
    \end{tabularx}
    \caption{All relevant boundary conditions and related quantities for Kirchhoff beams and Kirchhoff--Love shells. A hat $\hat{\bullet}$ indicates a prescribed quantity at the boundary.}
    \label{Tab:BoundaryConditions}
\end{table}
For beams, rotational boundary conditions, hence prescribed rotations or moments, require special considerations. From a user's perspective, we want to prescribe the rotations and moments globally (around the $z$-direction, pointing out of the two-dimensional plane), but the composition of these quantities is based on the local triad of $\vek{n}$, $\vek{q}$, and $\vek{t}$. In the context of beams, $\vek{t}$ is not a tangent vector but more a second normal vector perpendicular to $\vek{n}$ and the two-dimensional plane in which $\Omega$ lives, hence, it is either $\vek{t} = [0, 0, 1]^\text{T}$ or $\vek{t} = [0, 0, -1]^\text{T}$, depending on which boundary $\vek{t}$ is evaluated, as seen in Fig.~\ref{Fig:VectorFields2dIn3d}. Transformations between global and local quantities may then be realized by multiplying them with $t_z$, the third component of $\vek{t}$.

For the Kirchhoff--Love shell, a direct boundary condition for the rotation around the conormal direction $\omega_{\vek{q}}$ does not exist, as it is already determined for a prescribed $\vek{u}$. There is neither a conjugated boundary moment around the conormal direction $m_{\vek{q}}$, hence, it cannot be prescribed directly via a corresponding Neumann boundary condition. In Kirchhoff--Love plates and shells, this issue is addressed by linking $m_{\vek{q}}$ and the boundary force $\vek{p}$, resulting in the effective boundary force $\tilde{\vek{p}}$. The effect of Kirchhoff corner forces is not further discussed herein as we shall later focus on shells with smooth boundaries only, i.e., without corners. See \cite{Basar_1985a, Neumeyer_2025a} for further information on this topic.

\section{Bulk Trace FEM for shear-rigid beams and shells}\label{Sec:BTFForShearRigidBeams&Shells}
In Section~\ref{Sec:MechanicalModel}, the mechanical model in the strong form of a \emph{single} Kirchhoff beam and Kirchhoff--Love shell are expressed in a mixed formulation, see Eqs.~(\ref{Eq:StrongForm1}) and (\ref{Eq:StrongForm2}). The bending-related part of the equations is formulated in terms of the Hellinger--Reissner principle, resulting in the moment tensor $\mat{m}_\Gamma$ as a primary unknown, in addition to the displacement vector $\vek{u}$. Next, the governing equations are transformed to the weak form and extended to simultaneously apply for all beams/shells as implied by the level sets within some bulk domain. Then, the resulting numerical analysis based on the Bulk Trace FEM is outlined.

\subsection{Mixed continuous function spaces}\label{Sec:MixedContinuousFunctionSpaces}
Suitable trial and test function spaces for the displacements $\vek{u}$ and moment tensor $\mat{m}_\Gamma$ are given as 
\begin{gather}
    \label{Eq:FunctionSpaceContMoment}
        \mathcal{S}_{\mat{m}_\Gamma} = \mathcal{V}_{\mat{m}_\Gamma} =
     \begin{Bmatrix}
      \mat{V} \in \mathcal{H}(\text{div}, \Omega): \mat{V} = \mat{V}^\text{T}
     \end{Bmatrix}, \\
    \label{Eq:FunctionSpaceContDisplacementTrial}
        \mathcal{S}_{\vek{u}} =
     \begin{Bmatrix}
      \vek{v} \in [\mathcal{H}^1(\Omega)]^d: \vek{v} = \hat{\vek{u}} \:\: \text{on} \:\: \partial\Omega_{\text{D},\vek{u}}
     \end{Bmatrix},\\
    \label{Eq:FunctionSpaceContDisplacementTest}
     \mathcal{V}_{\vek{u}} =
     \begin{Bmatrix}
      \vek{v} \in [\mathcal{H}^1(\Omega)]^d: \vek{v} = \vek{0} \:\: \text{on} \:\: \partial\Omega_{\text{D},\vek{u}}
     \end{Bmatrix},
\end{gather}
where $\mathcal{H}^1(\Omega) = \{ v \in \mathcal{L}^2(\Omega): \nabla_\Gamma v \in [\mathcal{L}^2(\Omega)]^d \}$ is the Sobolev space, and $\mathcal{H}(\text{div}, \Omega) = \{ \mat{V} \in [\mathcal{L}^2(\Omega)]^{d \times d}: \text{div}_\Gamma \mat{V} \in [\mathcal{L}^2(\Omega)]^d \}$ with $\mathcal{L}^2$ being the Lebesgue space. Furthermore, the boundary condition
\begin{equation}\label{Eq:LagrangeMultiplier}
    m_{\vek{t}} = \hat{m}_{\vek{t}} \:\: \text{on} \:\: \partial\Gamma_{\text{N},\omega}^c.
\end{equation}
is enforced by means of a Lagrange multiplier which may mechanically be interpreted as the rotation $\omega_{\vek{t}}$. For the case of Kirchhoff \emph{beams}, the relationship $\hat{\omega}_{\vek{t}} = \hat{\omega}_z \, t_z$ and $\hat{m}_{\vek{t}} = \hat{m}_z \, t_z$ is used. This approach is especially convenient, as the Lagrange multiplier $\omega_{\vek{t}}$ is introduced anyway for the hybridization of the formulation later on. The corresponding function spaces are introduced as
\begin{gather}
    \label{Eq:FunctionSpaceContRotationTrial}
     \mathcal{S}_{\omega_{\vek{t}}} =
     \begin{Bmatrix}
      v \in \mathcal{L}^2(\partial\Omega): v = \hat{\omega}_{\vek{t}} \:\: \text{on} \:\: \partial\Omega_{\text{D},\omega}
     \end{Bmatrix},\\
     \label{Eq:FunctionSpaceContRotationTest}
     \mathcal{V}_{\omega_{\vek{t}}} =
     \begin{Bmatrix}
      v \in \mathcal{L}^2(\partial\Omega): v = 0 \:\: \text{on} \:\: \partial\Omega_{\text{D},\omega}
     \end{Bmatrix}.
\end{gather}

To obtain the continuous weak form of the Kirchhoff beam or the Kirchhoff--Love shell, we multiply the equations in strong form from Eqs.~(\ref{Eq:StrongForm1}), (\ref{Eq:StrongForm2}), and (\ref{Eq:LagrangeMultiplier}) with their corresponding test functions, respectively, and integrate over the level sets $\Gamma^c$ within the interval $[\phi^{\text{min}}, \phi^{\text{max}}]$. Using the coarea formula, i.e., Eq.~(\ref{Eq:CoareaFormula}), this double integral is converted to an integral over the bulk domain $\Omega$. Lastly, the integral theorems of Eqs.~(\ref{Eq:IntegralTheorem1}) and (\ref{Eq:IntegralTheorem2}) are applied. The task is then to find $\mat{m}_\Gamma \in \mathcal{S}_{\mat{m}_\Gamma}$, $\vek{u} \in \mathcal{S}_{\vek{u}}$, and $\omega_{\vek{t}} \in \mathcal{S}_{\omega_{\vek{t}}}$ such that this weak form is valid for all $(\mat{V}_{\mat{m}_\Gamma}, \vek{v}_{\vek{u}}, v_{\omega_{\vek{t}}}) \in \mathcal{V}_{\mat{m}_\Gamma} \times \mathcal{V}_{\vek{u}} \times \mathcal{V}_{\omega_{\vek{t}}}$. For brevity, we shall write down the weak form \emph{after} the discretization below.

\subsection{Mixed-hybrid discrete function spaces}\label{Sec:MixedHybridDiscretizedFunctionSpaces}
The advantage of the mixed formulation as described in Section \ref{Sec:MixedContinuousFunctionSpaces} is that $C^0$-continuous shape functions may be employed in the numerical analysis, i.e., classical FEM software is readily applicable. However, the mixed formulation results in significantly more  degrees of freedom in the FEM analysis compared to purely displacement-based formulations that require $C^1$-continuous shape functions in the analysis. In order to mitigate this problem, after the discretization, a \emph{hybridization} of the formulation is performed, following \cite{Arnold_1985a, Boffi_2013a, Cockburn_2004a, Cockburn_2009a, Kabaria_2015a}. That is, the continuity of the moment tensor field is broken between the interfaces of the elements and then weakly reinforced by the Lagrange multiplier $\omega_{\vek{t}}$ on element boundaries. In this context, $\omega_{\vek{t}}$ may also be referred to as \emph{hybridization variable} and is applied not only on the Neumann boundary $\partial\Omega_{\text{N},\omega}$, but also on the boundaries inbetween all elements.

\begin{figure}[ht!]
 \centering
 
 \subfigure[discretized domain $\Omega^h$ in $\mathbb{R}^2$]
 {\includegraphics[width=0.33\textwidth]{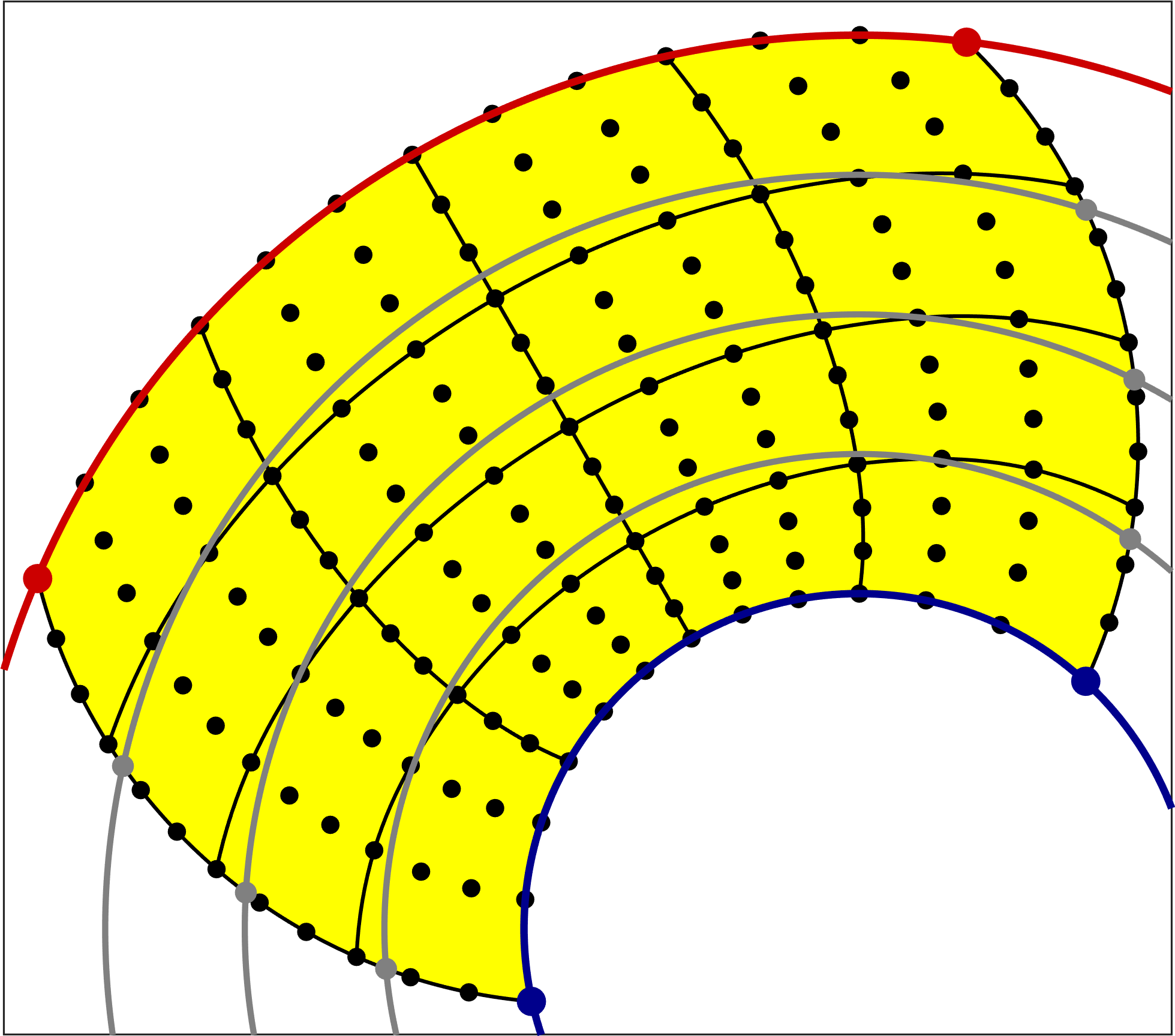}\label{Fig:MeshDomain2d}}\hfill
 \subfigure[discretized boundary $\partial\Omega^h$ in $\mathbb{R}^2$]
 {\includegraphics[width=0.33\textwidth]{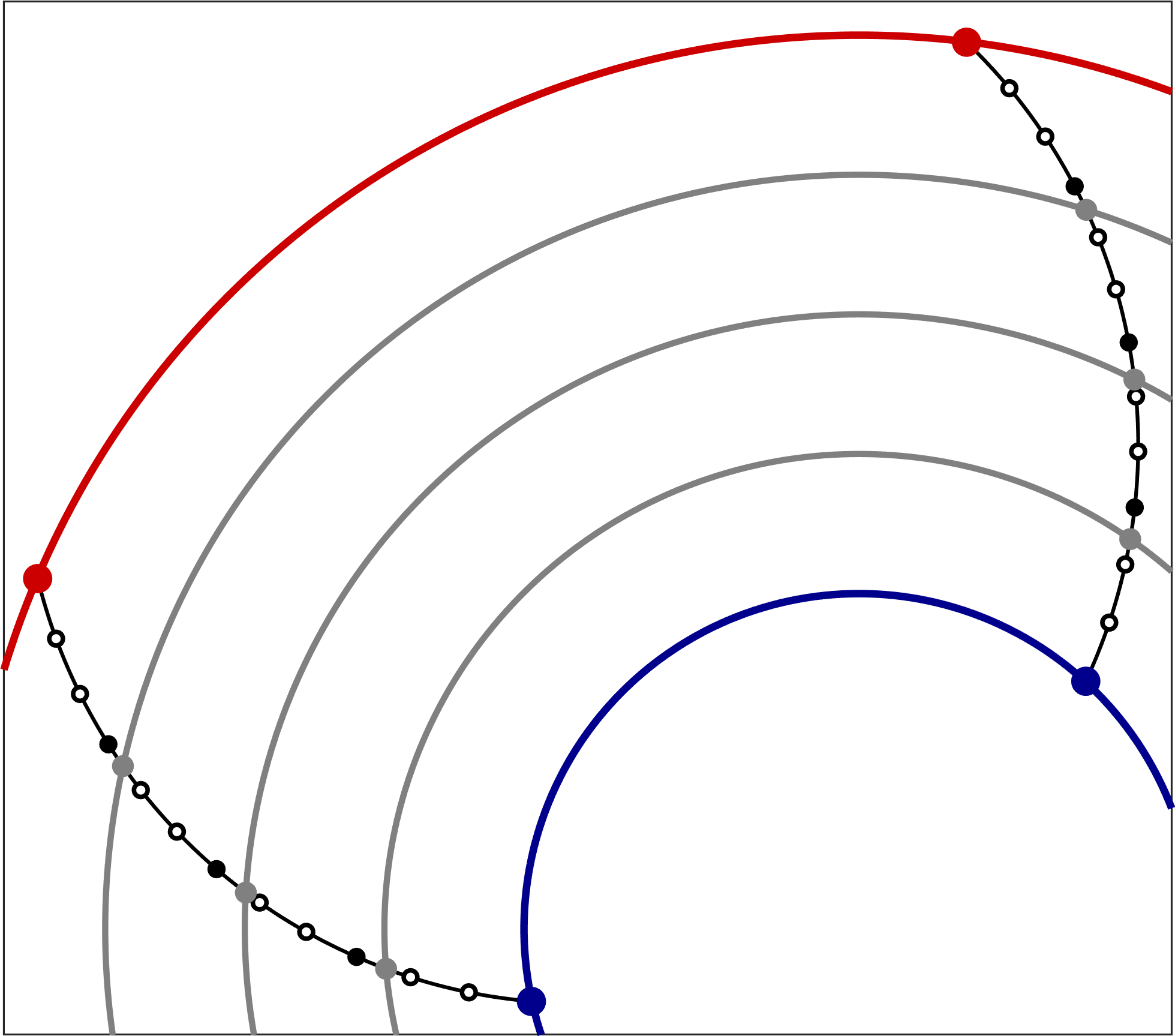}\label{Fig:MeshBoundary2d}}\hfill
    \subfigure[element interfaces $\Psi^h$ in $\mathbb{R}^2$]
 {\includegraphics[width=0.33\textwidth]{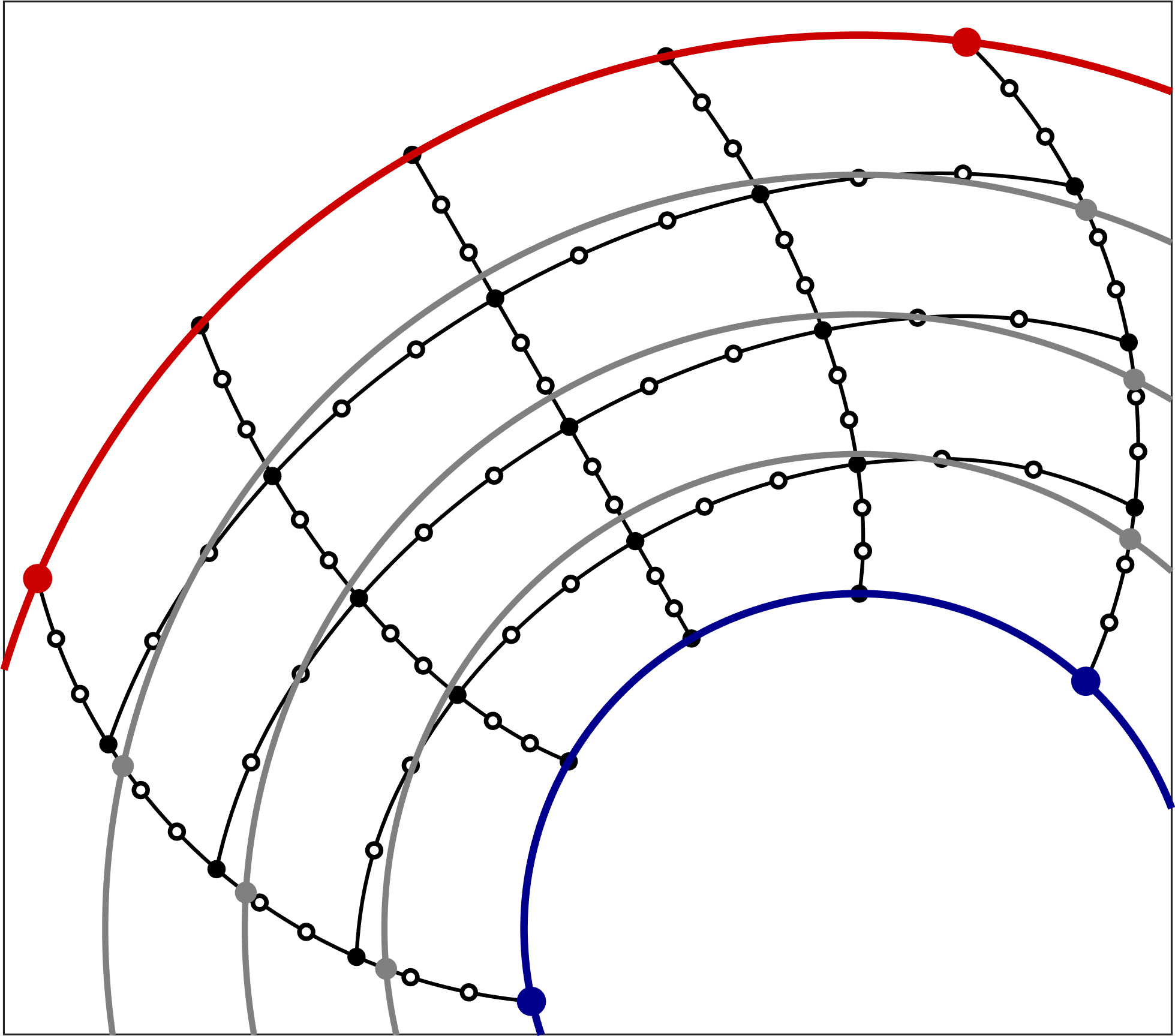}\label{Fig:MeshInterfaces2d}}\\
    \subfigure[discretized domain $\Omega^h$ in $\mathbb{R}^3$]
 {\includegraphics[width=0.33\textwidth]{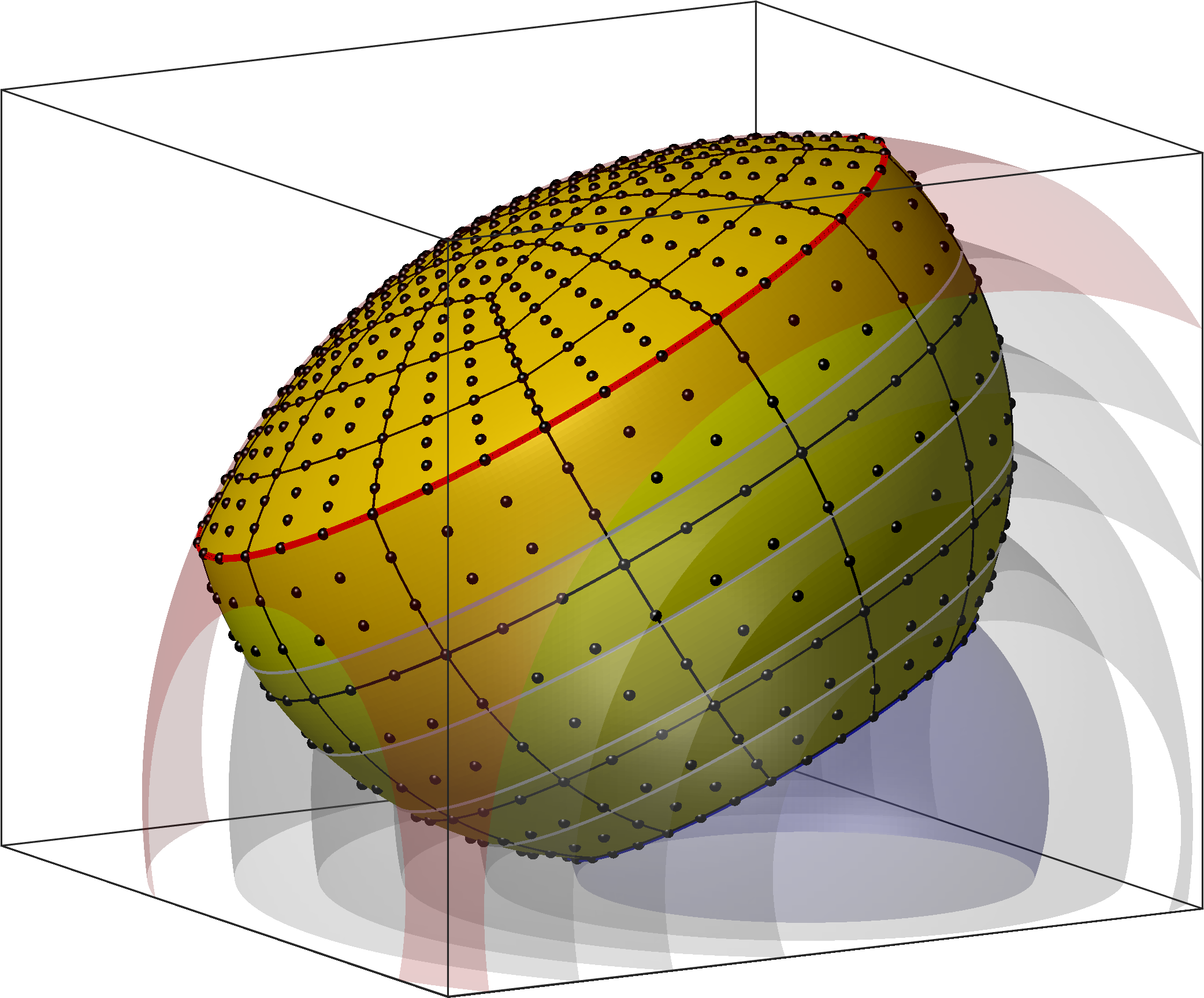}\label{Fig:MeshDomain3d}}\hfill
 \subfigure[discretized boundary $\partial\Omega^h$ in $\mathbb{R}^3$]
 {\includegraphics[width=0.33\textwidth]{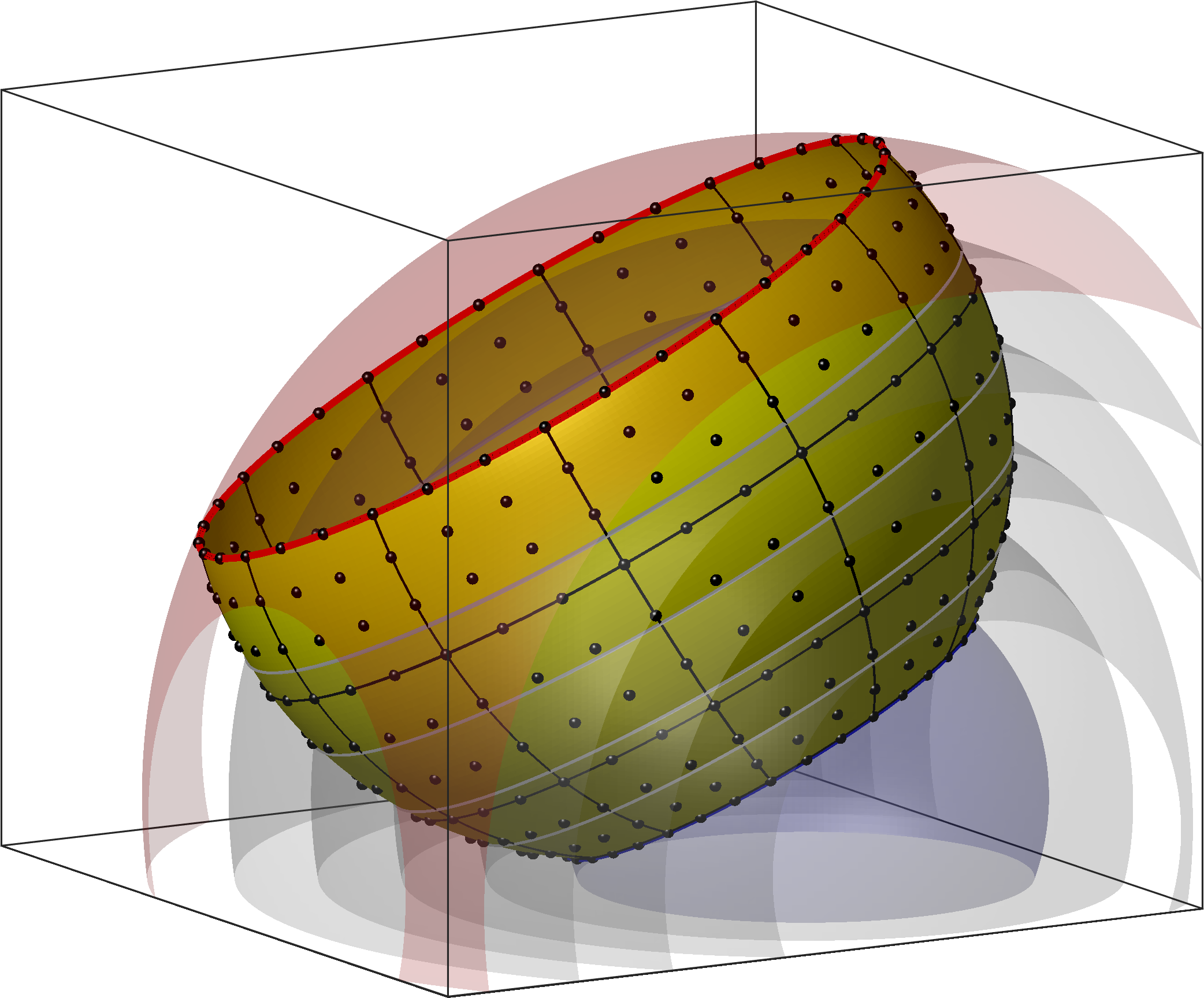}\label{Fig:MeshBoundary3d}}\hfill
    \subfigure[element interfaces $\Psi^h$ in $\mathbb{R}^3$]
 {\includegraphics[width=0.33\textwidth]{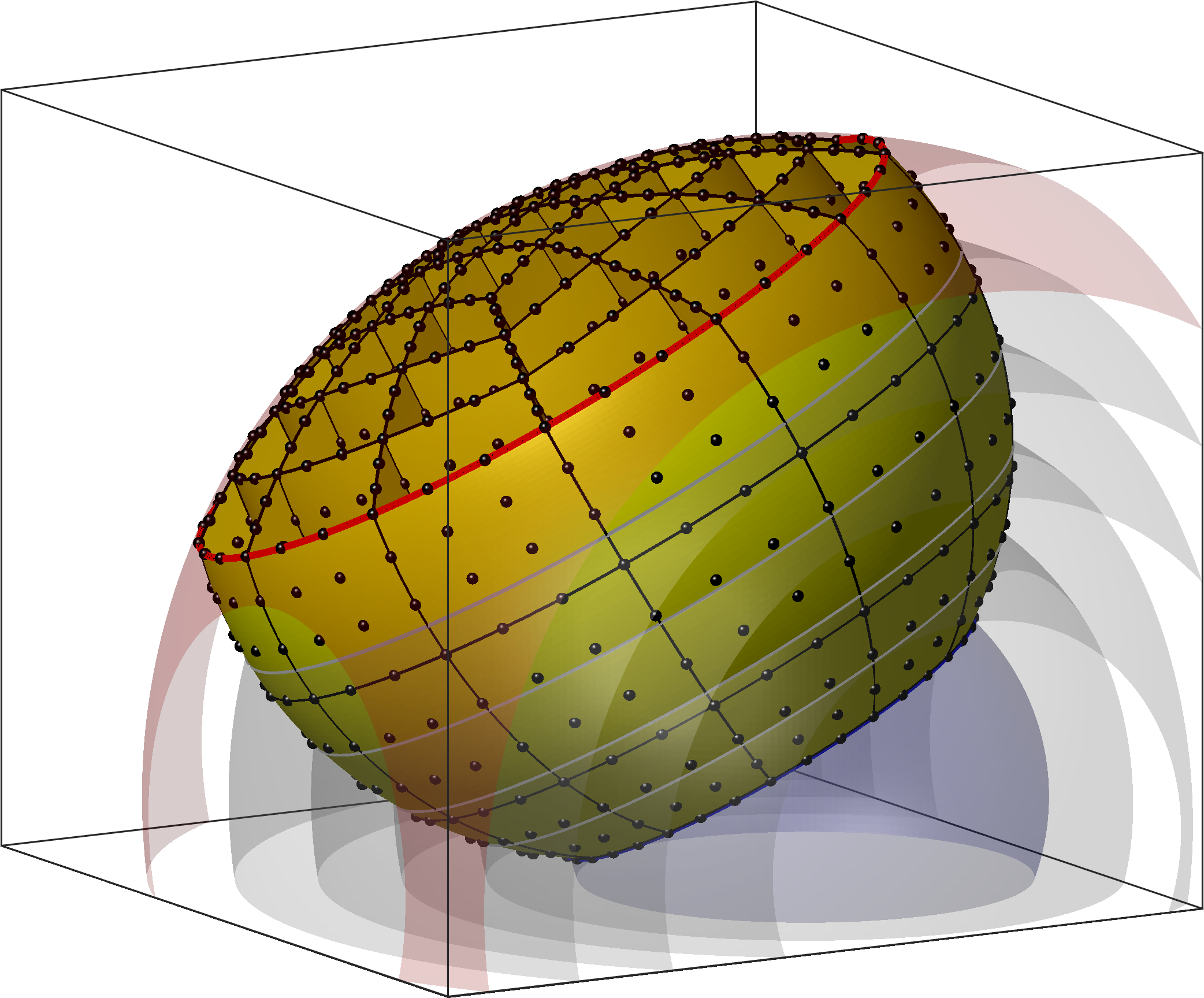}\label{Fig:MeshInterfaces3d}}
 
 \caption{\label{Fig:Meshes}For some bulk domain $\Omega$ in $\mathbb{R}^2$ and $\mathbb{R}^3$, (a) and (d) shows an example mesh $\Omega^h$ discretized by cubic elements, (b) and (e) the discretized boundary $\partial\Omega^h$, and (c) and (f) the interface mesh $\Psi^h$. The lower limit of the level set $\phi^{\text{min}}$ is depicted in blue, the upper limit of the level set $\phi^{\text{max}}$ in red, and some selected level sets of $\phi$ within these limits in gray.}
\end{figure}

Next, different meshes are introduced that imply the shape functions for the analysis. Let there be a conforming discretization of the bulk domain $\Omega^h$ and its boundary $\partial\Omega^h$. For a beam in $\mathbb{R}^2$, $\Omega^h$ consists of two-dimensional triangular or quadrilateral Lagrange elements, and its boundary $\partial\Omega^h$ of curved one-dimensional Lagrange (line-)elements. For a shell in $\mathbb{R}^3$, $\Omega^h$ is composed by three-dimensional tetrahedral or hexahedral Lagrange elements, the boundary $\partial\Omega^h$ is then discretized by (curved) two-dimensional triangular or quadrilateral Lagrange elements. An illustration of $\Omega^h$ in $\mathbb{R}^2$ and $\mathbb{R}^3$ can be seen in Figs.~\ref{Fig:MeshDomain2d} and \ref{Fig:MeshDomain3d} and of $\partial\Omega^h$ in Figs.~\ref{Fig:MeshBoundary2d} and \ref{Fig:MeshBoundary3d}.

For the hybridization variable $\omega_{\vek{t}}$, living only on the interfaces of the elements, a discretization of the set of interfaces of all the elements $\Psi^h$ is also necessary. Subsequently, Eq.~(\ref{Eq:LagrangeMultiplier}) is reformulated as
\begin{equation}\label{Eq:LagrangeMultiplierHybrid}
    m_{\vek{t}} = \hat{m}_{\vek{t}} \:\: \text{on} \:\: \Psi^h,
\end{equation}
as this condition now applies to $\Psi^h$ and not only to the Neumann boundary $\partial\Omega_{\text{N},\omega}$. For the Kirchhoff beams in $\mathbb{R}^2$, $\Psi^h$ consists of the edges and, for the Kirchhoff--Love shells in $\mathbb{R}^3$, of the faces of the elements in $\Omega^h$, as seen in Figs.~\ref{Fig:MeshInterfaces2d} and \ref{Fig:MeshInterfaces3d}, respectively. Similarly to $\partial\Omega^h$, $\Psi^h$ does not include the limits of the considered level-set interval from $\phi^{\text{min}}$ to $\phi^{\text{max}}$.

\begin{figure}[ht!]
    \centering
    \raisebox{-0.5\height}{%
      \subfigure[general situation in $\mathbb{R}^2$]{%
        \includegraphics[width=0.435\textwidth]{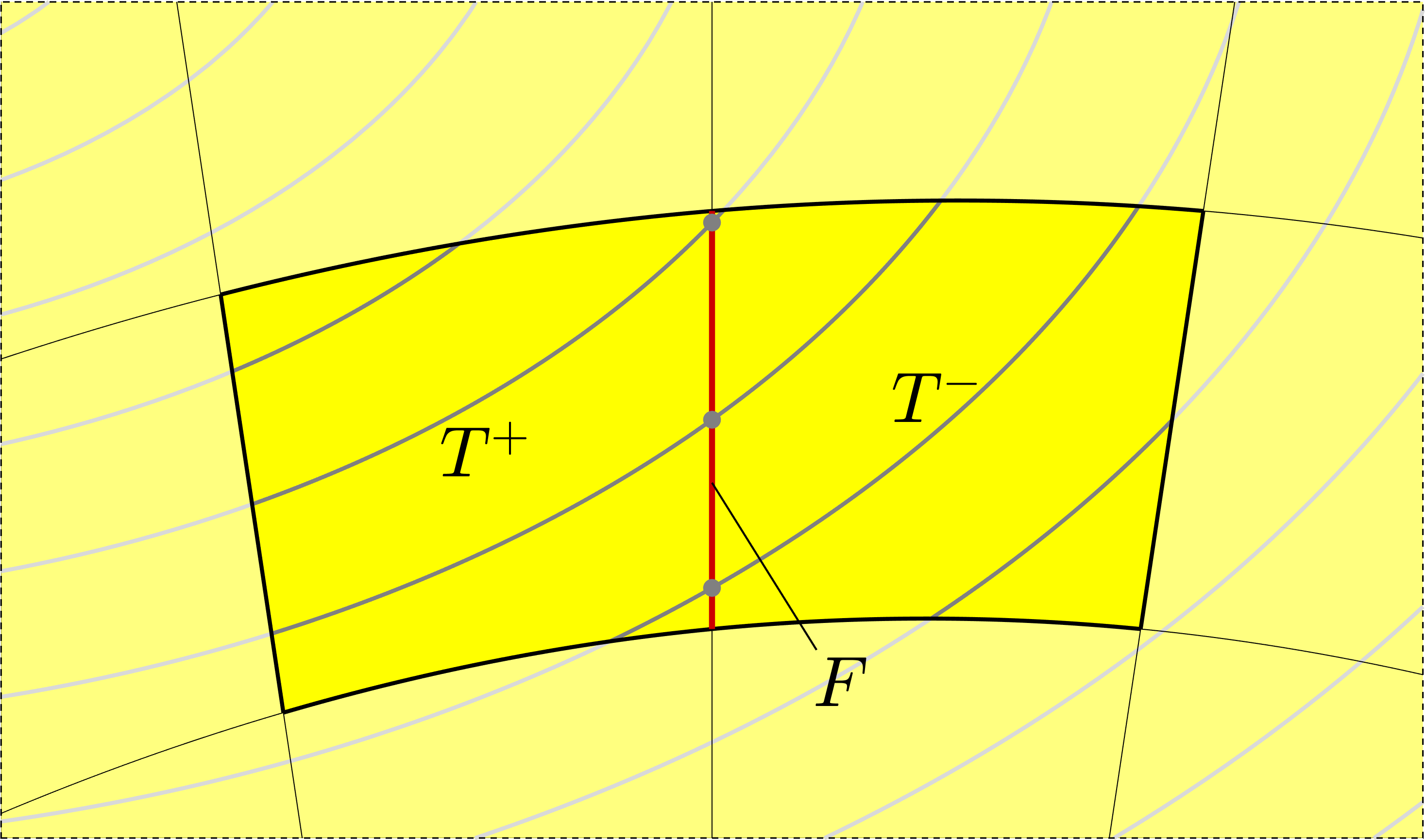}%
        \label{Fig:InterfaceSituation2d}
      }%
    }
    \raisebox{-0.5\height}{%
      \subfigure[orientation of vectors in $\mathbb{R}^2$]{%
        \includegraphics[width=0.535\textwidth]{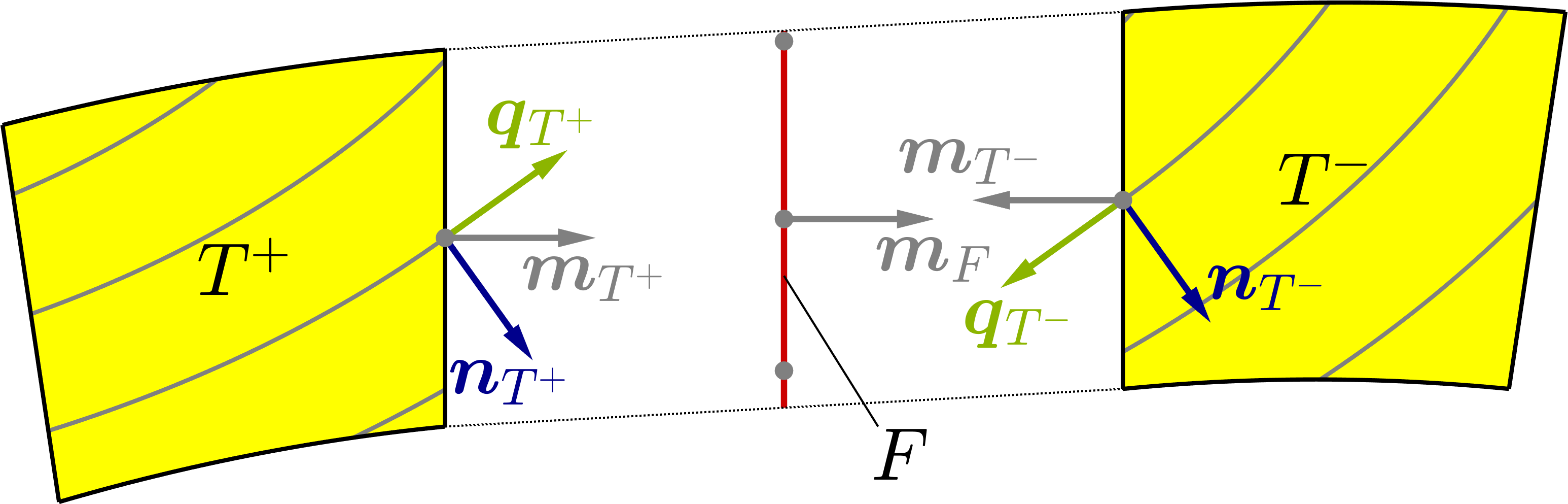}%
        \label{Fig:InterfaceOrientation2d}
      }%
    }\\
    \raisebox{-0.5\height}{%
      \subfigure[general situation in $\mathbb{R}^3$]{%
        \includegraphics[width=0.45\textwidth]{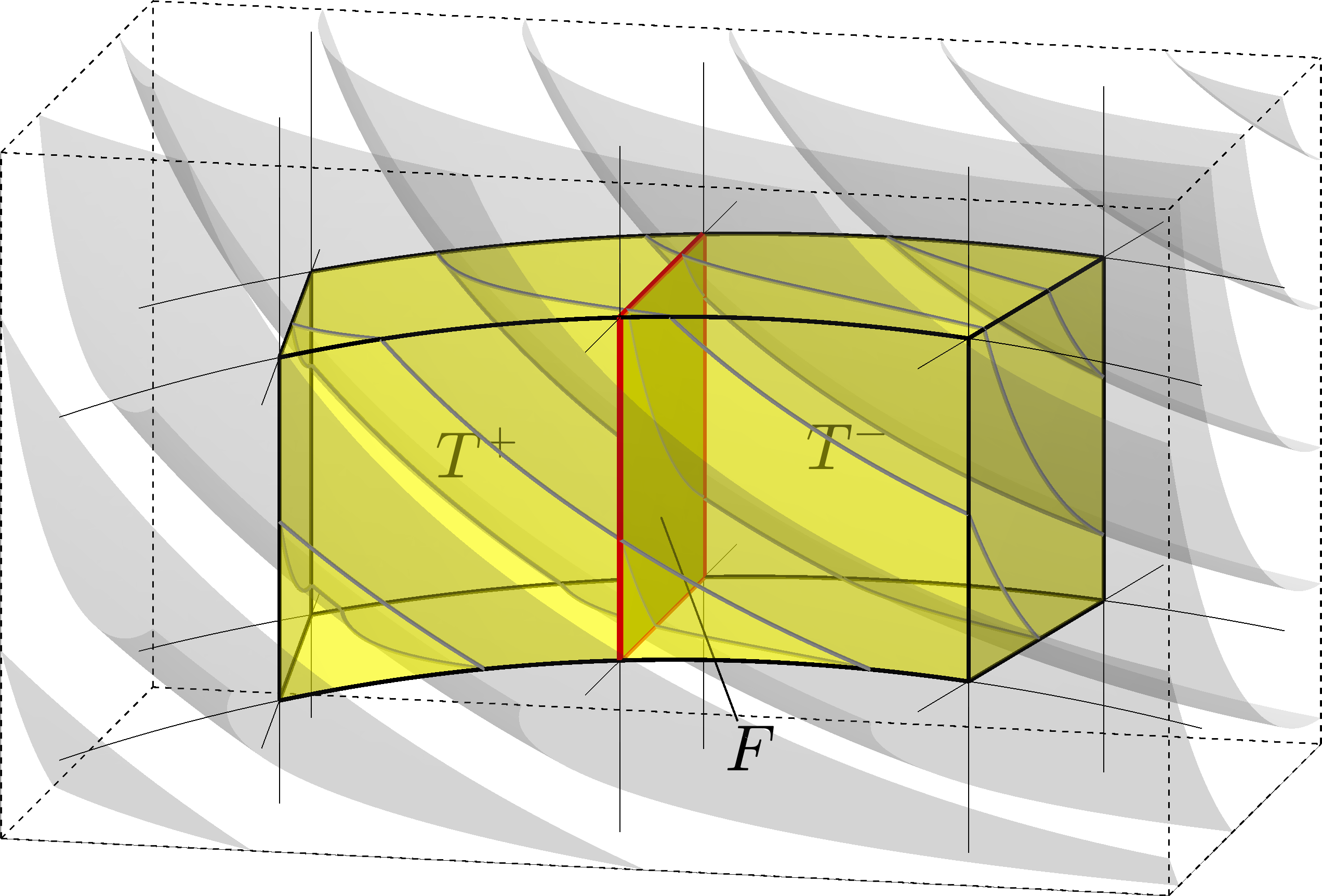}%
        \label{Fig:InterfaceSituation3d}
      }%
    }
    \raisebox{-0.5\height}{%
      \subfigure[orientation of vectors in $\mathbb{R}^3$]{%
        \includegraphics[width=0.52\textwidth]{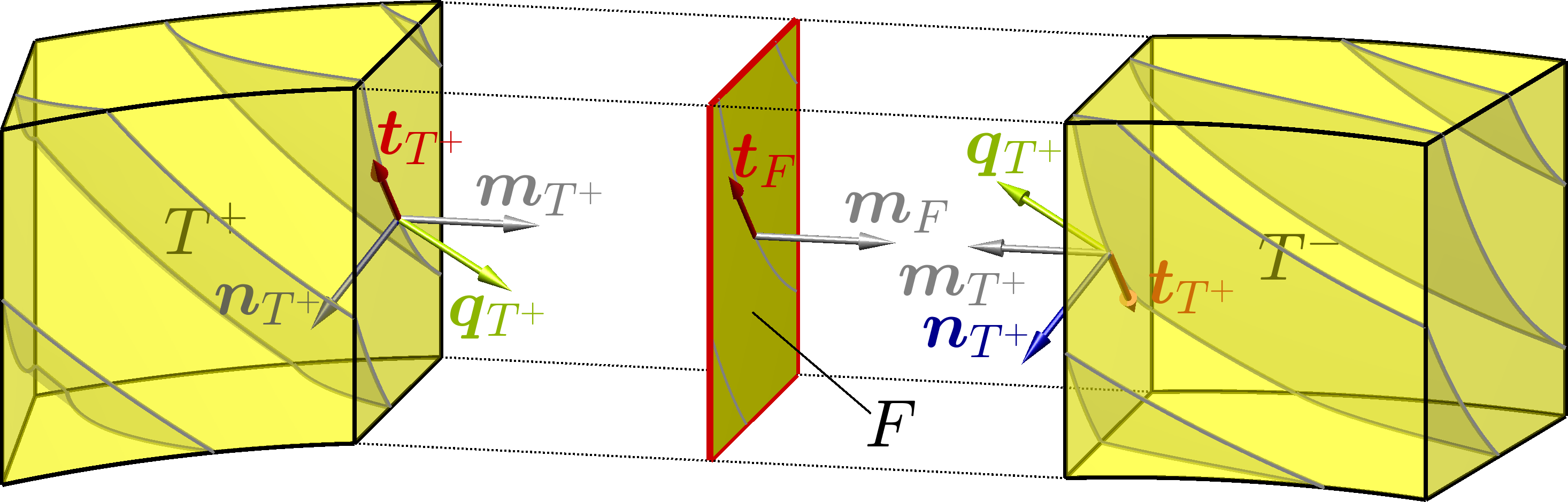}%
        \label{Fig:InterfaceOrientation3d}
      }%
    }\\
    \caption{\label{Fig:InterfaceOrientation}Neighboring elements $T^+$ and $T^-$ and the edge or face element $F$ located inbetween. Some selected level sets are indicated in gray.}
\end{figure}

Extending the boundary terms to the element interfaces means that the vector fields $\vek{q}$, $\vek{t}$, and $\vek{m}$ must also be evaluated there, i.e., on $\Psi^h$. Note that the orientation of these vectors is opposite between two neighboring elements, as seen in Fig.~\ref{Fig:InterfaceOrientation}. This change of orientation can be considered by using the jump operator,
\begin{equation}\label{Eq:JumpOperator}
    \jumpl f \jumpr = 
    \begin{cases}
        f|_{T^+} - f|_{T^-}& \text{on} \:\: \Psi^h \setminus \partial\Omega^h, \\
        f|_{T}&  \text{on} \:\: \Psi^h \cap \partial\Omega^h,
    \end{cases}
\end{equation}
in the weak form, with $T^+$ and $T^-$ being two neighboring elements.

For the finite element spaces, classic Lagrange elements with interpolating shape functions are used. The nodal coordinates in the mesh are labeled $\vek{x}_i$ with $i = 1,\dots,n$, where $n$ is the total number of nodes in some mesh composed of elements of order $p$. The corresponding nodal basis functions $N_i^{p}(\vek{x})$, again of a certain order $p$, which are later used as test and trial functions, are obtained straightforwardly by usual isoparametric mappings of one-, two-, or three-dimensional reference elements to the real, physical elements in the mesh, respectively. Note that \emph{equal-order} interpolations of the individual mechanical fields (e.g., for $\vek{u}$, $\mat{m}_{\Gamma}$, $\omega_{\vek{t}}$, etc.) are chosen for convenience although this is not strictly necessary and individual orders could have been chosen as well.
Finite element spaces for the respective fields may then be defined as
\begin{gather}
    \label{Eq:FESpaceMoment}
        \mathcal{Q}_{\Omega,\mat{m}_\Gamma}^h :=
        \begin{Bmatrix}
            v^h \in C^{-1}(\Omega^h): v^h = \sum_{i=1}^{n_{\mat{m}_\Gamma}} N_i^p(\vek{x}) \cdot \hat{v}_i \: \: \text{with} \: \: \hat{v}_i \in \mathbb{R}
        \end{Bmatrix}
            \subset \mathcal{L}^2(\Omega^h),\\
    \label{Eq:FESpaceDisplacement}
        \mathcal{Q}_{\Omega,\vek{u}}^h :=
        \begin{Bmatrix}
            v^h \in C^{0}(\Omega^h): v^h = \sum_{i=1}^{n_{\vek{u}}} N_i^p(\vek{x}) \cdot \hat{v}_i \: \: \text{with} \: \: \hat{v}_i \in \mathbb{R}
        \end{Bmatrix}
            \subset \mathcal{H}^1(\Omega^h),\\
    \label{Eq:FESpaceRotation}
    \mathcal{Q}_{\Psi,\omega_{\vek{t}}}^h :=
    \begin{Bmatrix}
        v^h \in C^{-1}(\Psi^h): v^h = \sum_{i=1}^{n_{\omega_{\vek{t}}}} N_i^p(\vek{x}) \cdot \hat{v}_i \: \: \text{with} \: \: \hat{v}_i \in \mathbb{R}
    \end{Bmatrix}
        \subset \mathcal{L}^2(\Psi^h).
\end{gather}
Note that although the involved shape functions refer to the same mesh of order $p$, individual indices $n_{\mat{m}_\Gamma}$, $n_{\vek{u}}$, $n_{\omega_{\vek{t}}}$, indicate that they are used in different contexts. That is, the set $\mathcal{Q}_{\Omega,\mat{m}_\Gamma}^h$ is a \emph{discontinuous}, element-local function space, $\mathcal{Q}_{\Omega,\vek{u}}^h$ is a (usual) $C^0$-continuous function space in $\Omega^h$, and $\mathcal{Q}_{\Psi,\omega_{\vek{t}}}^h$ is discontinuous, only living on element boundaries $\Psi^h$. Based on Eqs.~(\ref{Eq:FESpaceMoment})-(\ref{Eq:FESpaceRotation}), the discrete test and trial function spaces are now defined as
\begin{gather}
    \label{Eq:FunctionSpaceDiscMoment}
     \mathcal{S}_{\mat{m}_\Gamma}^h = \mathcal{V}_{\mat{m}_\Gamma}^h =
     \begin{Bmatrix}
      \mat{V}^h \in \big[\mathcal{Q}_{\Omega,\mat{m}_\Gamma}^h\big]^{3 \times 3}: \mat{V}^h = {\mat{V}^h}^{\text{T}}
     \end{Bmatrix},\\
    \label{Eq:FunctionSpaceDiscDisplacementTrial}
     \mathcal{S}_{\vek{u}}^h =
     \begin{Bmatrix}
      \vek{v}^h \in \big[\mathcal{Q}_{\Omega,\vek{u}}^h\big]^3: \vek{v}^h = \hat{\vek{u}} \:\: \text{on} \:\: \partial\Omega_{\text{D},\vek{u}}^h
     \end{Bmatrix},\\
     \label{Eq:FunctionSpaceDiscDisplacementTest}
     \mathcal{V}_{\vek{u}}^h =
     \begin{Bmatrix}
      \vek{v}^h \in \big[\mathcal{Q}_{\Omega,\vek{u}}^h\big]^3: \vek{v}^h = \vek{0} \:\: \text{on} \:\: \partial\Omega_{\text{D},\vek{u}}^h
     \end{Bmatrix},\\
    \label{Eq:FunctionSpaceDiscRotationTrial}
     \mathcal{S}_{\omega_{\vek{t}}}^h =
     \begin{Bmatrix}
      v^h \in \mathcal{Q}_{\Psi,\omega_{\vek{t}}}^h: v^h = \hat{\omega}_{\vek{t}} \:\: \text{on} \:\: \partial\Omega_{\text{D},\omega}^h
     \end{Bmatrix},\\
    \label{Eq:FunctionSpaceDiscRotationTest}
     \mathcal{V}_{\omega_{\vek{t}}}^h =
     \begin{Bmatrix}
      v^h \in \mathcal{Q}_{\Psi,\omega_{\vek{t}}}^h: v^h = 0 \:\: \text{on} \:\: \partial\Omega_{\text{D},\omega}^h
     \end{Bmatrix}.
\end{gather}

\subsection{Weak form of the Kirchhoff beam}
Based on the function spaces of Eqs.~(\ref{Eq:FunctionSpaceDiscMoment})-(\ref{Eq:FunctionSpaceDiscRotationTest}), the hybridized and discrete variational problem of the Kirchhoff \emph{beam} is defined as follows: Given the input data $\phi \in \mathbb{R}$, $E, I, A \in \mathbb{R}^+$, $\vek{f}^h \in \mathbb{R}^2$ in $\Omega^h$, $\hat{\vek{p}}^h \in \mathbb{R}^2$ on $\partial\Omega_{\text{N},\vek{u}}^h$, and $\hat{m}_z^h \in \mathbb{R}$ on $\partial\Omega_{\text{N},\vek{u}}^h$, find $\mat{m}_\Gamma^h \in \mathcal{S}_{\mat{m}_\Gamma}^h$, $\vek{u}^h \in \mathcal{S}_{\vek{u}}^h$, and $\omega_{\vek{t}}^h \in \mathcal{S}_{\omega_{\vek{t}}}^h$ such that for all test functions $(\mat{V}_{\mat{m}_\Gamma}^h, \vek{v}_{\vek{u}}^h, v_{\omega_{\vek{t}}}^h) \in \mathcal{V}_{\mat{m}_\Gamma}^h \times \mathcal{V}_{\vek{u}}^h \times \mathcal{V}_{\omega_{\vek{t}}}^h$, there holds
\begin{gather}
    \label{Eq:WeakFormBeam1}
    \begin{split}
        \int_{\Omega^h}& \mat{V}_{\mat{m}_\Gamma}^h \FrobProd
        \big(
            -\vek{\varepsilon}_{\Gamma,\text{Bend}}(\mat{m}_\Gamma^h) + \mat{H} \cdot \nabla^\text{dir}_\Gamma \vek{u}^h  
        \big) \, \|\nabla\phi\| \, \text{d}\Omega \\
        &+ \sum_{T \in \Omega^h} \int_T \text{div}_\Gamma\big(\mat{P} \cdot \mat{V}_{\mat{m}_\Gamma}^h \cdot \mat{P}\big) \ScalProd
        \big[
            (\nabla^\text{dir}_\Gamma \vek{u}^h)^\text{T} \cdot \vek{n}
        \big]
        \, \|\nabla\phi\| \, \text{d}\Omega \\
       &+ \int_{\Psi^h} \jumpl m_{\vek{t}}(\mat{V}_{\mat{m}_\Gamma}^h) \jumpr \, \omega_{\vek{t}}^h \, \big(\vek{q} \ScalProd \vek{m}\big) \, \|\nabla\phi\| \, \text{d}\Psi = 0,
    \end{split}\\
    \label{Eq:WeakFormBeam2}
    \begin{split}
        \int_{\Omega^h}&
        \big[
            \big(
                \mat{H} \cdot \nabla^\text{dir}_\Gamma \vek{v}_{\vek{u}}^h
            \big) \FrobProd \mat{m}_\Gamma^h 
            + \nabla^\text{dir}_\Gamma \vek{v}_{\vek{u}}^h \FrobProd \tilde{\mat{n}}_\Gamma(\vek{u}^h)
        \big] \, \|\nabla\phi\|
        \, \text{d}\Omega \\
        &+ \sum_{T \in \Omega^h} \int_T (\nabla^\text{dir}_\Gamma \vek{v}_{\vek{u}}^h)^\text{T} \cdot \vek{n} \ScalProd \text{div}_\Gamma \mat{m}_\Gamma^h \, \|\nabla\phi\| \, \text{d}\Omega \\
        &= \int_{\Omega^h} \vek{v}_{\vek{u}}^h \ScalProd \vek{f}^h \, \|\nabla\phi\| \, \text{d}\Omega + \int_{\partial\Omega_{\text{N},\vek{u}}^h}\!\!\!\!\!\!\! \vek{v}_{\vek{u}}^h \ScalProd \hat{\vek{p}}^h \, \big(\vek{q} \ScalProd \vek{m}\big) \, \|\nabla\phi\| \, \text{d}\partial\Omega,
    \end{split}\\
    \label{Eq:WeakFormBeam3}
        \int_{\Psi^h} v_{\omega_{\vek{t}}}^h \, \jumpl m_{\vek{t}}(\mat{m}_\Gamma^h) \jumpr \, \big(\vek{q} \ScalProd \vek{m}\big) \, \|\nabla\phi\| \, \text{d}\Psi = \int_{\partial\Omega_{\text{N},\omega}^h}\!\!\!\!\!\!\! v_{\omega_{\vek{t}}}^h \, \hat{m}_z^h \, t_z \, \big(\vek{q} \ScalProd \vek{m}\big) \, \|\nabla\phi\| \, \text{d}\partial\Omega.
\end{gather}
For brevity, the superscript $h$ was not added to geometric quantities such as $\vek{n} \rightarrow \vek{n}^h$ and differential operators such as $\nabla_\Gamma \rightarrow \nabla_\Gamma^h$, even though they are also discretized, i.e., based on the mesh.

\subsection{Weak form of the Kirchhoff--Love shell}
The hybridized and discrete variational problem of the Kirchhoff--Love \emph{shell} is defined as: Given the input data $\phi \in \mathbb{R}$, $E, t \in \mathbb{R}^+$, $\nu \in [0, 0.5)$, $\vek{f}^h \in \mathbb{R}^3$ in $\Omega^h$, $\hat{\tilde{\vek{p}}}^h \in \mathbb{R}^3$ on $\partial\Omega_{\text{N},\vek{u}}^h$, and $\hat{m}_{\vek{t}}^h \in \mathbb{R}$ on $\partial\Omega_{\text{N},\vek{u}}^h$, find $\mat{m}_\Gamma^h \in \mathcal{S}_{\mat{m}_\Gamma}^h$, $\vek{u}^h \in \mathcal{S}_{\vek{u}}^h$, and $\omega_{\vek{t}}^h \in \mathcal{S}_{\omega_{\vek{t}}}^h$ such that for all test functions $(\mat{V}_{\mat{m}_\Gamma}^h, \vek{v}_{\vek{u}}^h, v_{\omega_{\vek{t}}}^h) \in \mathcal{V}_{\mat{m}_\Gamma}^h \times \mathcal{V}_{\vek{u}}^h \times \mathcal{V}_{\omega_{\vek{t}}}^h$, there holds
\begin{gather}
    \label{Eq:WeakFormShell1}
    \begin{split}
        \int_{\Omega^h}& \mat{V}_{\mat{m}_\Gamma}^h \FrobProd
        \big(
            -\vek{\varepsilon}_{\Gamma,\text{Bend}}(\mat{m}_\Gamma^h) + \mat{H} \cdot \nabla^\text{dir}_\Gamma \vek{u}^h  
        \big) \, \|\nabla\phi\| \, \text{d}\Omega \\
        &+ \sum_{T \in \Omega^h} \int_T \text{div}_\Gamma\big(\mat{P} \cdot \mat{V}_{\mat{m}_\Gamma}^h \cdot \mat{P}\big) \ScalProd
        \big[
            (\nabla^\text{dir}_\Gamma \vek{u}^h)^\text{T} \cdot \vek{n}
        \big]
        \, \|\nabla\phi\| \, \text{d}\Omega \\
       &+ \int_{\Psi^h} \big(m_{\vek{q}}(\mat{V}_{\mat{m}_\Gamma}^h) \, \omega_{\vek{q}}(\vek{u}^h)
        + \jumpl m_{\vek{t}}(\mat{V}_{\mat{m}_\Gamma}^h) \jumpr \, \omega_{\vek{t}}^h\big) \, \big(\vek{q} \ScalProd \vek{m}\big) \, \|\nabla\phi\| \, \text{d}\Psi = 0,
    \end{split}\\
    \label{Eq:WeakFormShell2}
    \begin{split}
        \int_{\Omega^h}&
        \big[
            \big(
                \mat{H} \cdot \nabla^\text{dir}_\Gamma \vek{v}_{\vek{u}}^h
            \big) \FrobProd \mat{m}_\Gamma^h 
            + \nabla^\text{dir}_\Gamma \vek{v}_{\vek{u}}^h \FrobProd \tilde{\mat{n}}_\Gamma(\vek{u}^h)
        \big] \, \|\nabla\phi\|
        \, \text{d}\Omega \\
        &+ \sum_{T \in \Omega^h} \int_T (\nabla^\text{dir}_\Gamma \vek{v}_{\vek{u}}^h)^\text{T} \cdot \vek{n} \ScalProd \text{div}_\Gamma \mat{m}_\Gamma^h \, \|\nabla\phi\| \, \text{d}\Omega \\
        &+ \int_{\Psi^h} \omega_{\vek{q}}(\vek{v}_{\vek{u}}^h) \, m_{\vek{q}}(\mat{m}_\Gamma^h) \, \big(\vek{q} \ScalProd \vek{m}\big) \, \|\nabla\phi\| \, \text{d}\Psi \\
        &= \int_{\Omega^h} \vek{v}_{\vek{u}}^h \ScalProd \vek{f}^h \, \|\nabla\phi\| \, \text{d}\Omega + \int_{\partial\Omega_{\text{N},\vek{u}}^h}\!\!\!\!\!\!\! \vek{v}_{\vek{u}}^h \ScalProd \hat{\tilde{\vek{p}}}^h \, \big(\vek{q} \ScalProd \vek{m}\big) \, \|\nabla\phi\| \, \text{d}\partial\Omega,
    \end{split}\\
    \label{Eq:WeakFormShell3}
        \int_{\Psi^h} v_{\omega_{\vek{t}}}^h \cdot \jumpl m_{\vek{t}}(\mat{m}_\Gamma^h) \jumpr \, \big(\vek{q} \ScalProd \vek{m}\big) \, \|\nabla\phi\| \, \text{d}\Psi = \int_{\partial\Omega_{\text{N},\omega}^h}\!\!\!\!\!\!\! v_{\omega_{\vek{t}}}^h \, \hat{m}_{\vek{t}}^h \, \big(\vek{q} \ScalProd \vek{m}\big) \, \|\nabla\phi\| \, \text{d}\partial\Omega.
\end{gather}
For a detailed derivation of the mixed-hybrid weak form of a \emph{single} Kirchhoff--Love shell, see \cite{Neumeyer_2025a}. Here, the result is obtained for a whole family of shells embedded in a bulk domain by integrating over the considered level-set interval, for which the coarea-formulas in Eqs.~(\ref{Eq:CoareaFormula}) and (\ref{Eq:CoareaFormulaBound}) are employed.

\subsection{Static condensation}
Eqs.~(\ref{Eq:WeakFormBeam1})-(\ref{Eq:WeakFormBeam3}) and Eqs.~(\ref{Eq:WeakFormShell1})-(\ref{Eq:WeakFormShell3}) lead to a linear system of equations with a large number of degrees of freedom (DOFs). An element-wise static condensation is performed to reduce the number of DOFs, hence, to improve the computational performance of the method. With the local stiffness matrix $\mat{K}^\text{el}$ and the local right-hand side $\vek{b}^\text{el}$
\begin{equation}\label{Eq:LocalStiffMat&RHS}
    \mat{K}^\text{el} =
    \begin{bmatrix}
        \mat{K}_{\mat{m}\mat{m}}^\text{el} & \mat{K}_{\mat{m}\vek{u}}^\text{el} & \mat{K}_{\mat{m}\omega}^\text{el} \\
        \mat{K}_{\vek{u}\mat{m}}^\text{el} & \mat{K}_{\vek{u}\vek{u}}^\text{el} & \mat{0} \\
        \mat{K}_{\omega\mat{m}}^\text{el} & \mat{0} & \mat{0}
    \end{bmatrix}, \quad
    \vek{b}^\text{el} =
    \begin{bmatrix}
        \vek{0} \\
        \vek{b}_{\vek{u}}^\text{el} \\
        \vek{b}_{\omega}^\text{el} 
    \end{bmatrix},
\end{equation}
the now entirely local field of the moment tensor may be eliminated locally by static condensation via the Schur complement, following, e.g., \cite{Arnold_1985a, Cockburn_2004a, Boffi_2013a},
\begin{equation}\label{Eq:StaticCondensation}
    \tilde{\mat{K}}^\text{el} =
    \begin{bmatrix}
        \mat{K}_{\vek{u}\vek{u}}^\text{el} & \mat{0} \\
        \mat{0} & \mat{0}
    \end{bmatrix} -
    \begin{bmatrix}
        \mat{K}_{\vek{u}\mat{m}}^\text{el} \\
        \mat{K}_{\omega\mat{m}}^\text{el}
    \end{bmatrix} \cdot
    {
        \mat{K}_{\mat{m}\mat{m}}^\text{el} 
    }^{-1} \cdot
    \begin{bmatrix}
        \mat{K}_{\mat{m}\vek{u}}^\text{el} & \mat{K}_{\mat{m}\omega}^\text{el}
    \end{bmatrix}, \quad
    \tilde{\vek{b}}^\text{el} =
    \begin{bmatrix}
        \vek{b}_{\vek{u}}^\text{el} \\
        \vek{b}_{\omega}^\text{el} 
    \end{bmatrix}.
\end{equation}
$\tilde{\mat{K}}^\text{el}$ and $\tilde{\vek{b}}^\text{el}$ are the condensed versions of the local stiffness matrix and right-hand side, respectively. Assembling the entries of $\tilde{\mat{K}}^\text{el}$ and $\tilde{\vek{b}}^\text{el}$ into the global, condensed stiffness matrix $\tilde{\mat{K}}$ and right-hand side $\tilde{\vek{b}}$ results in a linear system of equations
\begin{equation}\label{Eq:CondensedSOE}
    \tilde{\mat{K}} \cdot
    \begin{bmatrix}
        \underline{\vek{u}} \\
        \vek{\omega_{\vek{t}}}
    \end{bmatrix}
    = \tilde{\vek{b}},
\end{equation}
where only the sought nodal components of the displacement vector ($\underline{\vek{u}} = [\vek{u}, \vek{v}]^\text{T}$ for beams in $\mathbb{R}^2$ and $\underline{\vek{u}} = [\vek{u}, \vek{v}, \vek{w}]^\text{T}$ for shells in $\mathbb{R}^3$) and the Lagrange multiplier ($\vek{\omega_{\vek{t}}}$) remain as primary unknowns. The condensed stiffness matrix offers the advantage of not only being smaller but also positive-definite, which is beneficial regarding the efficient solvability.
For the Kirchhoff beam in $\mathbb{R}^2$, we may substitute $\omega_{\vek{t}}$ with $\omega_z \, t_z$ on $\partial\Omega_{\text{D},\omega}^h$ prescribing the boundary rotation globally.

It is possible to recover the nodal components of the moment tensor ($\underline{\underline{\vek{m_\Gamma}}} = [\vek{m}_{\Gamma,11}, \vek{m}_{\Gamma,22},\\ \vek{m}_{\Gamma,12}]^\text{T}$ for $\mathbb{R}^2$ and $\underline{\underline{\vek{m_\Gamma}}} = [\vek{m}_{\Gamma,11}, \vek{m}_{\Gamma,22}, \vek{m}_{\Gamma,33}, \vek{m}_{\Gamma,12}, \vek{m}_{\Gamma,13}, \vek{m}_{\Gamma,23}]^\text{T}$ for $\mathbb{R}^3$) element-wise in a post-processing step as
\begin{equation}\label{Eq:PostProcessingMoment}
    \underline{\underline{\vek{m_\Gamma^\text{el}}}} =
    -{\mat{K}_{\mat{m}\mat{m}}^\text{el}}^{-1} \cdot
    \begin{bmatrix}
        \mat{K}_{\mat{m}\vek{u}}^\text{el} & \mat{K}_{\mat{m}\omega}^\text{el} \\
    \end{bmatrix} \cdot
    \begin{bmatrix}
        \underline{\vek{u}}^\text{el} \\
        \vek{\omega_{\vek{t}}^\text{el}}
    \end{bmatrix}.
\end{equation}

\section{Numerical results}\label{Sec:NumericalResults}
In this section, a variety of test cases for the simultaneous analysis of families of Kirchhoff beams and Kirchhoff--Love shells over bulk domains are shown. In order to confirm higher-order convergence of the proposed methodology, different error measures are assessed, such as the $\mathcal{L}^2$-error, residual errors, and the stored energy error.

For the $\mathcal{L}^2$-error, an analytical solution of a test case must be available. In this case, the relative $\mathcal{L}^2$-error of a certain quantity $f$ in a bulk domain can be computed as
\begin{equation}\label{Eq:L2Error}
    \varepsilon_{\mathcal{L}^2,\text{rel}}^2(f) = \frac{\int_{\Omega^h} (f^{\text{ex}} - f^h)^2 \, \|\nabla\phi\| \, \text{d}\Omega}{\int_{\Omega^h} (f^{\text{ex}})^2 \, \|\nabla\phi\| \, \text{d}\Omega},
\end{equation}
where the superscripts "ex" and "$h$" indicate the analytical and numerical solution, respectively.

If an analytical solution is not available, other error measures may be considered. In the context of higher-order finite element analysis, residual errors are an adequate method. Here, the approximated solution of the primary variables $\mat{m}_\Gamma^h$ and $\vek{u}^h$ is inserted in the strong form, see Eqs.~(\ref{Eq:StrongForm1}) and (\ref{Eq:StrongForm2}), and the residuals are evaluated. Integration over all elements $T$ leads to the two residual errors
\begin{equation}\label{Eq:ResidualError1}
    \begin{gathered}
        \varepsilon_{\text{res},1}^2 = \sum_{T \in \Omega^h} \int_T \mathfrak{r}_1(\mat{m}_\Gamma^h,\vek{u}^h) \FrobProd \mathfrak{r}_1(\mat{m}_\Gamma^h,\vek{u}^h) \, \|\nabla\phi\| \, \text{d}\Omega,\\
        \text{with} \quad \mathfrak{r}_1(\mat{m}_\Gamma^h,\vek{u}^h) = -\vek{\varepsilon}_{\Gamma,\text{Bend}}(\mat{m}_\Gamma^h) + \vek{\varepsilon}_{\Gamma,\text{Bend}}(\vek{u}^h),
    \end{gathered}  
\end{equation}
\begin{equation}\label{Eq:ResidualError2}
    \begin{gathered}
        \varepsilon_{\text{res},2}^2 = \sum_{T \in \Omega^h} \int_T \mathfrak{r}_2(\mat{m}_\Gamma^h,\vek{u}^h) \ScalProd \mathfrak{r}_2(\mat{m}_\Gamma^h,\vek{u}^h) \, \|\nabla\phi\| \, \text{d}\Omega,\\
        \text{with} \quad \mathfrak{r}_2(\mat{m}_\Gamma^h,\vek{u}^h) = \text{div}_\Gamma \mat{n}_\Gamma^\text{real}(\mat{m}_\Gamma^h,\vek{u}^h) + \vek{n} \cdot \text{div}_\Gamma \big(\mat{P} \cdot \text{div}_\Gamma \mat{m}_\Gamma^h\big) + \mat{H} \cdot \text{div}_\Gamma \mat{m}_\Gamma^h + \vek{f}^h.
    \end{gathered}  
\end{equation}
By dividing the second residual error from Eq.~(\ref{Eq:ResidualError2}) by $\sum_{T \in \Omega^h} \int_T {\vek{f}^h}^2 \, \|\nabla\phi\| \, \text{d}\Omega$ for the case that $\vek{f}^h \neq \vek{0}$, a relative version of the error may be computed. Generally, a convergence rate of $\mathcal{O}(p-1)$ can be expected for both residual errors, as second-order derivatives of the primary unknowns are required.

The stored energy error is defined as
\begin{equation}\label{Eq:EnergyError}
    \varepsilon_{\mathfrak{e}} = \frac{|\mathfrak{e}_{\text{ref}} - \mathfrak{e}^h|}{\mathfrak{e}_{\text{ref}}},
\end{equation}
where $\mathfrak{e}_{\text{ref}}$ is the reference solution of the stored elastic energy integrated over all manifolds in a bulk domain, and $\mathfrak{e}^h$ the approximated energy, computed as
\begin{equation}\label{Eq:StoredEnergy}
    \mathfrak{e}^h=\mathfrak{e}(\mat{m}_\Gamma^h,\vek{u}^h) = \frac{1}{2} \, \int_{\Gamma^h} \big(\vek{\varepsilon}_{\Gamma,\text{Memb}}(\vek{u}^h) \FrobProd \tilde{\mat{n}}_\Gamma(\vek{u}^h) + \vek{\varepsilon}_{\Gamma,\text{Bend}}(\mat{m}_\Gamma^h) \FrobProd \mat{m}_\Gamma^h\big) \, \|\nabla\phi\| \, \text{d}\Gamma.
\end{equation}
When the exact energy is not available, the reference energy $\mathfrak{e}_{\text{ref}}$ may also be determined via an overkill computation using an extremely fine discretization with higher-order elements. The expected convergence rate in this energy error measure is $\mathcal{O}(2\,p)$ when integrating Eq.~(\ref{Eq:StoredEnergy}) exactly \cite{Zienkiewicz_2013a}. However, note that this may interfere with the error in \emph{numerical} integration in an isoparametric FEM setting which, for general integrands, is reported as $\mathcal{O}(p+1)$ for odd element orders and $\mathcal{O}(p+2)$ for even element orders. Therefore, depending on the impact of the numerical integration error, rates of $\mathcal{O}(2\,p)$ and $\mathcal{O}(p+1\;\mathrm{or}\;2)$ are found in the literature, see, e.g.,  \cite{Kaiser_2025a,Neumeyer_2025a,Fries_2020a,Fries_2023a}. The advantage of using the stored energy in an error measure is that $\mathfrak{e}^h$ is a characteristic global scalar value for a given test case and mesh, which is obtained by using first-order derivatives that are accessible in any FEM-code. The reference energies $\mathfrak{e}_{\text{ref}}$ are easily provided when proposing new benchmark test cases as done herein and may then be used for verification.

\subsection{Family of arc-shaped beams in \texorpdfstring{$\mathbb{R}^2$}{R2}}\label{Sec:TC1Arcs}
\begin{figure}[ht!]
 \centering
 
 \subfigure[geometry]
 {\includegraphics[width=0.452\textwidth]{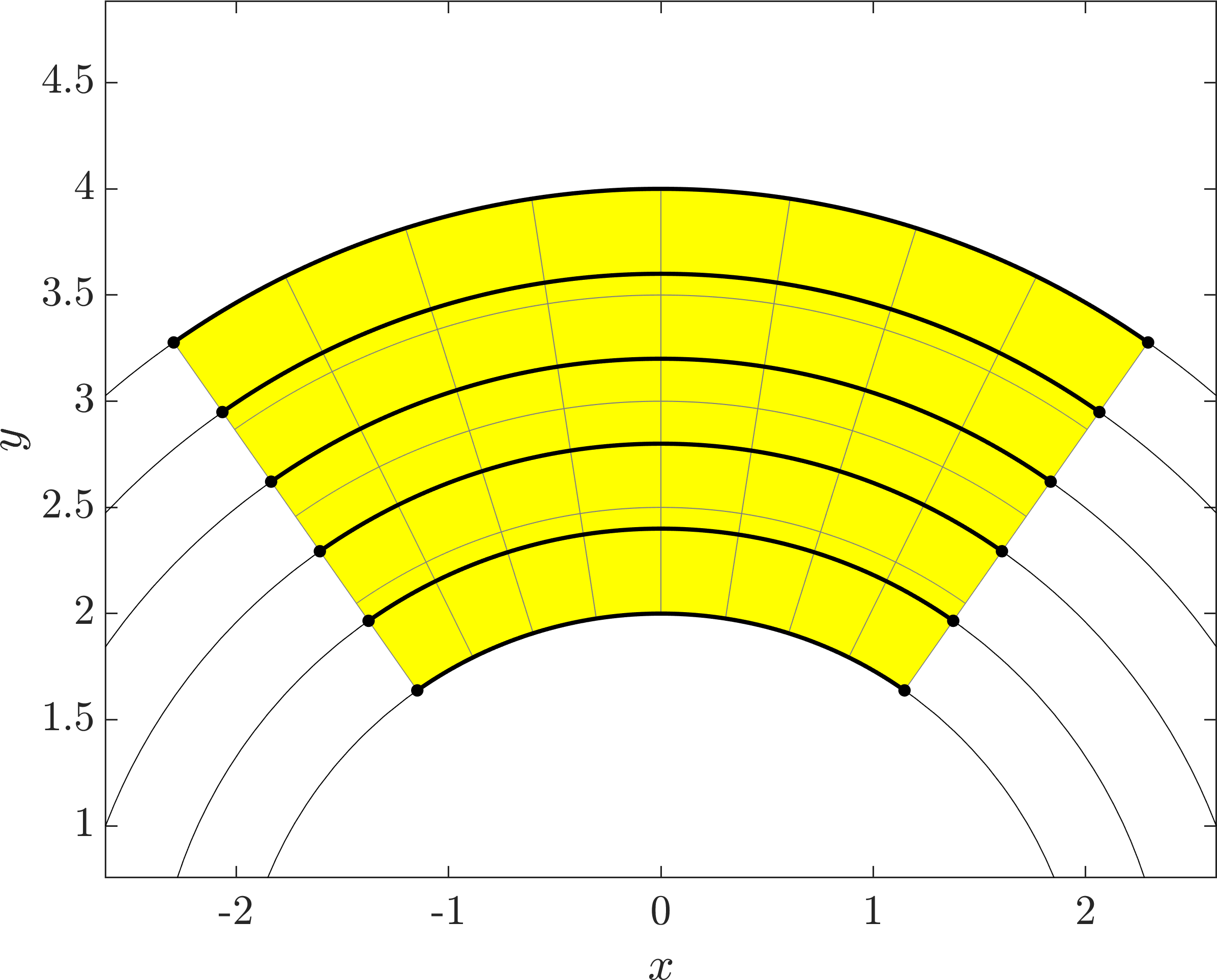}\label{Fig:TC1ArcsGeometry}}\hfill
 \subfigure[displacements]
 {\includegraphics[width=0.528\textwidth]{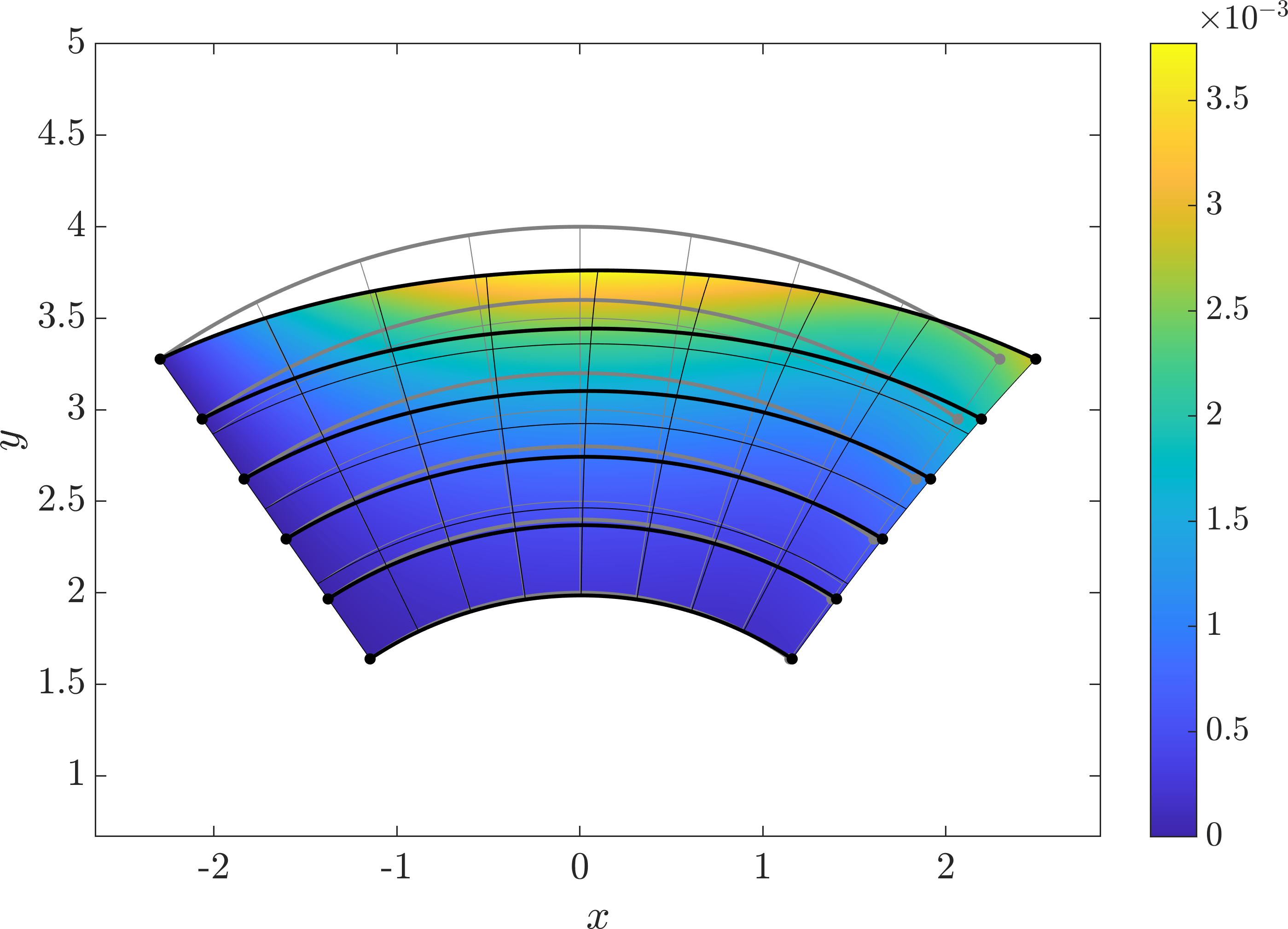}\label{Fig:TC1ArcsDisplacement}}
 
 \caption{\label{Fig:TC1Arcs}Setup of the family of arc-shaped beams in $\mathbb{R}^2$: (a) Geometry of the discretized bulk domain $\Omega^h$ and some selected beams $\Gamma^c$. (b) Scaled deformed configuration with the Euclidean norm of the displacements $\| \vek{u} \|$ as a color plot over the bulk domain. The gray mesh lines and level sets indicate the undeformed configuration.}
\end{figure}

Based on a test case for a single arc in \cite{Kaiser_2023a, Neumeyer_2025a}, the first test case adapts this geometry to a family of arcs in a bulk domain. The geometry of the individual beams is described by the level-set function
\begin{equation}\label{Eq:TC1Arcs_LSF}
    \phi(\vek{x}) = \sqrt{x^2 + y^2},
\end{equation}
representing the implicit definition of a circle. With the level-set function $\psi(\vek{x}) = \arctan(\frac{y}{x})$ describing a circular sector, the geometry of the bulk domain can be described as
\begin{equation}\label{Eq:TC1Arcs_BD}
    \Omega = \big\{ \vek{x} \in \mathbb{R}^2: \tfrac{\pi - \theta}{2} \leq \psi(\vek{x}) \leq \tfrac{\pi + \theta}{2}, \, 2 < \phi(\vek{x}) < 4 \big\},
\end{equation}
where $\theta = \frac{7\,\pi}{18}$ is the central angle. The geometry of the discretized bulk domain $\Omega^h$ and the level-set function $\phi$ is depicted in Fig.~\ref{Fig:TC1ArcsGeometry}.

The input parameters are defined as the load vector $\vek{f} = [0, -10]^\text{T}$, Young's modulus $E = 2.1 \cdot 10^8$, the area $A = b\,h$, and the area moment of inertia $I = \frac{b\,h^3}{12}$ of the cross section, where $b = 1$ and $h = 0.1$. The left boundary ($\psi(\vek{x}) = \frac{\pi + \theta}{2}$) is simply supported, whereas, on the right boundary ($\psi(\vek{x}) = \frac{\pi - \theta}{2}$), only vertical displacements are prevented, i.e., $u \neq 0$, $v = \hat{v} = 0$.

\begin{figure}[ht!]
 \centering
 
 \subfigure[moment over $\Omega$]
 {\includegraphics[width=0.329\textwidth]{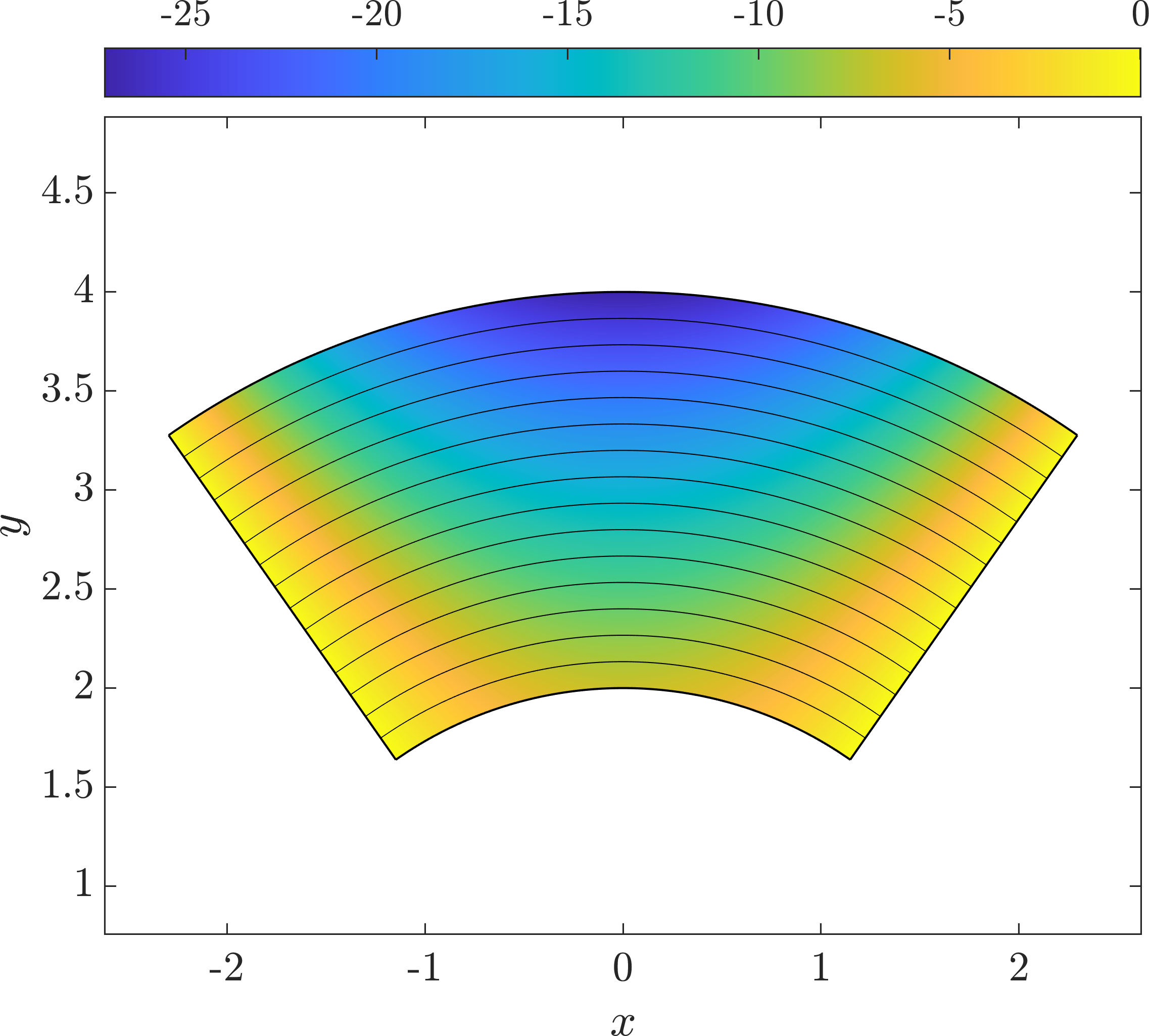}\label{Fig:TC1ArcsMomentBD}}\hfill
 \subfigure[physical normal force over $\Omega$]
 {\includegraphics[width=0.329\textwidth]{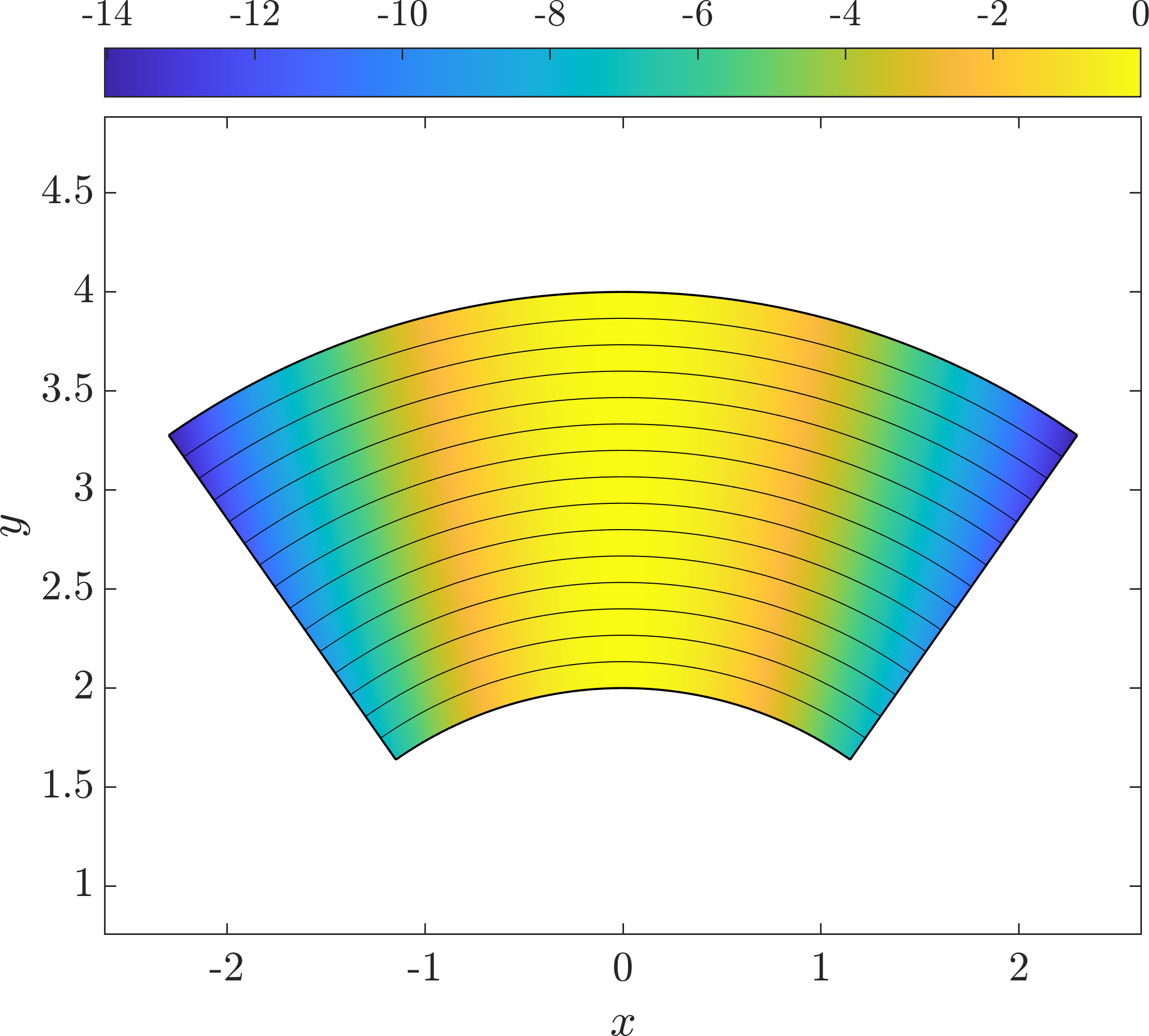}\label{Fig:TC1ArcsNormalForceBD}}\hfill
    \subfigure[shear force over $\Omega$]
 {\includegraphics[width=0.3315\textwidth]{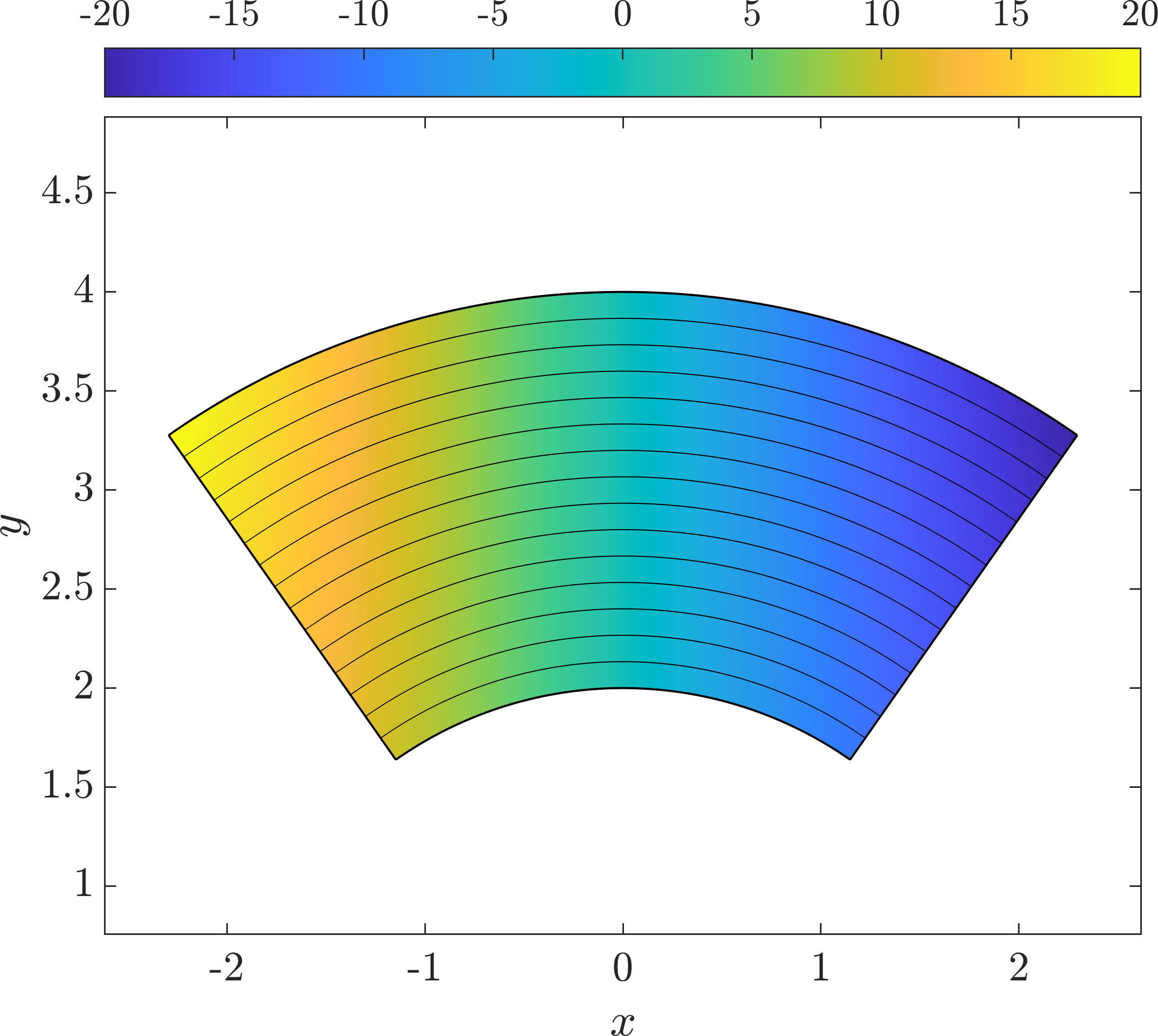}\label{Fig:TC1ArcsShearForceBD}}\\
    \subfigure[moment over $\Gamma^c$]
 {\includegraphics[width=0.329\textwidth]{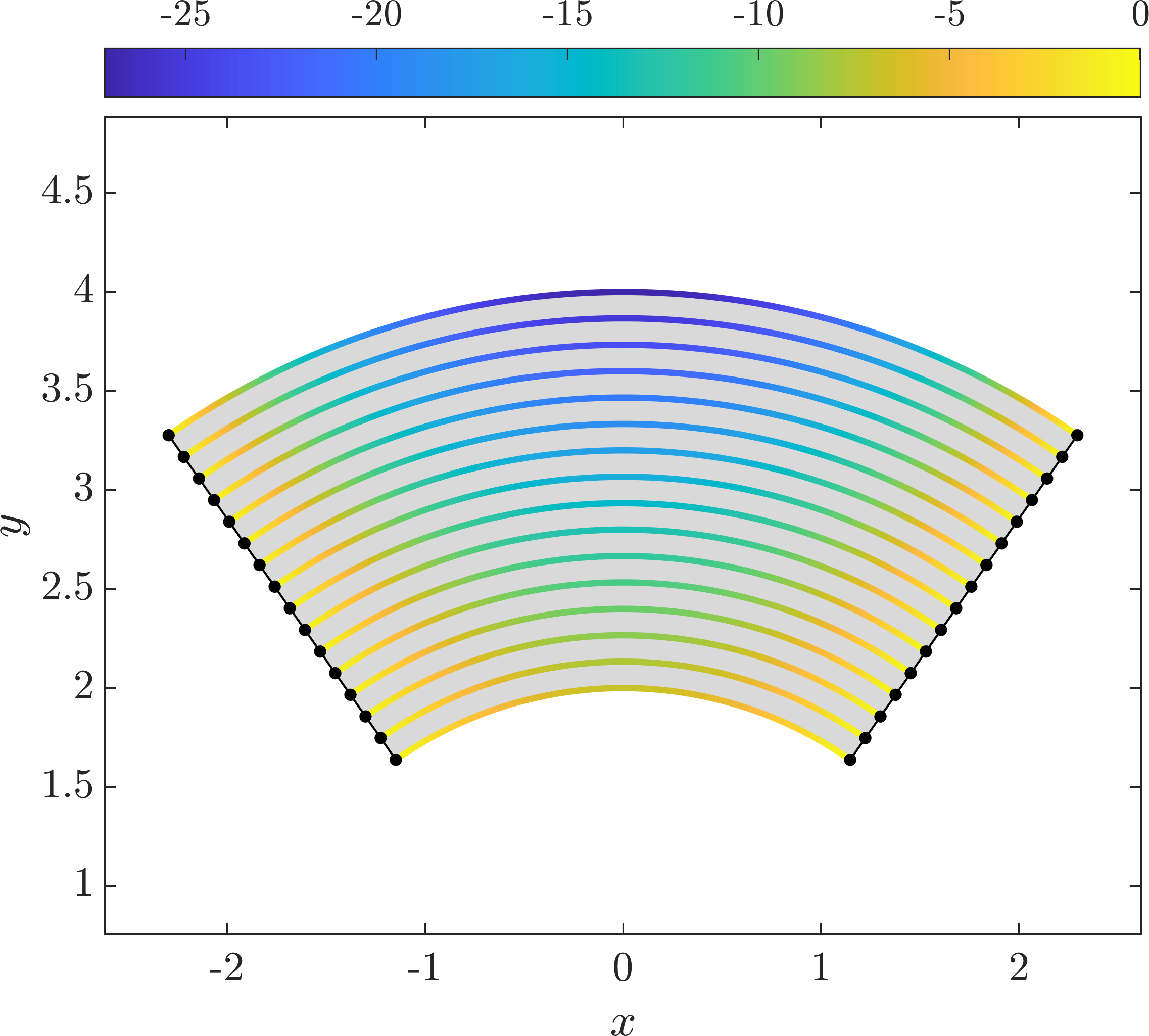}\label{Fig:TC1ArcsMomentLS}}\hfill
 \subfigure[physical normal force over $\Gamma^c$]
 {\includegraphics[width=0.329\textwidth]{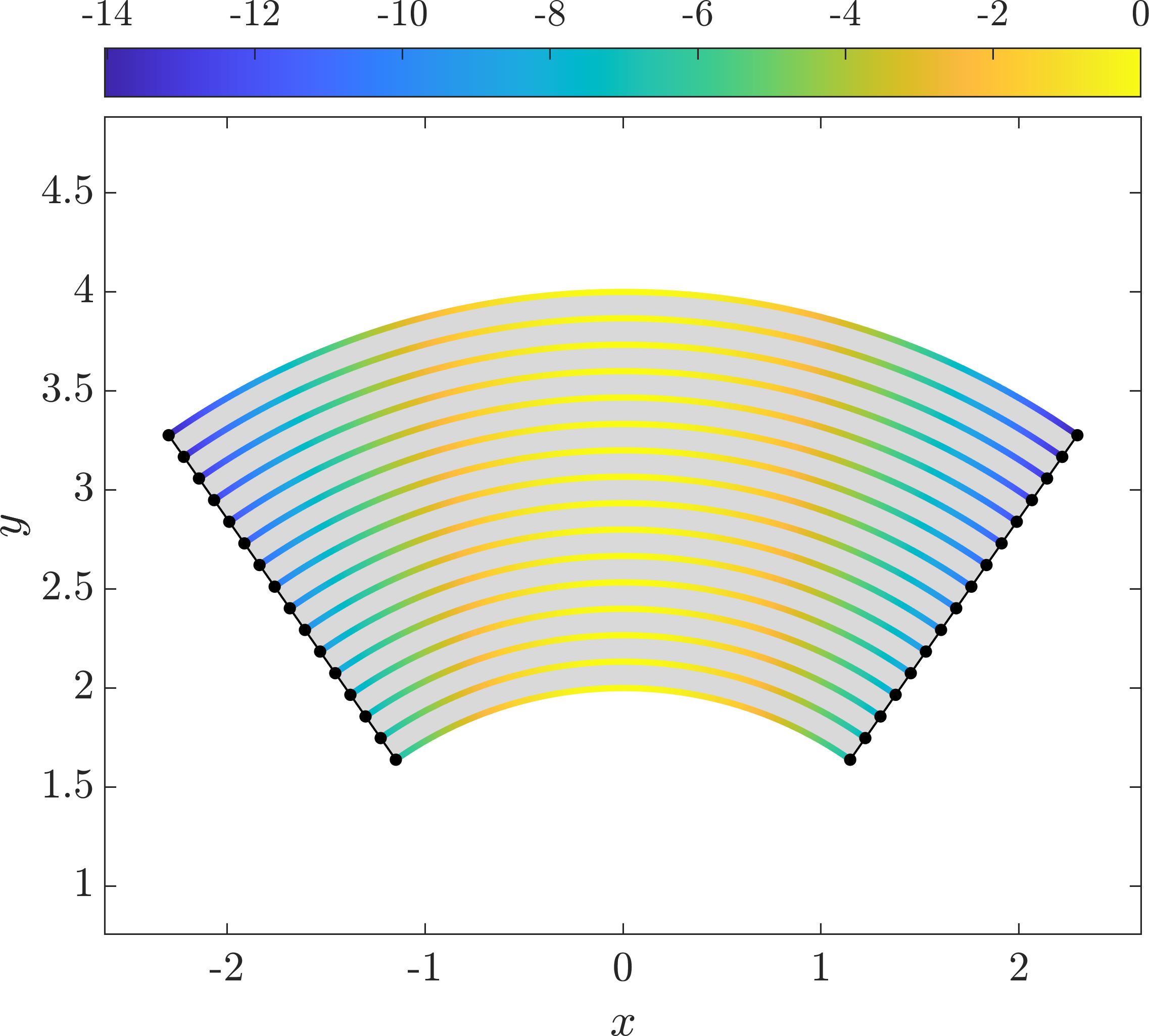}\label{Fig:TC1ArcsNormalForceLS}}\hfill
    \subfigure[shear force over $\Gamma^c$]
 {\includegraphics[width=0.3315\textwidth]{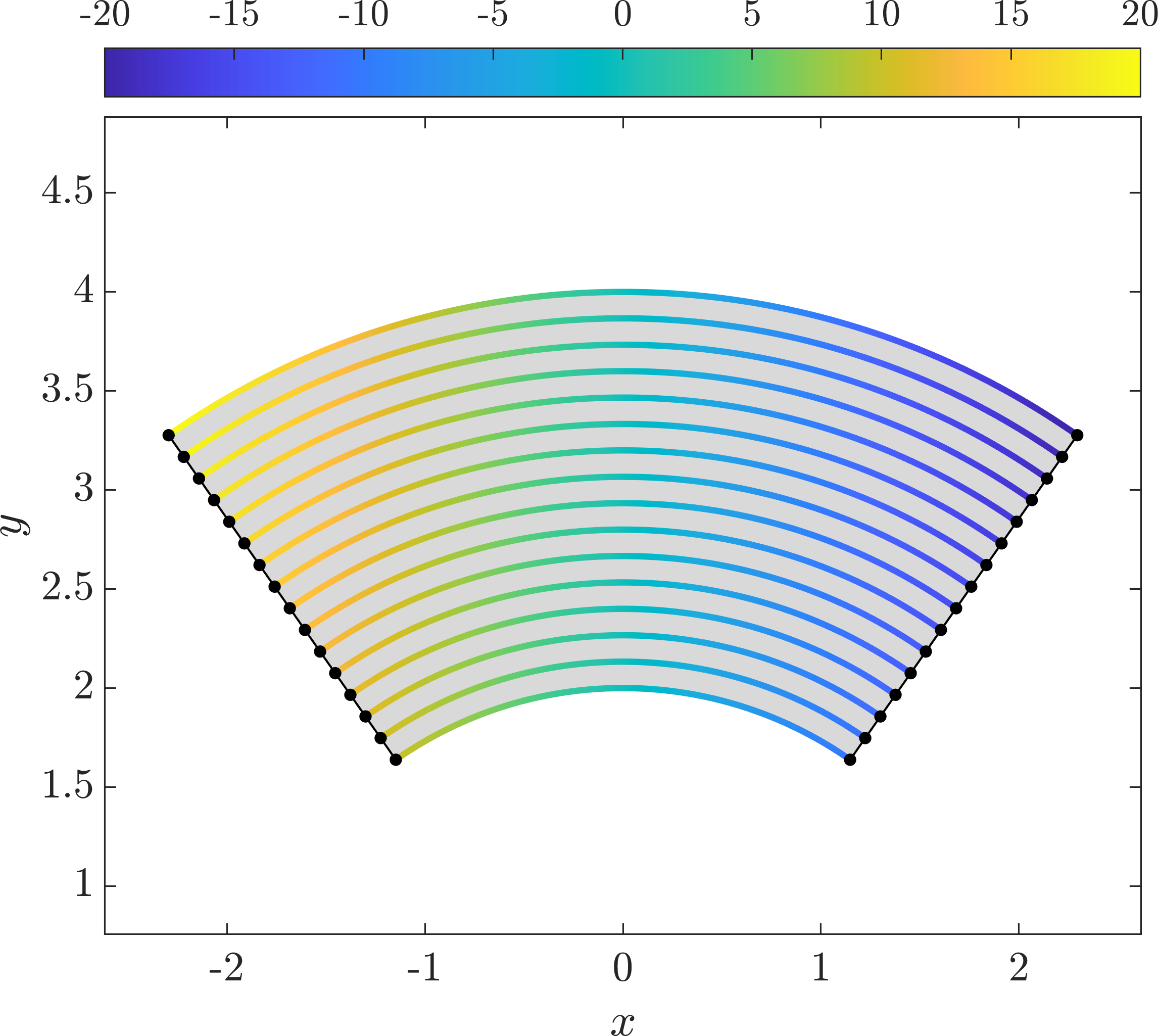}\label{Fig:TC1ArcsShearForceLS}}
 
 \caption{\label{Fig:TC1ArcsStressResultants}Plots of different stress resultants of the family of arc-shaped beams in $\mathbb{R}^2$ over $\Omega$ and $\Gamma^c$. (a) and (d) show the only nonzero principal component of the moment tensor $m_\Gamma^{(1)}$, (b) and (e) the only nonzero principal component of the physical normal force tensor $n_{\Gamma}^{\text{real}(1)}$ and (c) and (f) the shear force $q$.}
\end{figure}

An analytical solution for an individual arc can be found in \cite{Neumeyer_2025a}. There, not only are the solutions for the components of the displacements provided, but also the solutions of the stress resultants, i.e., the moment, the (physical) normal force, and the shear force. This analytical solution can then be compared to the only nonzero principal components of the moment tensor and normal force tensor, i.e., $m_\Gamma^{(1)}$ and $n_{\Gamma}^{\text{real}(1)}$. The shear force can be computed as $q = [-n_y, n_x]^\text{T} \ScalProd (\mat{P} \cdot \text{div}_\Gamma \mat{m}_\Gamma)$. Solutions for the displacements are depicted in Fig.~\ref{Fig:TC1ArcsDisplacement} and for the stress resultants in Fig.~\ref{Fig:TC1ArcsStressResultants}.

\begin{figure}[ht!]
 \centering
 
 \subfigure[convergence in $\varepsilon_{\mathcal{L}^2,\text{rel}}(\vek{u})$]
 {\includegraphics[width=0.32\textwidth]{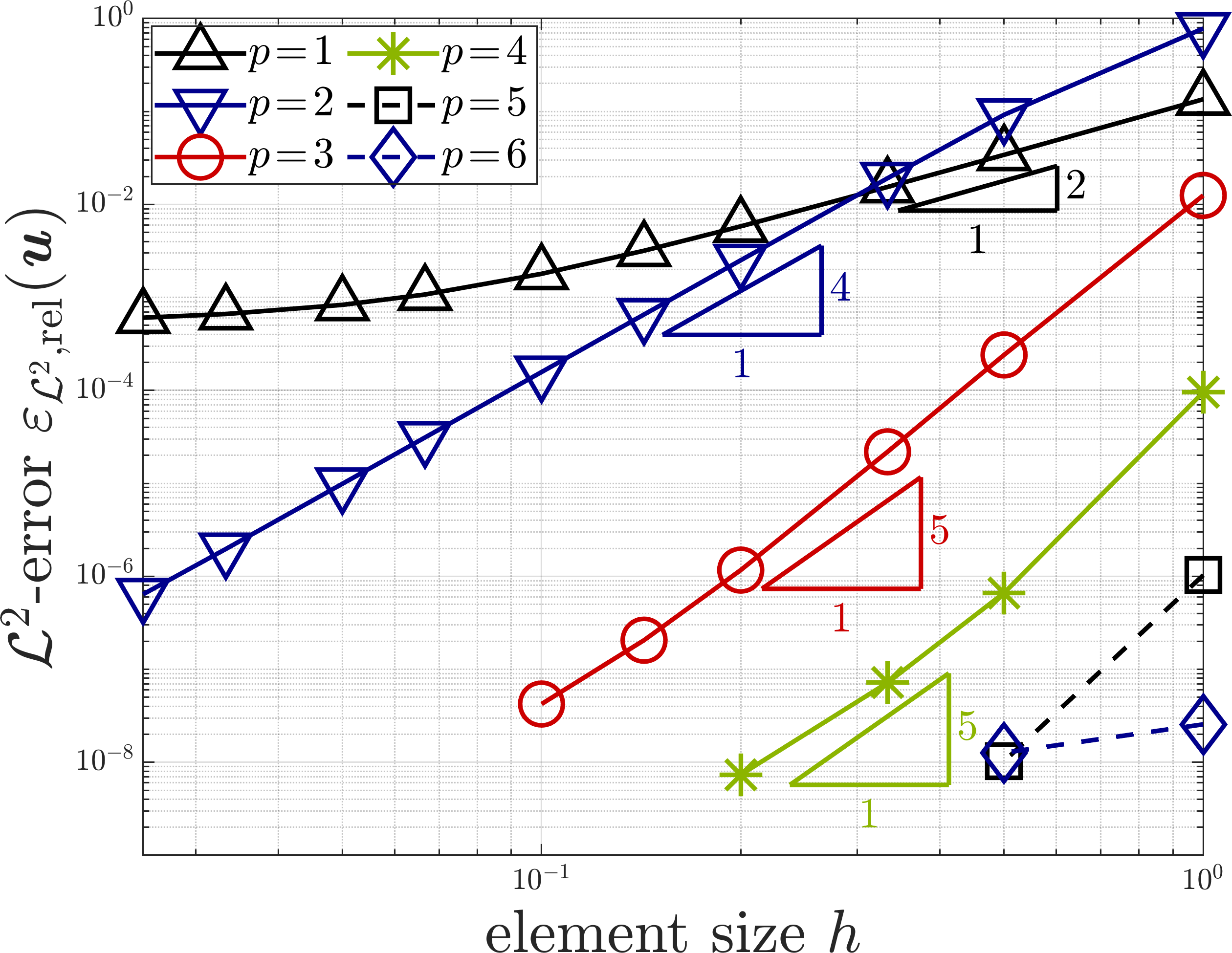}\label{Fig:TC1ArcsL2ErrorDisplacement}}
 \subfigure[convergence in $\varepsilon_{\mathcal{L}^2,\text{rel}}(m_\Gamma^{(1)})$]
 {\includegraphics[width=0.32\textwidth]{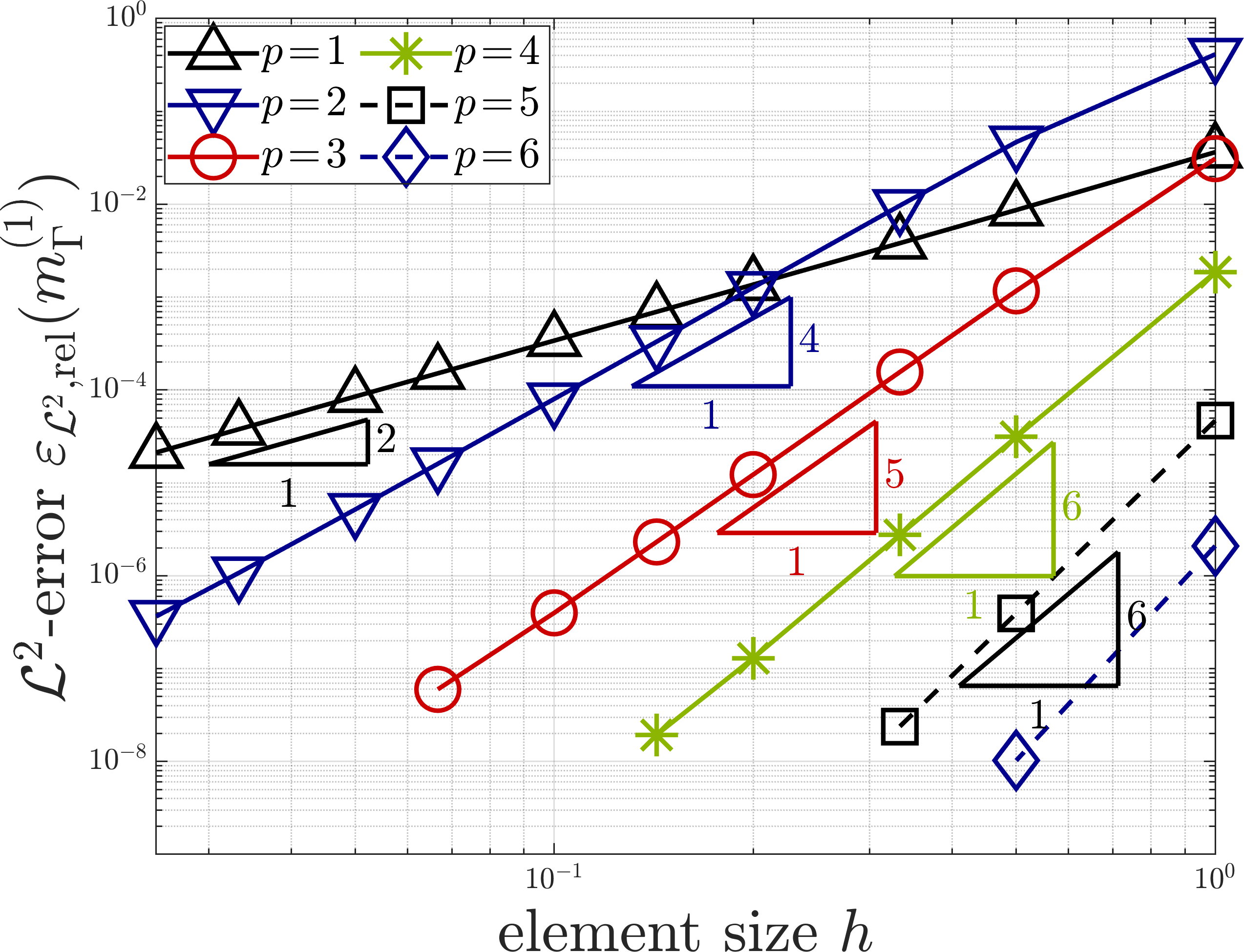}\label{Fig:TC1ArcsL2ErrorMoment}}\\
 \subfigure[convergence in $\varepsilon_{\mathcal{L}^2,\text{rel}}(n_{\Gamma}^{\text{real}(1)})$]
 {\includegraphics[width=0.32\textwidth]{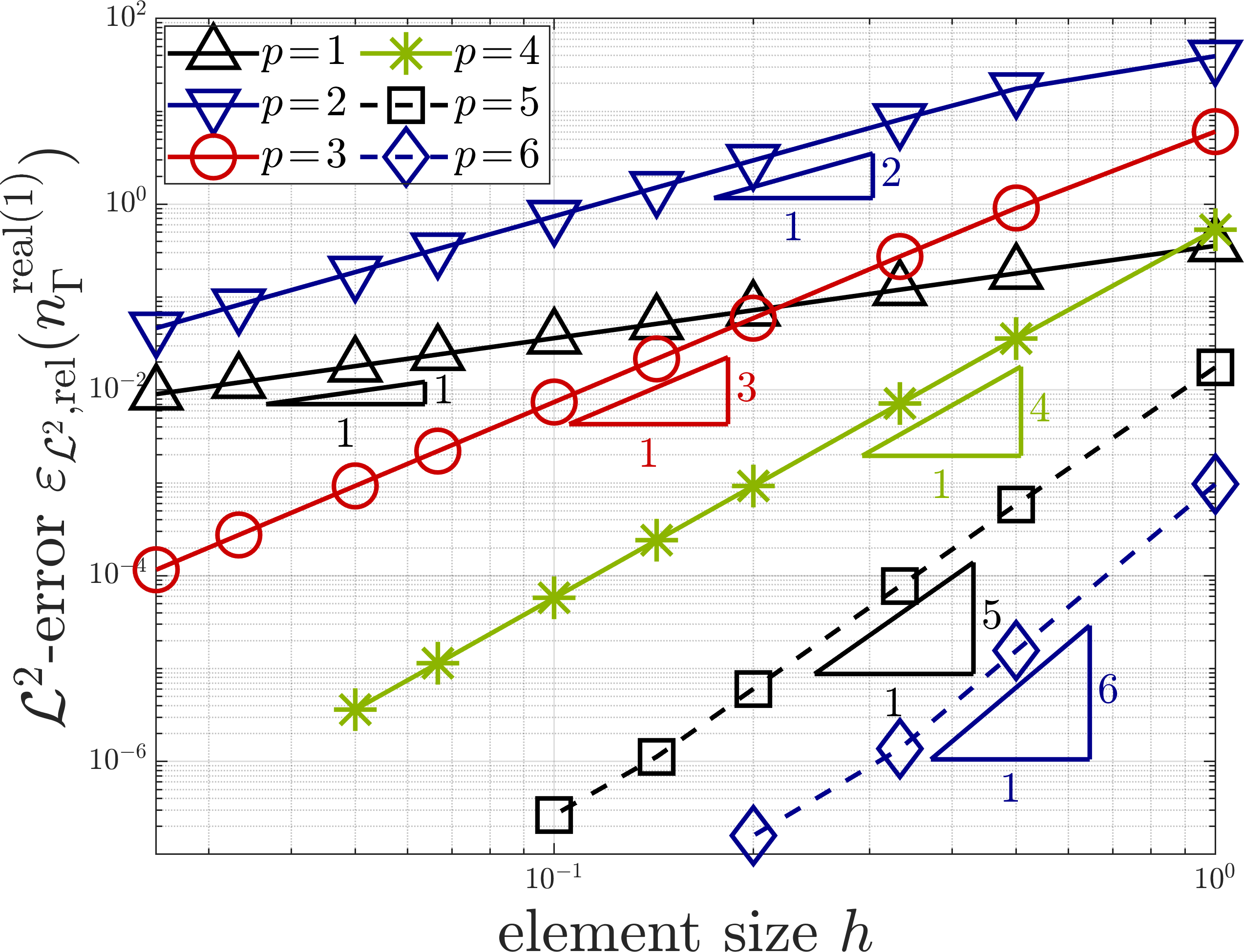}\label{Fig:TC1ArcsL2ErrorNormalForce}}
 \subfigure[convergence in $\varepsilon_{\mathcal{L}^2,\text{rel}}(q)$]
 {\includegraphics[width=0.32\textwidth]{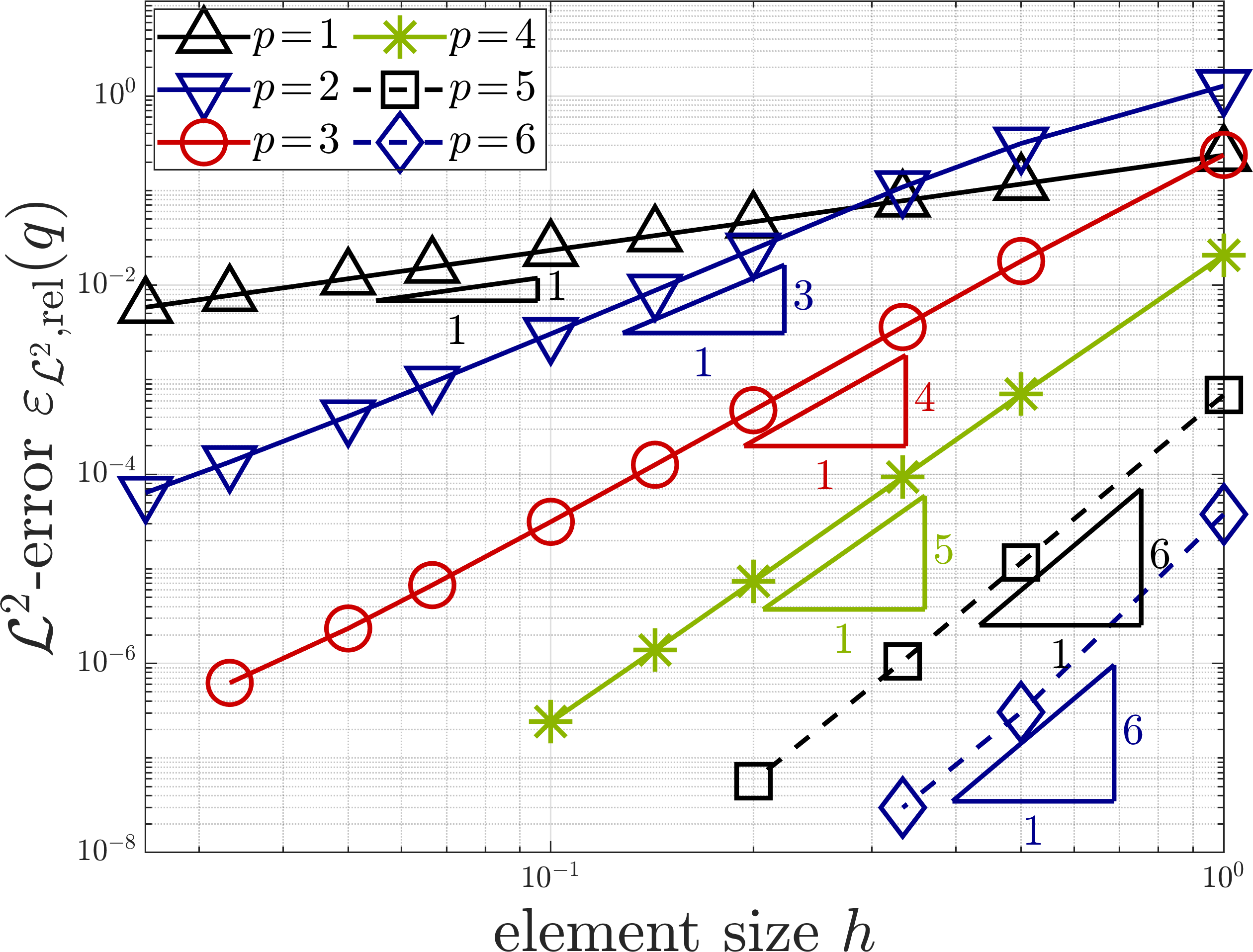}\label{Fig:TC1ArcsL2ErrorShearForce}}
  \subfigure[convergence in $\varepsilon_{\mathfrak{e}}$]
 {\includegraphics[width=0.32\textwidth]{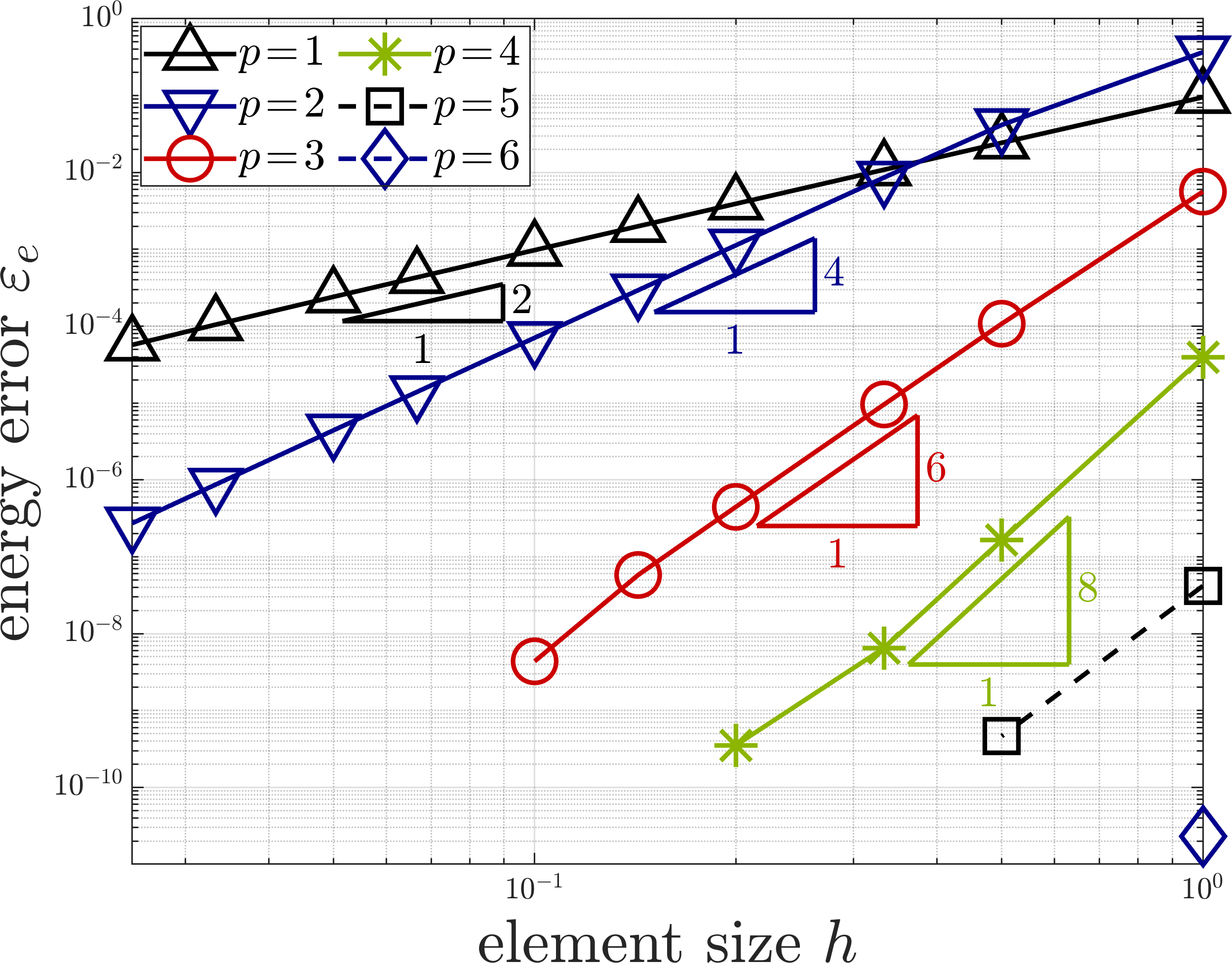}\label{Fig:TC1ArcsEnergyError}}
 
 \caption{\label{Fig:TC1ArcsError}Convergence studies for the family of arc-shaped beams in $\mathbb{R}^2$. Convergence rates in the $\mathcal{L}^2$-error for (a) the displacements $\vek{u}$, (b) the only nonzero principal component of the moment tensor $m_\Gamma^{(1)}$, (c) the only nonzero principal component of the physical normal force tensor $n_{\Gamma}^{\text{real}(1)}$, and (d) the shear force $q$ and (e) the relative stored energy error $\varepsilon_{\mathfrak{e}}$.}
\end{figure}

Based on the analytical solution, convergence studies in the relative $\mathcal{L}^2$-error, as defined in Eq.~(\ref{Eq:L2Error}), are performed and are depicted in Fig.~\ref{Fig:TC1ArcsL2ErrorDisplacement}-\ref{Fig:TC1ArcsL2ErrorShearForce}. The convergence studies in the primary unknowns $\vek{u}$ and $\mat{m}_\Gamma$ show optimal convergence rates of at least $\mathcal{O}(p+1)$ with some instances of super convergence for some orders, as seen in Figs.~\ref{Fig:TC1ArcsL2ErrorDisplacement} and \ref{Fig:TC1ArcsL2ErrorMoment}. For the derived quantities, i.e., $\mat{n}_\Gamma^\text{real}$ and $q$, the expected convergence of $\mathcal{O}(p)$ is obtained; see Figs.~\ref{Fig:TC1ArcsL2ErrorNormalForce} and \ref{Fig:TC1ArcsL2ErrorShearForce}. Instances of super convergence can also be seen in the error of the shear force $q$. A convergence study for the stored energy error from Eq.~(\ref{Eq:EnergyError}) is shown in Fig.~\ref{Fig:TC1ArcsEnergyError}. Here, a convergence rate of $\mathcal{O}(2\,p)$ is observed. The reference energy for this error is calculated analytically with $\mathfrak{e}_{\text{ref}} = 3.49511413801986 \cdot 10^{-2}$.

\subsection{Family of circular beams over a more general bulk domain in \texorpdfstring{$\mathbb{R}^2$}{R2}}\label{Sec:TC2BeamsInCircularBD}
\begin{figure}[ht!]
 \centering
 
 \subfigure[geometry]
 {\includegraphics[width=0.457\textwidth]{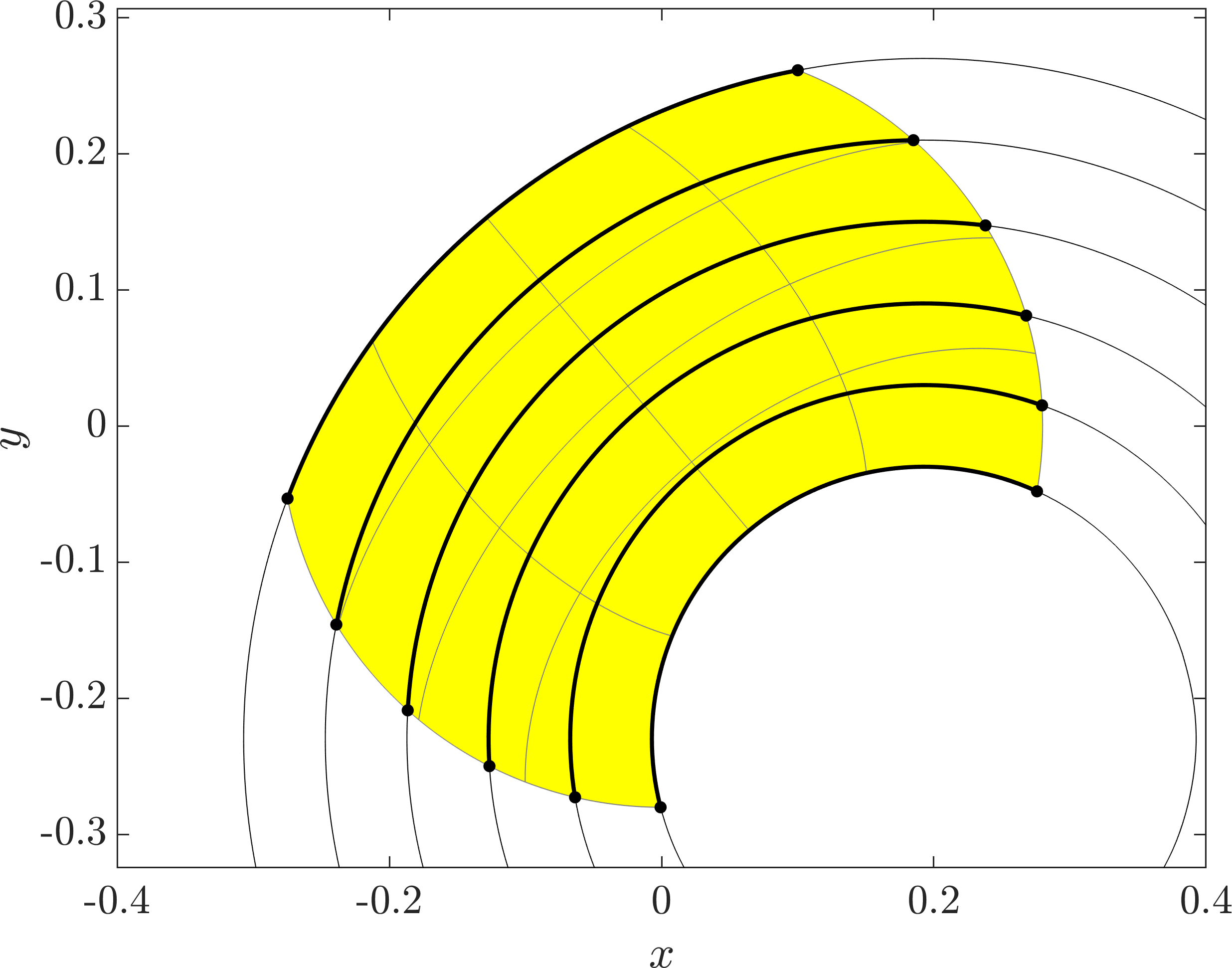}\label{Fig:TC2BeamsInCircularBDGeometry}}\hfill
 \subfigure[displacements]
 {\includegraphics[width=0.523\textwidth]{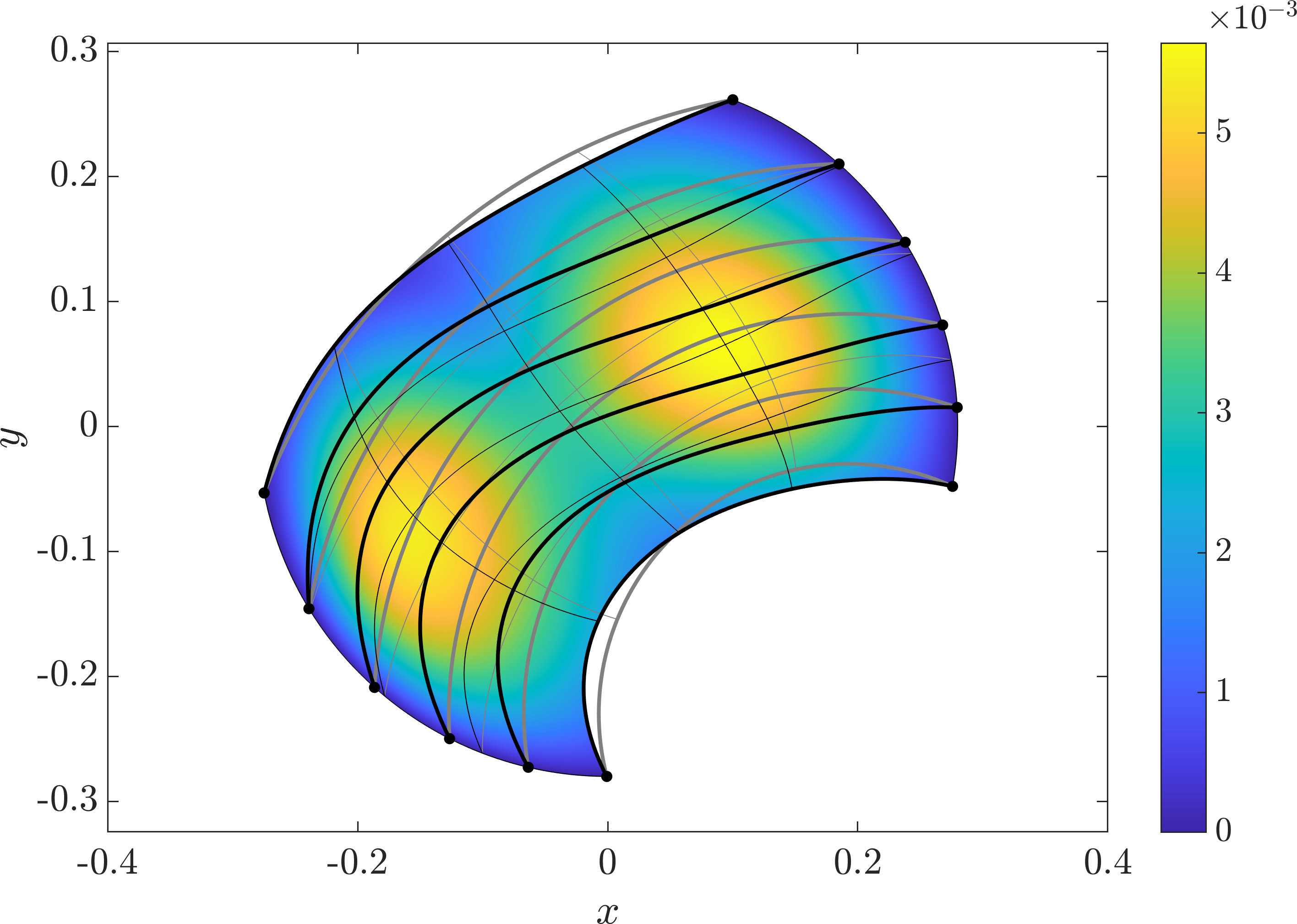}\label{Fig:TC2BeamsInCircularBDDisplacement}}
 
 \caption{\label{Fig:TC2BeamsInCircularBD}Setup of the family of circular beams over a more general bulk domain in $\mathbb{R}^2$: (a) Geometry of the discretized bulk domain $\Omega^h$ and some selected beams $\Gamma^c$. (b) Scaled deformed configuration with the Euclidean norm of the displacements $\| \vek{u} \|$ as a color plot over the bulk domain. The gray mesh lines and level sets indicate the undeformed configuration.}
\end{figure}

The geometric setup is reproduced from \cite{Kaiser_2024b} where a Bulk Trace FEM for Timoshenko-Ehrenfest beams was considered. Again, there is a family of circular beams, however, over a more general and unsymmetrical bulk domain, see Fig.~\ref{Fig:TC2BeamsInCircularBDGeometry}. The geometry of the beams is defined via the level-set function
\begin{equation}\label{Eq:TC2BeamsInCircularBD_LSF}
    \phi(\vek{x}) = \sqrt{(x + x_c)^2 + (y + y_c)^2}-r_c,
\end{equation}\\
implying circles centered at $x_c = -r_c \, \sin(\frac{11 \, \pi}{9})$ and $y_c = r_c \, \cos(\frac{11 \, \pi}{9})$ using $r_c = 0.3$. With the level-set function $\psi(\vek{x}) = \sqrt{x^2 + y^2}$, the geometry of the bulk domain is defined as
\begin{equation}\label{Eq:TC2BeamsInCircularBD_BD}
    \Omega = \big\{ \vek{x} \in \mathbb{R}^2: \psi(\vek{x}) \leq 0.28 , \, -0.1 < \phi(\vek{x}) < 0.2 \big\}.
\end{equation}
The geometric situation is visualized in Fig.~\ref{Fig:TC2BeamsInCircularBDGeometry}.

The input parameters of the test case are as follows: the load vector $\vek{f} = [0, -100]^\text{T}$, Young's modulus $E = 2.1\cdot10^8$, the area $A = b\,h$, and the area moment of inertia $I = \frac{b\,h^3}{12}$, with $b = 0.01$ and $h = 0.02$. The boundary conditions are simply supported on the whole boundary $\partial\Omega$. The solution of the displacements is depicted in Fig.~\ref{Fig:TC2BeamsInCircularBDDisplacement}.

\begin{figure}[ht!]
 \centering
 
 \subfigure[convergence in $\varepsilon_{\text{res},1}$]
 {\includegraphics[width=0.33\textwidth]{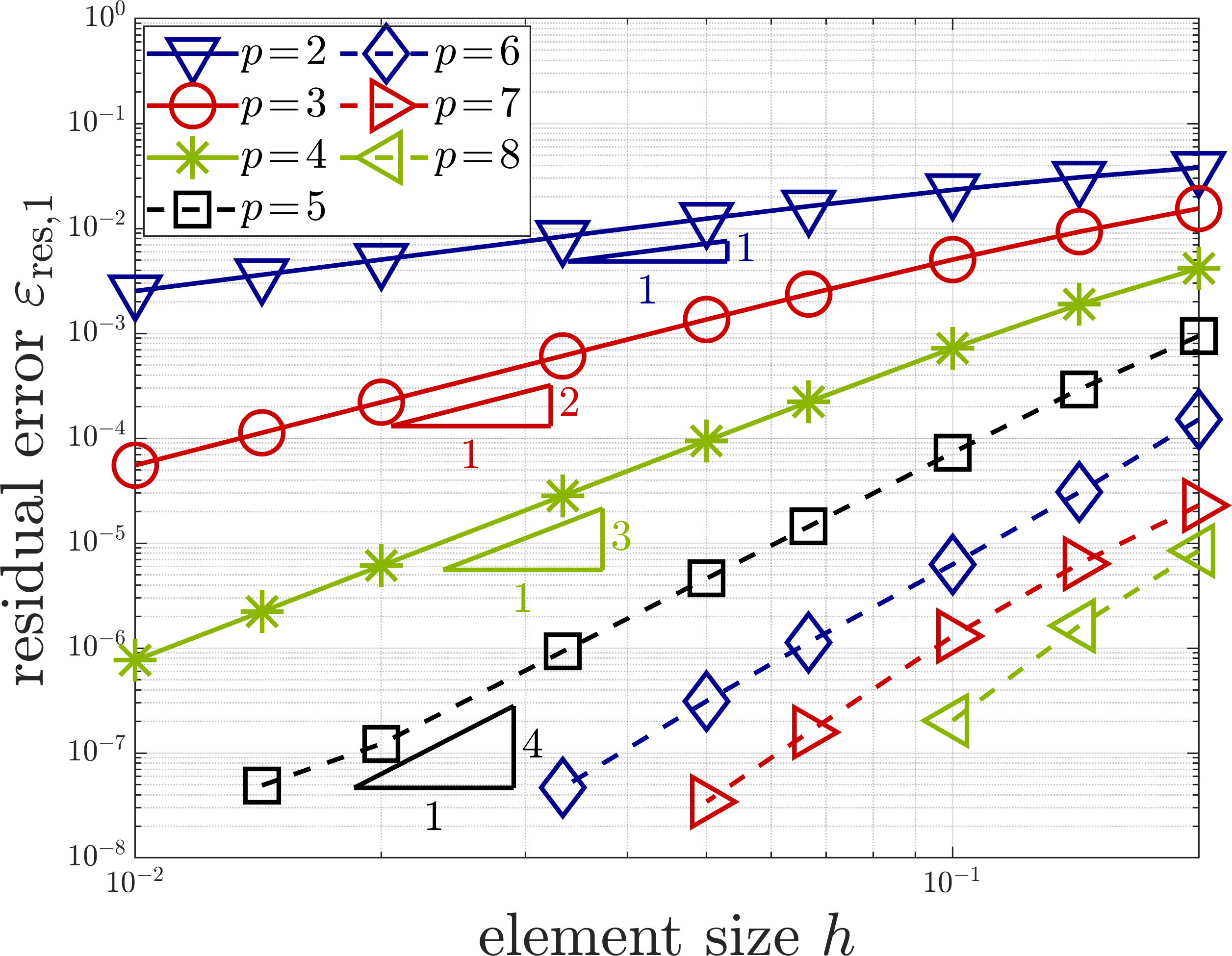}\label{Fig:TC2BeamsInCircularBDResError1}}\hfill
 \subfigure[convergence in $\varepsilon_{\text{res},2}$]
 {\includegraphics[width=0.33\textwidth]{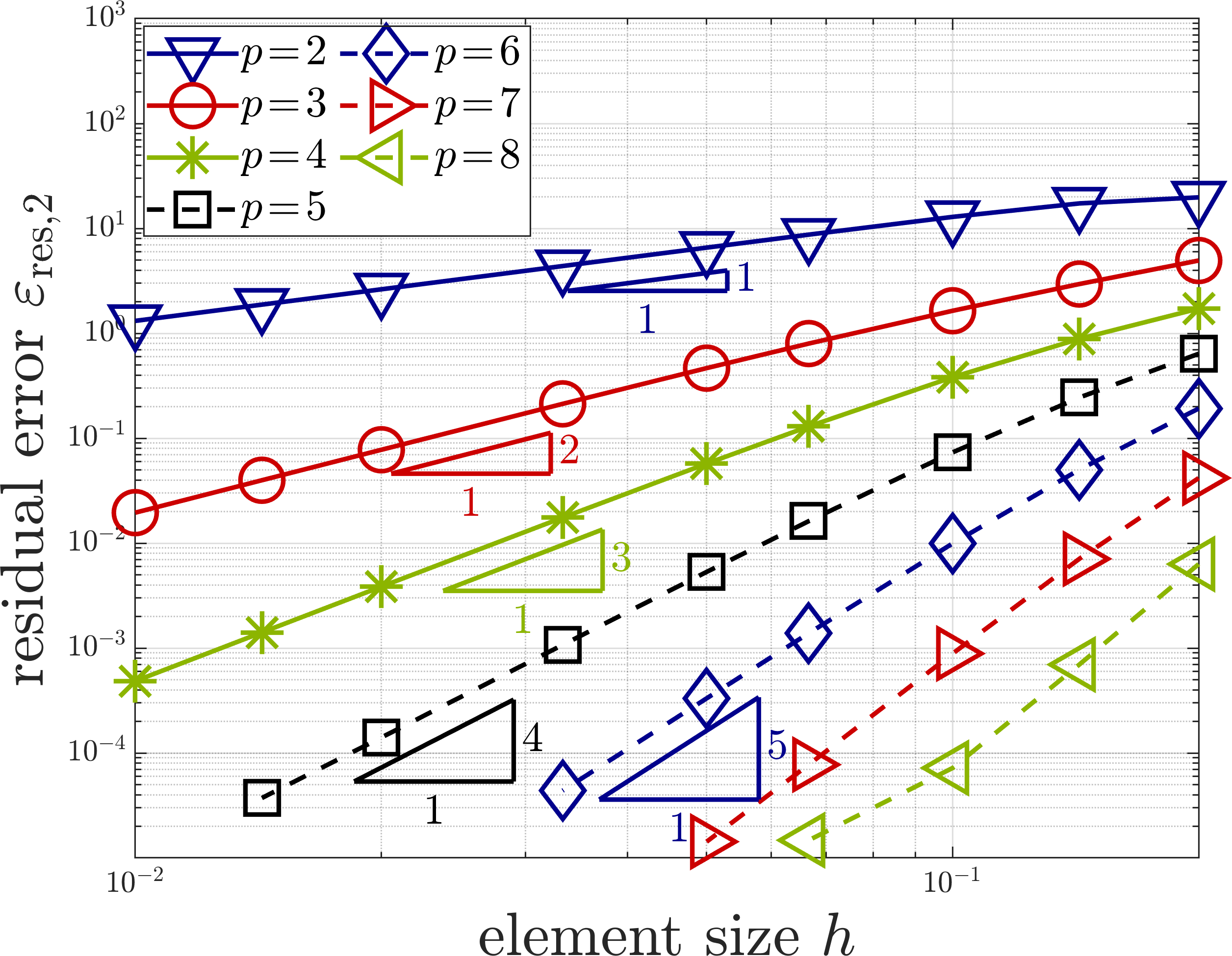}\label{Fig:TC2BeamsInCircularBDResError2}}\hfill
 \subfigure[convergence in $\varepsilon_{\mathfrak{e}}$]
 {\includegraphics[width=0.33\textwidth]{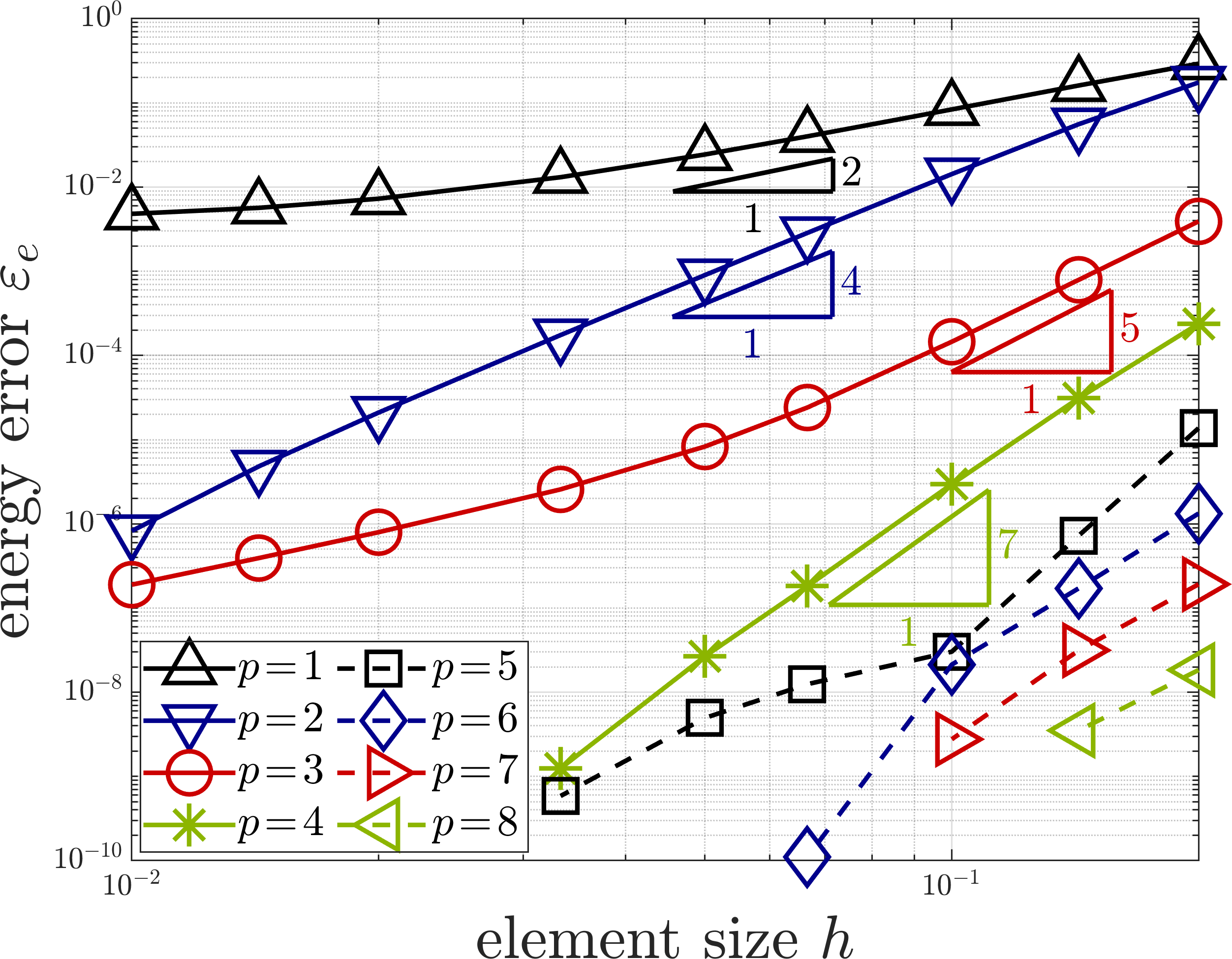}\label{Fig:TC2BeamsInCircularBDEnergyError}}
 
 \caption{\label{Fig:TC2BeamsInCircularBDError}Convergence studies for the family of circular beams over a more general bulk domain in $\mathbb{R}^2$. Convergence rates in (a) the absolute residual error $\varepsilon_{\text{res},1}$, (b) the relative residual error $\varepsilon_{\text{res},2}$, and (c) the relative stored energy error $\varepsilon_{\mathfrak{e}}$.}
\end{figure}

As the analytical solution is not available, we aim to confirm higher-order convergence rates using the residual and stored energy errors from Eqs.~(\ref{Eq:ResidualError1})-(\ref{Eq:EnergyError}). For the convergence analysis in the residual errors depicted in Figs.~\ref{Fig:TC2BeamsInCircularBDResError1} and \ref{Fig:TC2BeamsInCircularBDResError2}, the optimal convergence rate of $\mathcal{O}(p-1)$ can be observed. Higher-order curves are slightly sub-optimal, indicating a pre-asymptotic regime. For the stored energy error analysis, depicted in Fig.~\ref{Fig:TC2BeamsInCircularBDEnergyError}, a reference energy of $\mathfrak{e}_{\text{ref}} = 1.36582967 \cdot 10^{-2}$ is used. Although higher-order accuracy is confirmed in this error measure as well, the optimal rate of $\mathcal{O}(2\,p)$ is not achieved. For example, some suboptimal behavior is observed for the orders $p = 1$ and $p = 3$, where the curves flatten for finer meshes.

\subsection{Family of simply supported shells in \texorpdfstring{$\mathbb{R}^3$}{R2}}\label{Sec:TC3SimplySupGenShells}
\begin{figure}[ht!]
 \centering

 \subfigure[geometry]
 {\includegraphics[width=0.455\textwidth]{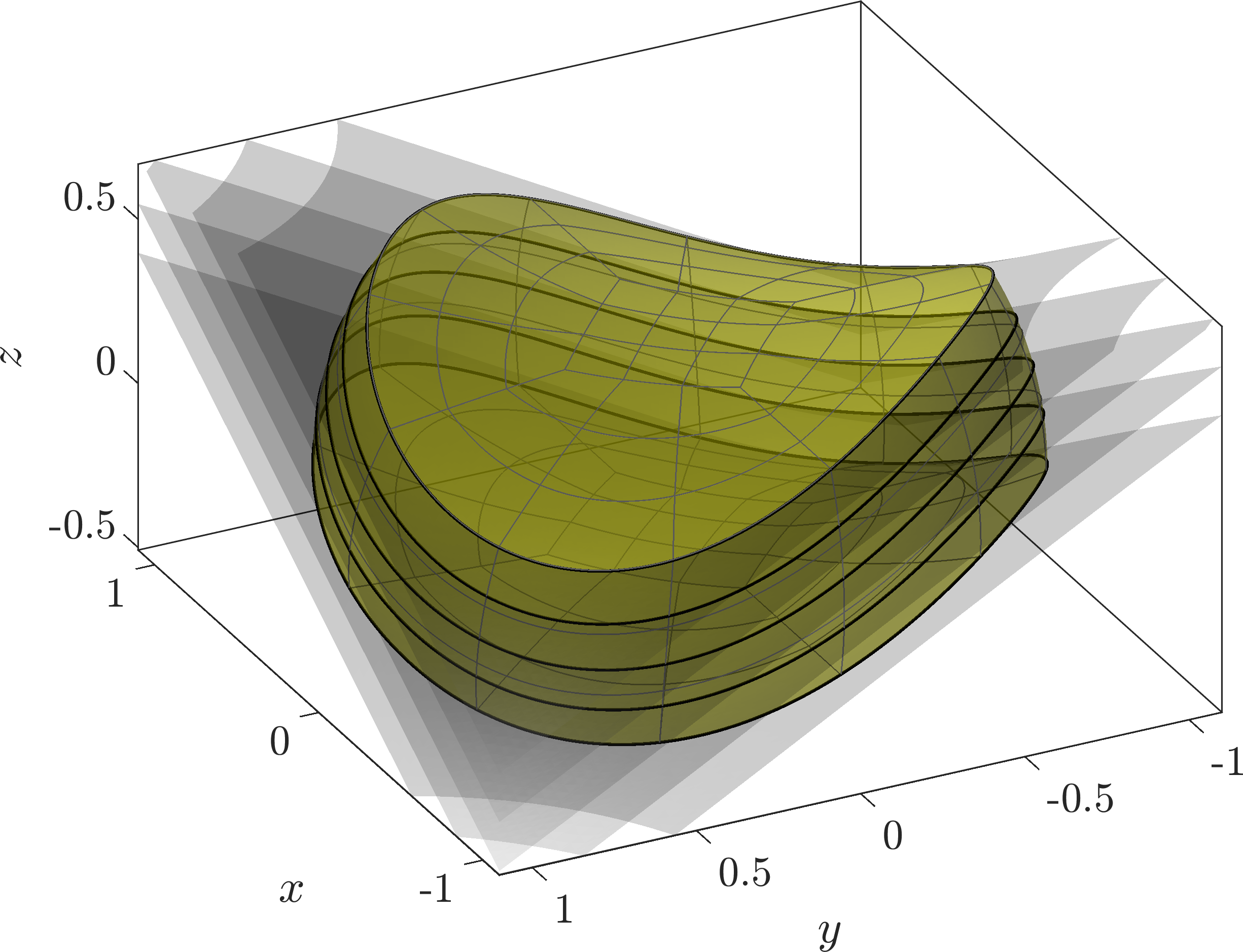}\label{Fig:TC3SimplySupGenShellsGeometry}}\hfill
 \subfigure[displacements]
 {\includegraphics[width=0.525\textwidth]{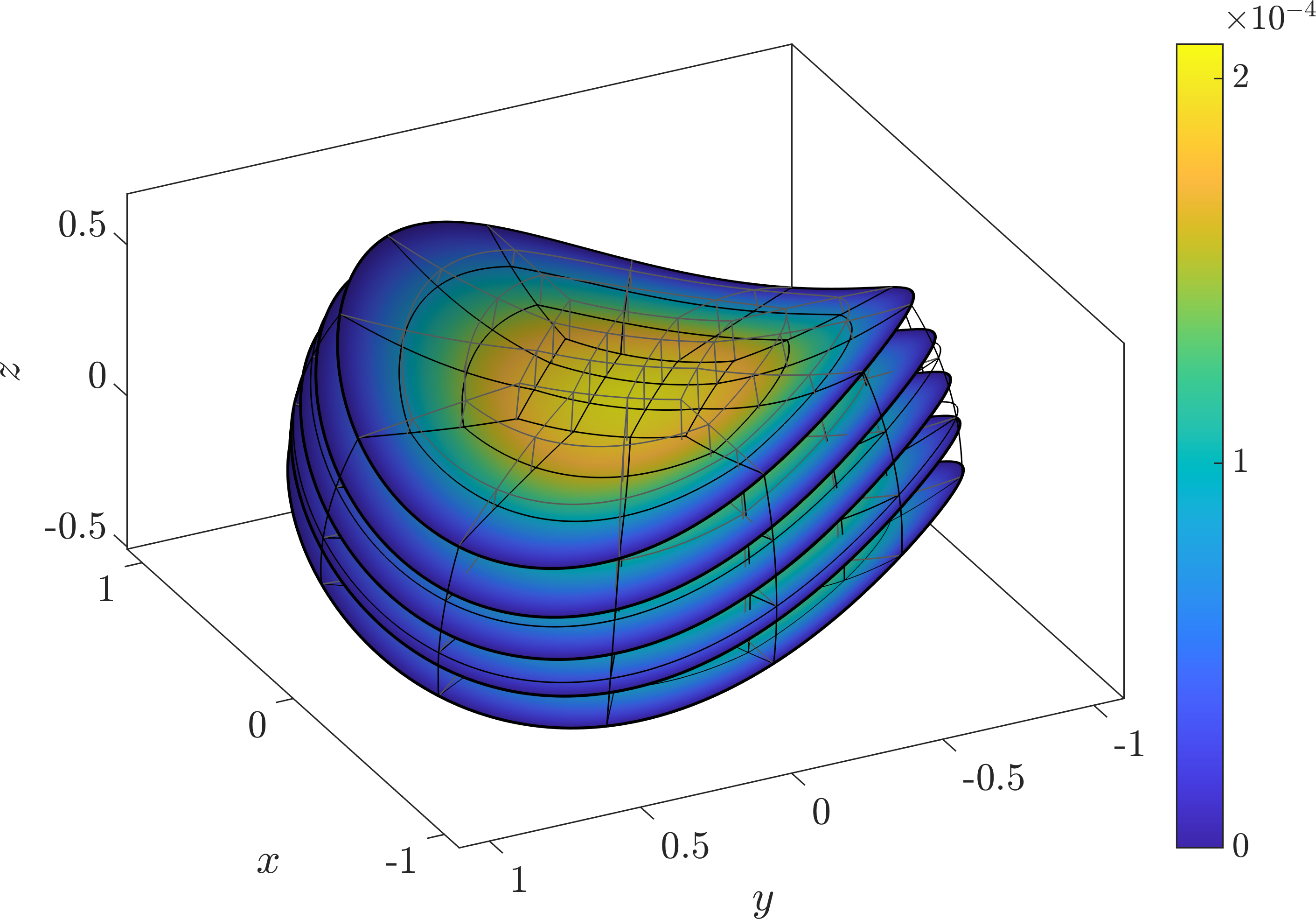}\label{Fig:TC3SimplySupGenShellsDisplacement}}
 
 \caption{\label{Fig:TC3SimplySupGenShells}Setup of the family of simply supported shells in $\mathbb{R}^3$: (a) Geometry of the discretized bulk domain $\Omega^h$ and some selected shells $\Gamma^c$. (b) Scaled deformed configuration with the Euclidean norm of the displacements $\| \vek{u} \|$ as a contour plot over some selected shells. The gray mesh lines imply the undeformed configuration.}
\end{figure}

The family of simply supported shells follows the outlines of a test case in \cite{Kaiser_2024a}. Although the test case was originally designed for a Bulk Trace FEM for Reissner--Mindlin shells, the results are well comparable to Kirchhoff--Love shells, as shear deformations only contribute a small share to the overall results. The geometry of the shells is described by the level-set function
\begin{equation}\label{Eq:TC3SimplySupGenShells_LSF}
    \phi(\vek{x}) = z - 2 \, \sin(\tfrac{1}{4}\,x\,y).
\end{equation}
The bulk domain is given by
\begin{equation}\label{Eq:TC3SimplySupGenShells_BD}
    \Omega = \big\{ \vek{x} \in \mathbb{R}^2: \psi(\vek{x}) \leq 1 , \, -0.2 < \phi(\vek{x}) < 0.4 \big\},
\end{equation}
where $\psi(\vek{x}) = \sqrt{x^2 + y^2 + z^2}$ is a level-set function implying a sphere. In Fig.~\ref{Fig:TC3SimplySupGenShellsGeometry}, the geometry of the discretized bulk domain and some level sets are shown.

Regarding the test case parameters, the load vector is set to $\vek{f} = [0, 0, -100]^\text{T}$, Young's modulus $E = 2.1 \cdot 10^7$, Poisson's ratio $\nu = 0.3$, and the thickness of the shells $t = 0.1$. The whole boundary $\partial\Omega$ is simply supported. Based on these input parameters, the solution of the displacements is depicted in Fig.~\ref{Fig:TC3SimplySupGenShellsDisplacement}.

\begin{figure}[ht!]
 \centering
 
 \subfigure[convergence in $\varepsilon_{\text{res},1}$]
 {\includegraphics[width=0.33\textwidth]{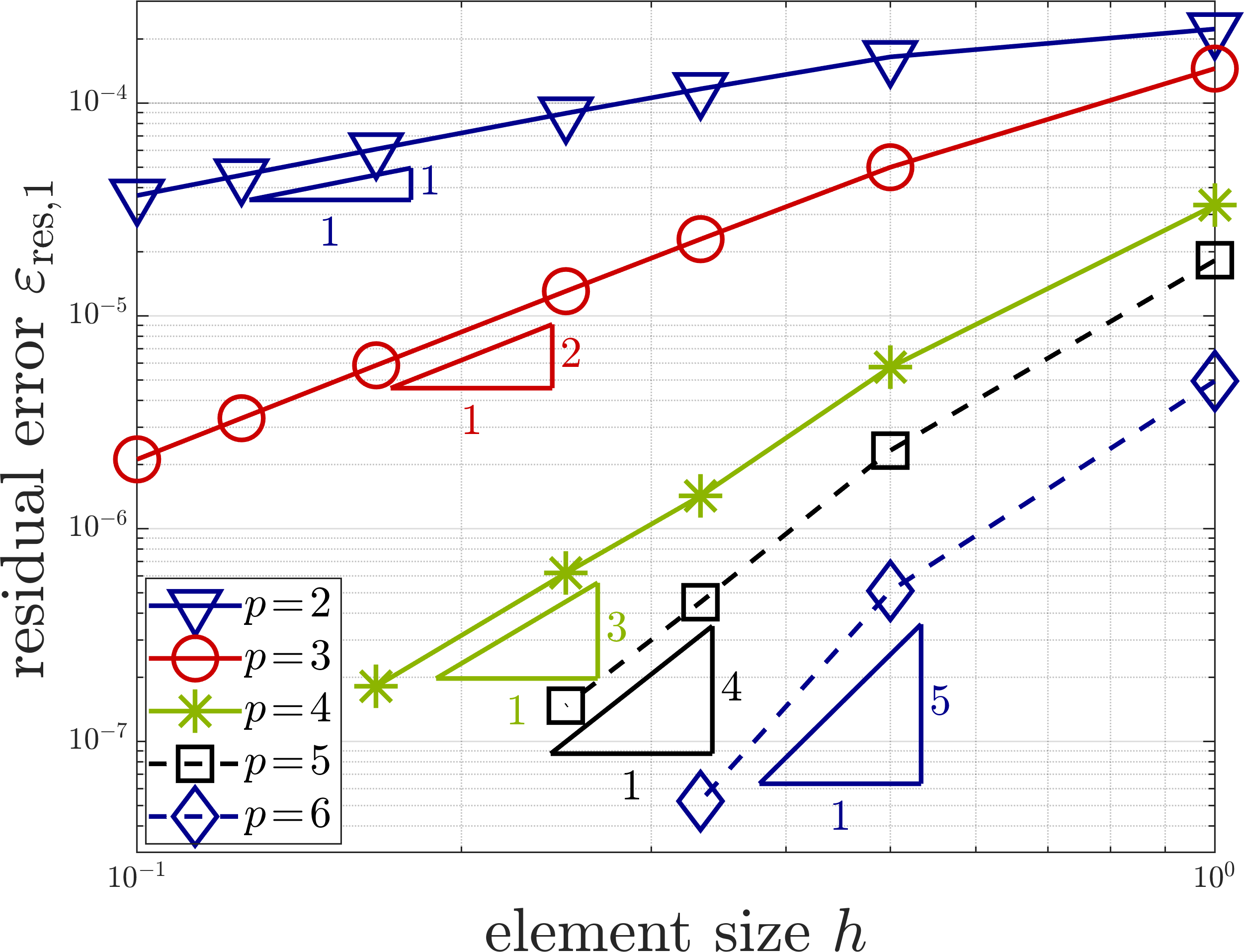}\label{Fig:TC3SimplySupGenShellsResError1}}\hfill
 \subfigure[convergence in $\varepsilon_{\text{res},2}$]
 {\includegraphics[width=0.33\textwidth]{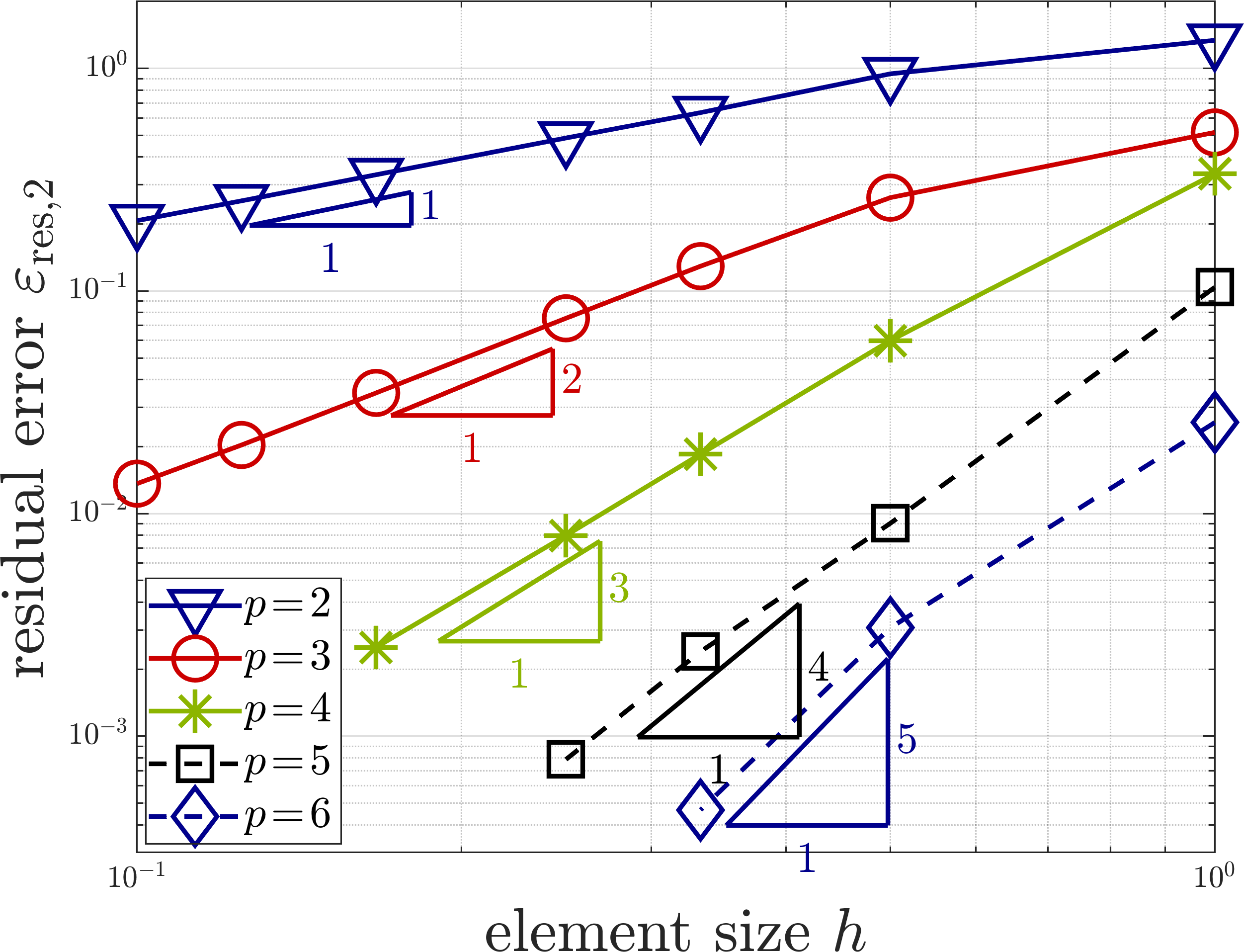}\label{Fig:TC3SimplySupGenShellsResError2}}\hfill
 \subfigure[convergence in $\varepsilon_{\mathfrak{e}}$]
 {\includegraphics[width=0.33\textwidth]{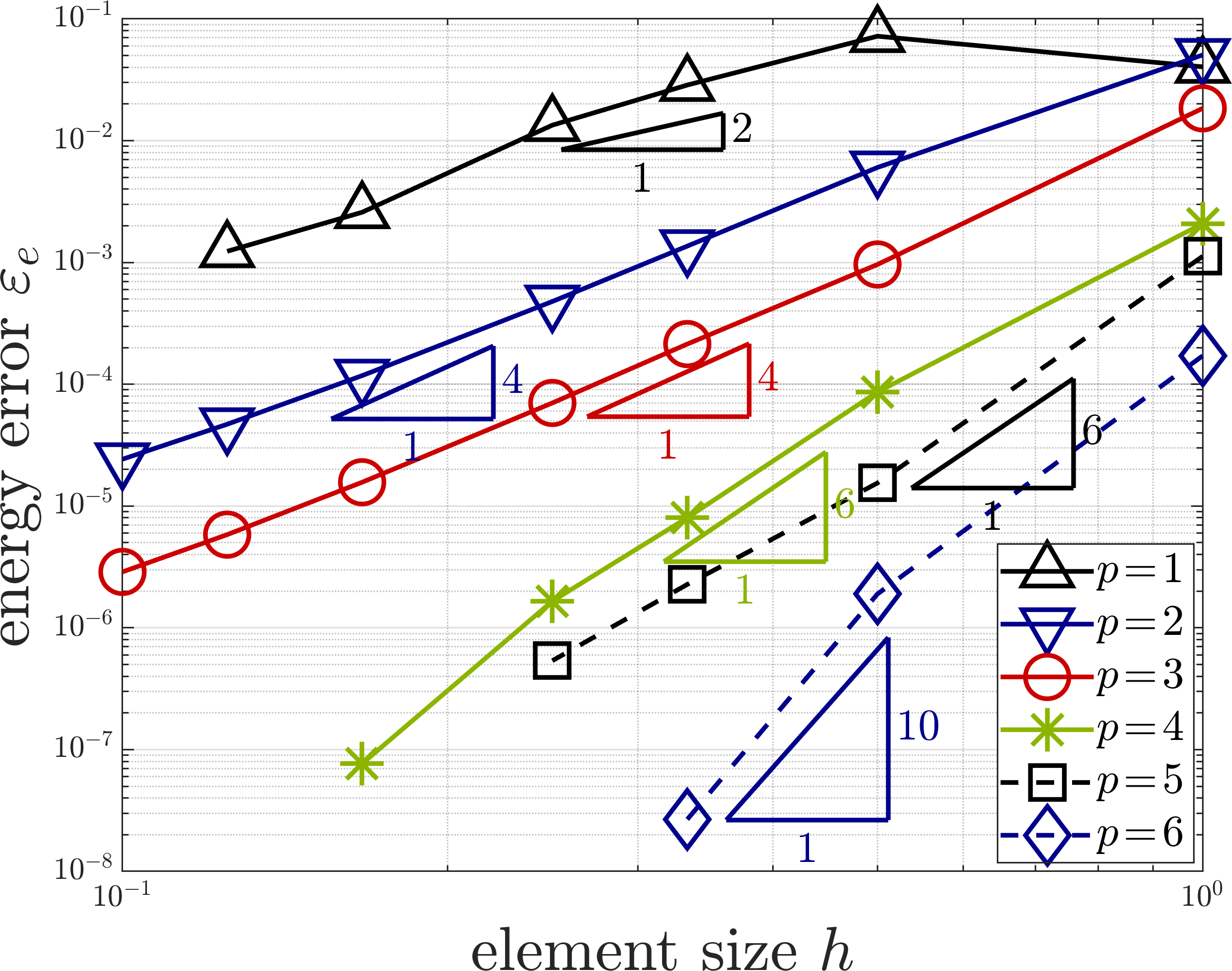}\label{Fig:TC3SimplySupGenShellsEnergyError}}
 
 \caption{\label{Fig:TC3SimplySupGenShellsError}Convergence studies for the family of simply supported shells in $\mathbb{R}^3$. Convergence rates in (a) the absolute residual error $\varepsilon_{\text{res},1}$, (b) the relative residual error $\varepsilon_{\text{res},2}$, and (c) the relative stored energy error $\varepsilon_{\mathfrak{e}}$.}
\end{figure}

In the residual errors, shown in Figs.~\ref{Fig:TC3SimplySupGenShellsResError1} and \ref{Fig:TC3SimplySupGenShellsResError2}, optimal convergence of $\mathcal{O}(p-1)$ is achieved. Due to higher computational efforts for shells in the Bulk Trace FEM---now performed on three-dimensional meshes---we restrict the results up to 6-th order elements and reduce the total number of elements considered when compared to the beam studies from before. Therefore, distinctive pre-asymptotic regimes are visible in the convergence plots, however, it can be seen how the curves approach the optimal asymptotic range. This effect can be seen in Fig.~\ref{Fig:TC3SimplySupGenShellsEnergyError}, where convergence in the stored energy error is analyzed. Here, a convergence rate of $\mathcal{O}(p+1)$ for odd $p$ and $\mathcal{O}(p+2)$ for even $p$ is observed, a behavior which has also been found in \cite{Fries_2020a, Fries_2023a} and traced back to errors in numerical integration in isoparametric $p$-FEM for general integrands. For the reference energy, a value of $\mathfrak{e}_{\text{ref}} = 9.894785 \cdot 10^{-3}$ has been determined, which is in alignment with \cite{Kaiser_2024a} with a reference energy $\mathfrak{e}_{\text{ref}} = 9.917787434703 \cdot 10^{-3}$. The small deviation of the values can be explained by the shear contribution, which is considered in the Reissner--Mindlin shell theory, however, neglected for Kirchhoff--Love shells as considered here.

\subsection{Family of clamped cupolas on a curved surface in \texorpdfstring{$\mathbb{R}^3$}{R2}}\label{Sec:TC4ClampedCopulasOnCurvedSurf}
\begin{figure}[ht!]
 \centering

 \subfigure[geometry]
 {\includegraphics[width=0.451\textwidth]{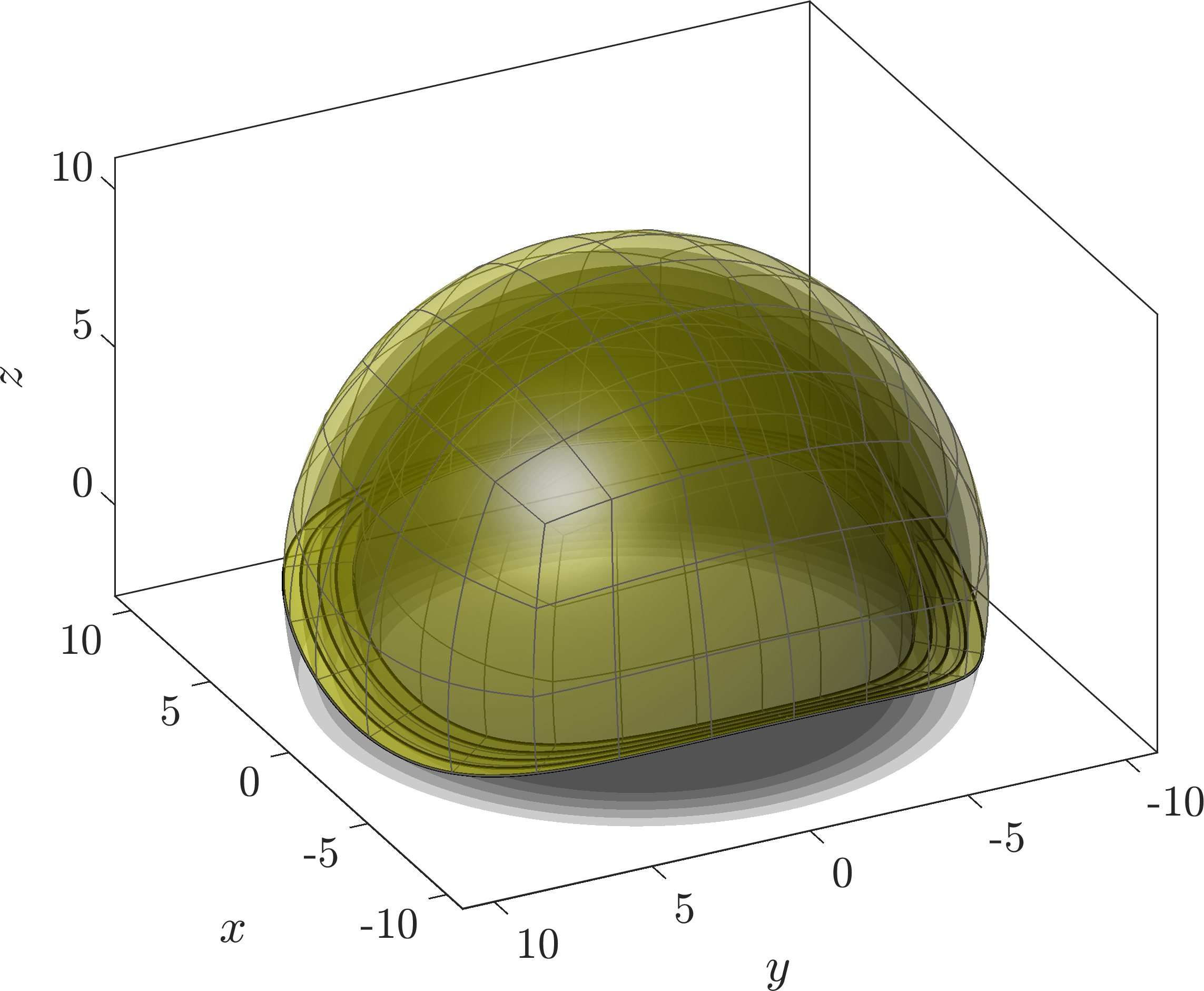}\label{Fig:TC4ClampedCopulasOnCurvedSurfGeometry}}\hfill
 \subfigure[displacements]
 {\includegraphics[width=0.529\textwidth]{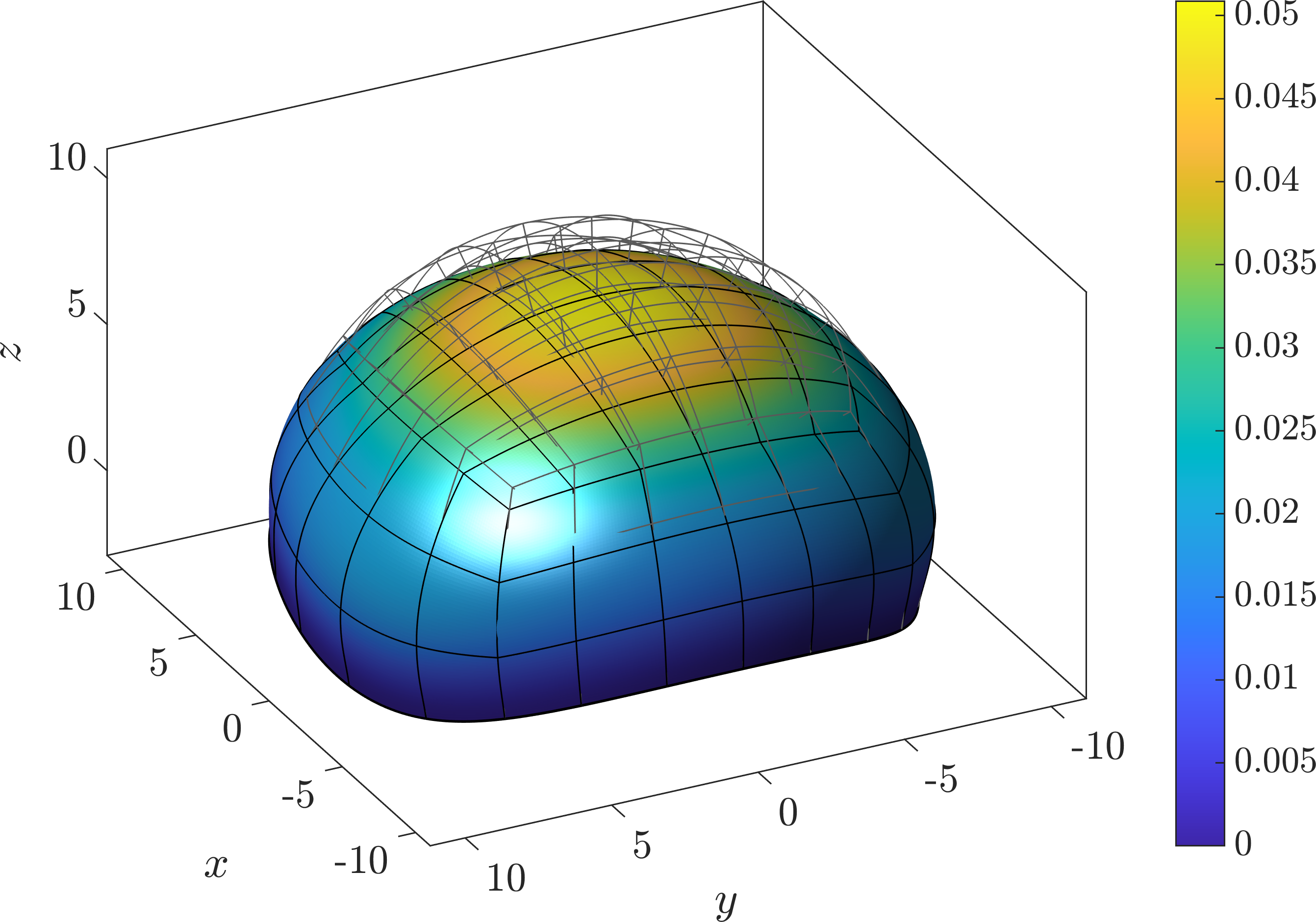}\label{Fig:TC4ClampedCopulasOnCurvedSurfDisplacement}}
 
 \caption{\label{Fig:TC4ClampedCopulasOnCurvedSurf}Setup of the family of clamped cupolas on a curved surface in $\mathbb{R}^3$: (a) Geometry of the discretized bulk domain $\Omega^h$ and some selected shells $\Gamma^c$. (b) Scaled deformed configuration with the Euclidean norm of the displacements $\| \vek{u} \|$ as a contour plot over some selected shells. The gray mesh lines imply the undeformed configuration.}
\end{figure}

Adopted from \cite{Kaiser_2024a}, the last test case represents a family of spherical cupolas, described by the level-set function
\begin{equation}\label{Eq:TC4ClampedCopulasOnCurvedSurf_LSF}
    \phi(\vek{x}) = \sqrt{x^2 + y^2 + z^2}.
\end{equation}
The cupolas are placed on a curved surface defined by $\psi(\vek{x}) = z - \sin(\frac{x}{3}) - \cos(\frac{y}{6}) + 1$. We define the bulk domain as
\begin{equation}\label{Eq:TC4ClampedCopulasOnCurvedSurf_BD}
    \Omega = \big\{ \vek{x} \in \mathbb{R}^2: \psi(\vek{x}) \leq 0, \, 8 < \phi(\vek{x}) < 10 \big\}.
\end{equation}
Fig.~\ref{Fig:TC4ClampedCopulasOnCurvedSurfGeometry} for the geometry of the bulk domain and the level-sets and Fig.~\ref{Fig:TC4ClampedCopulasOnCurvedSurfDisplacement} for the displacement results. The loading vector is defined as $\vek{f} = [0, 0, -10]^\text{T}$, Young's modulus as $E = 3.0 \cdot 10^4$, Poisson's ratio as $\nu = 0.3$, and the thickness of the shell as $t = 1.1$. The boundary is fully clamped.

\begin{figure}[ht!]
 \centering
 
 \subfigure[convergence in $\varepsilon_{\text{res},1}$]
 {\includegraphics[width=0.33\textwidth]{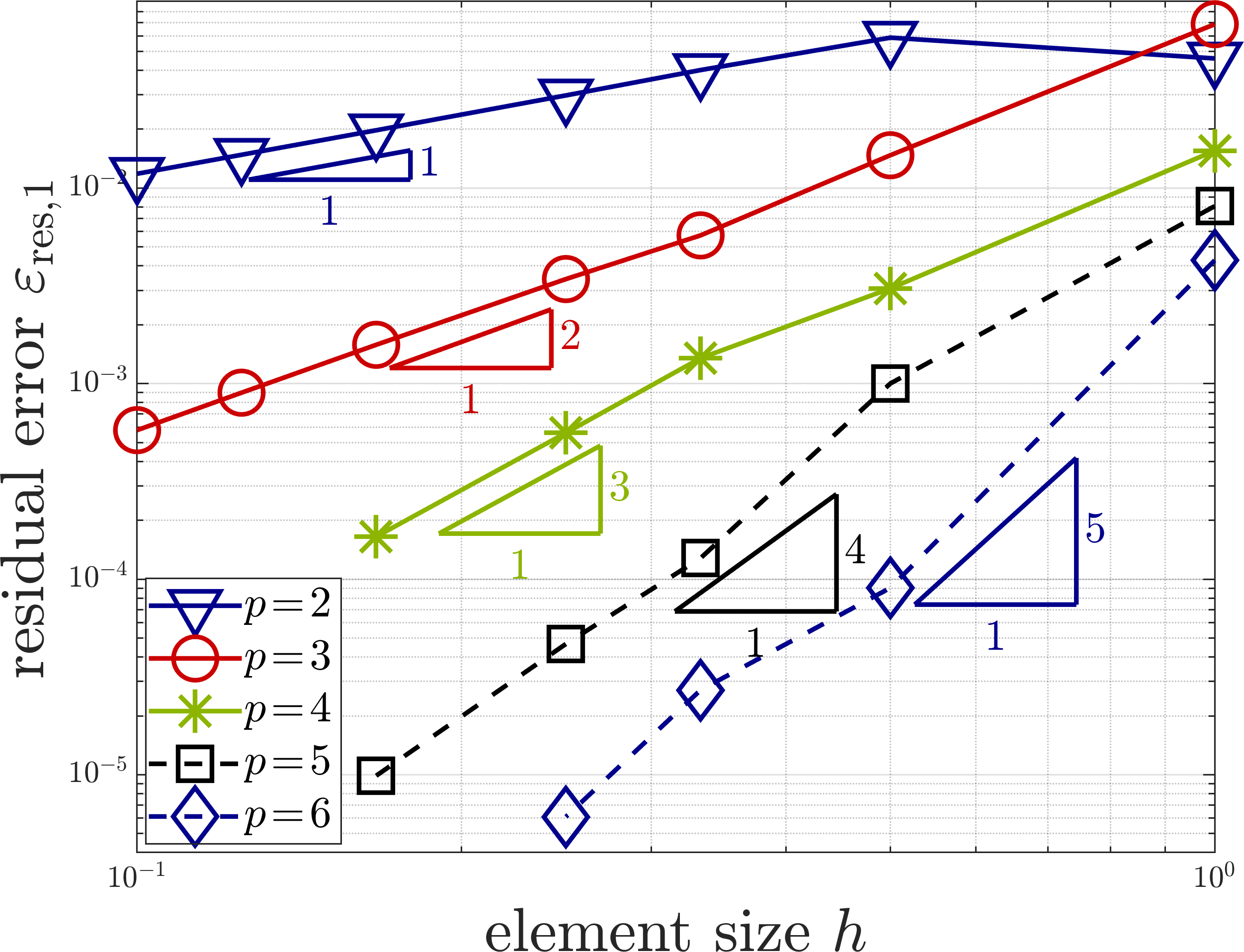}\label{Fig:TC4ClampedCopulasOnCurvedSurfResError1}}\hfill
 \subfigure[convergence in $\varepsilon_{\text{res},2}$]
 {\includegraphics[width=0.33\textwidth]{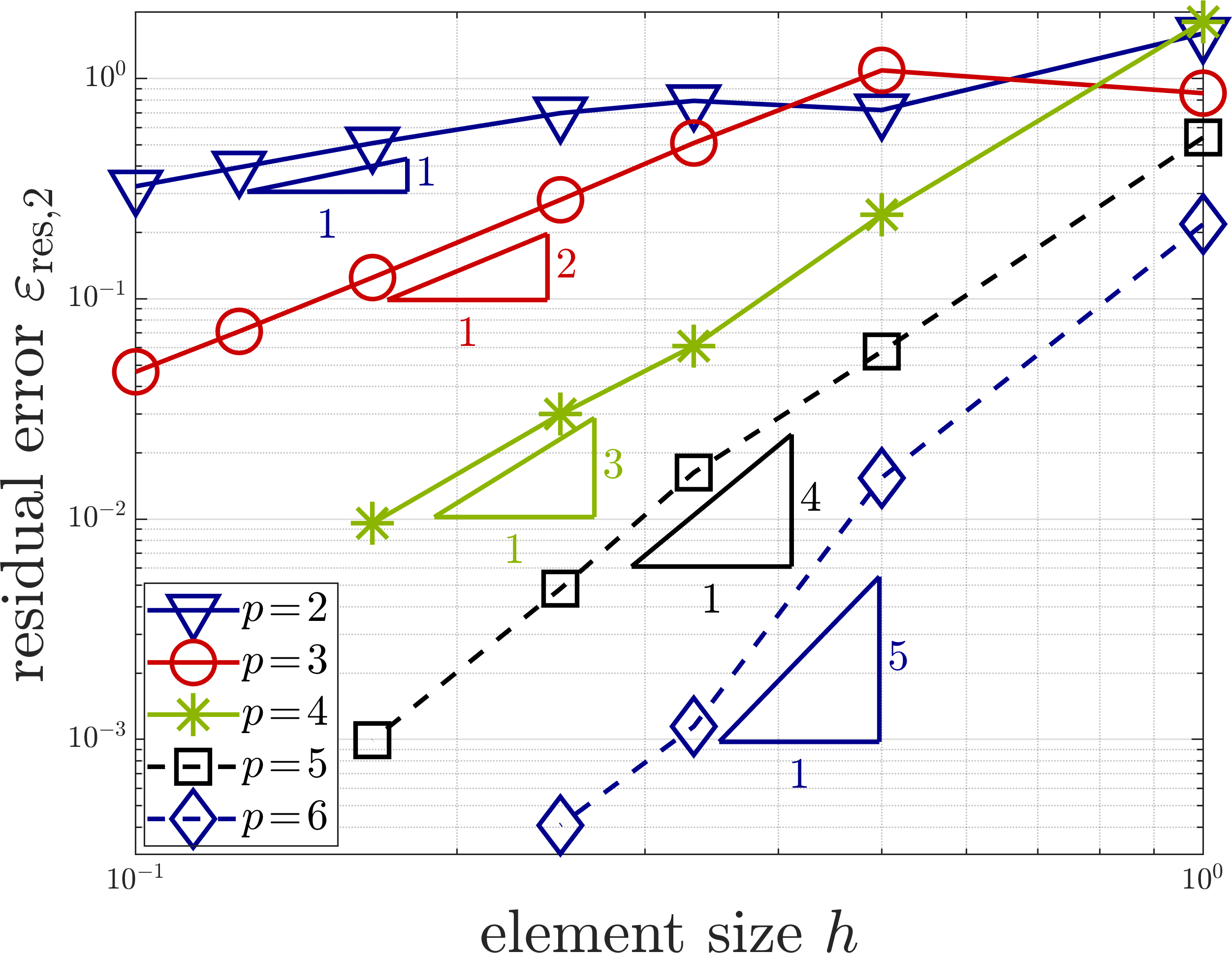}\label{Fig:TC4ClampedCopulasOnCurvedSurfResError2}}\hfill
 \subfigure[convergence in $\varepsilon_{\mathfrak{e}}$]
 {\includegraphics[width=0.33\textwidth]{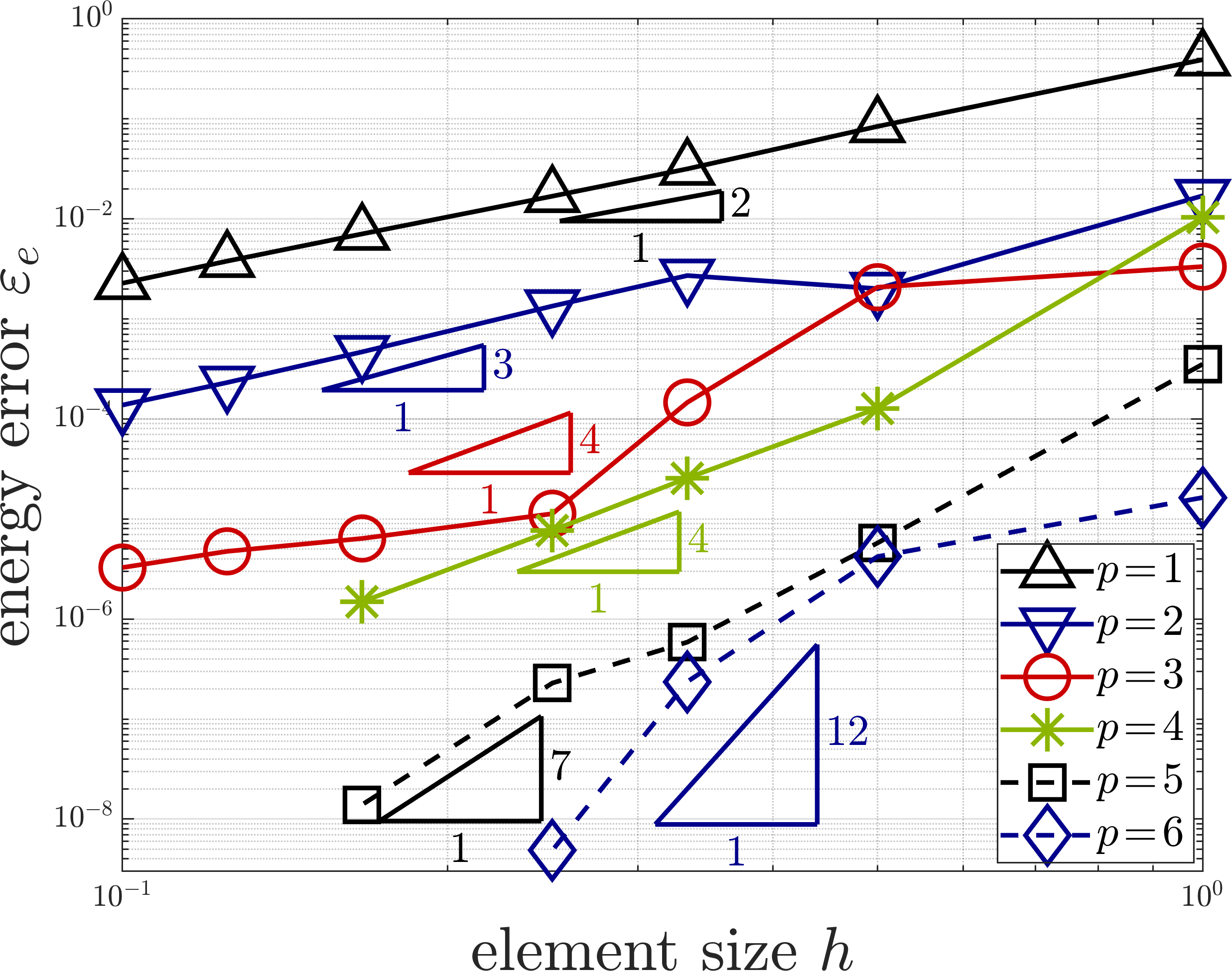}\label{Fig:TC4ClampedCopulasOnCurvedSurfEnergyError}}
 
 \caption{\label{Fig:TC4ClampedCopulasOnCurvedSurfError}Convergence studies for the family of clamped cupolas on a curved surface in $\mathbb{R}^3$. Convergence rates in (a) the absolute residual error $\varepsilon_{\text{res},1}$, (b) the relative residual error $\varepsilon_{\text{res},2}$, and (c) the relative stored energy error $\varepsilon_{\mathfrak{e}}$.}
\end{figure}

Fig.~\ref{Fig:TC4ClampedCopulasOnCurvedSurfError} shows the convergence rates in the residual and stored energy errors. The convergence rates in the residual errors achieve the expected rates, confirming optimal results in these error measures. The stored energy error also converges with higher-order, yet suboptimal rates, featuring a somewhat less regular pattern in the curves than the other error measures.

\section{Conclusions and outlook}\label{Sec:Conclusions&Outlook}
In this work, a mechanical model and a corresponding numerical method are proposed for the simultaneous analysis of geometrically linear Kirchhoff beams and Kirchhoff--Love shells on all level sets over a two- or three-dimensional bulk domain, respectively. The models are developed with a coordinate-free formulation, enabling an implicit definition of the manifolds and permitting the governing equations to be expressed by a global Cartesian coordinate system. 

For numerical analyses with the finite element method, the bulk domain is discretized by a higher-order mesh. This mesh conforms to the boundaries of the bulk domain, hence also to the boundaries of the structures. However, inside the bulk domain, it does not generally conform to the geometry of the structures described by the level sets. The resulting Bulk Trace FEM extends prior work, see \cite{Fries_2018a, Kaiser_2024a, Kaiser_2024b, Kaiser_2025a}, to shear-rigid structures. A possible field of application of the method are optimization tasks, e.g., determining an optimal shape of a structure, accelerating the design process. A further use case is the modeling of reinforced materials, especially with beams, physically and continuously embedded in a single shell.

A central challenge in the Kirchhoff--Love theory are the numerical properties of shear-rigid structures, whose displacement-based formulations involve fourth-order derivatives in the strong form, requiring $C^1$-continuous shape functions in FEM-based analyses. To overcome this limitation, we adopt a mixed-hybrid formulation extending our previous work in \cite{Neumeyer_2025a} for single shells. Here, the moment tensor is introduced as an additional primary unknown, reducing the continuity requirements to $C^0$ and, thereby, permitting the use of standard Lagrange elements which, however, increases the number of DOFs significantly. Therefore, a hybridization scheme is performed, where the inter-element continuity of the field of the moment tensor is broken and weakly reinforced by a Lagrange multiplier on the element interfaces. This allows for an element-wise static condensation of the components of the moment tensor, which substantially reduces the DOFs, lowering the computational effort.

Several numerical test cases confirm the accuracy and higher-order convergence behavior of the proposed method. These include general geometries for beams in $\mathbb{R}^2$ and for shells in $\mathbb{R}^3$. Resulting in smooth solutions, these test cases enable higher-order convergence in the observed error norms, such as the $\mathcal{L}^2$-error, residual errors, and the stored energy error. Optimal convergence rates are obtained for the beam examples. Although pre-asymptotic effects are observed for higher-order elements in the shell examples, the overall convergence in the residual errors aligns with the theoretical expectations and verifies the implementations. The stored energy error for the shells also converges with higher orders, however, sometimes with somewhat suboptimal rates.

Concluding, this work successfully presents a methodology for the simultaneous modeling and analysis of families of shear-rigid beams and shells over bulk domains, further establishing the Bulk Trace FEM in structural mechanics. Future research may focus on incorporating dynamic effects into the model, similarly to \cite{Fries_2024c} for ropes and membranes. A more application-oriented direction may also lead to the development of new anisotropic material models by embedding beams and shells into bulk materials, following \cite{Fries_2023a, Fries_2024b} for embedded ropes and membranes.


\bibliographystyle{schanz}
\addcontentsline{toc}{section}{\refname}\bibliography{NeumeyerRefs}

@string{AAM = "Arch. Appl. Mech."}

@string{CMAME = "Comput. Methods Appl. Mech. Eng."}

@string{CM = "Comput. Mech."}

@string{CS = "Comput. Struct."}

@string{ES = "Eng. Struct."}

@string{IJNME = "Int. J. Numer. Methods Eng."}

@string{JAM = "ASME J. Appl. Mech."}

@string{MathComp = "Math. Comput."}

@string{NumerMath = "Numer. Math."}

@string{IJNMF = "Int. J. Numer. Methods Fluids"}

@string{JCP = "J. Comput. Phys."}

@string{SIAM_JNA = "SIAM J. Numer. Anal."}

@string{AN = "Acta Numer."}

@string{RAIRO = "RAIRO Anal. Numér."}

@string{JRAM = "J. reine angew. Math."}

@string{PTRSA = "Philos. Trans. R. Soc. A"}

@string{IMA_JNA = "IMA J. Numer. Anal."}

@string{ESAIM = "ESAIM Math. Model. Numer. Anal."}

@string{JDE = "J. Differ. Equ."}

@string{PAMM = "Proc. Appl. Math. Mech."}

@string{IFB = "Interfaces Free Bound."}

@string{CVS = "Comput. Vis. Sci."}

@string{LEDPMJS = "Lond. Edinb. Dublin Philos. Mag. J. Sci."}

@string{Wiley = {{John Wiley \& Sons}}}

@string{Wiley:adr = {Chichester}}

@string{Spg = {{Sprin\-ger}}}

@string{Spg:adr = {Berlin}}

@string{CUP = {Cambridge University Press}}

@string{CUP:adr = {Cambridge}}

@string{SIAM = {SIAM}}

@string{SIAM:adr = {Philadelphia, PA}}

@string{BH = {Butterworth-Heinemann}}

@string{BH:adr = {Oxford}}

@string{ICE = {ICE Publishing}}

@string{ICE:adr = {London}}

@article{Arnold_1985a,
   author = "D.N. Arnold and F. Brezzi",
   title = "Mixed and nonconforming finite element methods: implementation, postprocessing and error estimates",
   journal = ESAIM,
   volume = 19,
   pages = "7--32",
   year = 1985,
   doi = "10.1051/m2an/1985190100071",
}

@book{Basar_1985a,
    author = "Y. Başar and W.B. Krätzig",
    title = "Mechanik der {F}lächentragwerke",
    publisher = "Vieweg$+$Teubner Verlag",
    address = "Braunschweig",
    year = 1985,
    doi = "10.1007/978-3-322-93983-8",
    isbn = "978-3-322-93984-5"
}

@incollection{Bischoff_2017a,
    author = "M. Bischoff and E. Ramm and J. Irslinger",
    title = "Models and Finite Elements for Thin-Walled Structures",
    booktitle = "Encyclopedia of Computational Mechanics",
    editor = "E. Stein and R. Borst and T.J. Hughes",
    publisher = Wiley,
    address = Wiley:adr,
    edition = 2,
    year = 2017,
    doi = "10.1002/9781119176817.ecm2026",
    isbn = "9781119176817"
}

@book{Boffi_2013a,
    author = "D. Boffi and F. Brezzi and M. Fortin",
    title = "Mixed Finite Element Methods and Applications",
    series = "Springer Series in Computational Mathematics",
    publisher = Spg,
    address = Spg:adr,
    volume = 44,
    year = 2013,
    doi = "10.1007/978-3-642-36519-5",
    isbn = "978-3-642-36518-8",
    issn = "0179-3632"
}

@article{Brezzi_1974a,
    author = "F. Brezzi",
    title = "On the existence, uniqueness and approximation of saddle-point problems arising from {L}agrange multipliers",
    journal = RAIRO,
    volume = 8,
    pages = "129--151",
    year = 1974,
    doi = "10.1051/m2an/197408R201291"
}

@article{Burger_2009a,
    author = "M. Burger",
    title = "Finite element approximation of elliptic partial differential equations on implicit surfaces",
    journal = CVS,
    volume = 12,
    pages = "87--100",
    year = 2009,
    doi = "10.1007/s00791-007-0081-x"
}

@article{Burman_2015a,
    author = "E. Burman and P. Hansbo and M.G. Larson",
    title = "A stabilized cut finite element method for partial differential equations on surfaces: The {L}aplace-{B}eltrami operator",
    journal = CMAME,
    volume = 285,
    pages = "188--207",
    year = 2015,
    doi = "10.1016/j.cma.2014.10.044"
}

@article{Burman_2018a,
    author = "E. Burman and P. Hansbo and M.G. Larson and A. Massing",
    title = "Cut finite element methods for partial differential equations on embedded manifolds of arbitrary codimensions",
    journal = ESAIM,
    volume = 52,
    pages = "2247--2282",
    year = 2018,
    doi = "10.1051/m2an/2018038"
}

@article{Cenanovic_2016a,
    author = "M. Cenanovic and P. Hansbo and M.G. Larson",
    title = "Cut finite element modeling of linear membranes",
    journal = CMAME,
    volume = 310,
    pages = "98--111",
    year = 2016,
    doi = "10.1016/j.cma.2016.05.018"
}

@article{Chapelle_1998a,
    author = "D. Chapelle and K.J. Bathe",
    title = "Fundamental considerations for the finite element analysis of shell structures",
    journal = CS,
    volume = 66,
    pages = "19--36",
    year = 1998,
    doi = "10.1016/S0045-7949(97)00078-3"
}

@article{Cockburn_2004a,
    author = "B. Cockburn and J. Gopalakrishnan",
    title = "A Characterization of Hybridized Mixed Methods for Second Order Elliptic Problems",
    journal = SIAM_JNA,
    volume = 42,
    pages = "283--301",
    year = 2004,
    doi = "10.1137/S0036142902417893"
}

@article{Cockburn_2009a,
    author = "B. Cockburn and J. Gopalakrishnan and R. Lazarov",
    title = "Unified Hybridization of Discontinuous {G}alerkin, Mixed, and Continuous {G}alerkin Methods for Second Order Elliptic Problems",
    journal = SIAM_JNA,
    volume = 47,
    pages = "1319--1365",
    year = 2009,
    doi = "10.1137/070706616"
}

@article{Comodi_1989a,
    author = "M.I. Comodi",
    title = "The {H}ellan-{H}errmann-{J}ohnson Method: some new error estimates and postprocessing",
    journal = MathComp,
    volume = 52,
    pages = "17--29",
    year = 1989,
    doi = "10.1090/s0025-5718-1989-0946601-7"
}

@article{Delfour_1995a,
    author = "M.C. Delfour and J.P. Zolésio",
    title = "A Boundary Differential Equation for Thin Shells",
    journal = JDE,
    volume = 119,
    pages = "426--449",
    year = 1995,
    doi = "10.1006/jdeq.1995.1097"
}

@article{Delfour_1996a,
    author = "M.C. Delfour and J.P. Zolésio",
    title = "Tangential Differential Equations for Dynamical Thin/Shallow Shells",
    journal = JDE,
    volume = 128,
    pages = "125--167",
    year = 1996,
    doi = "10.1006/jdeq.1996.0092"
}

@book{Delfour_2011a,
    author = "M.C. Delfour and J.P. Zolésio",
    title = "Shapes and geometries: Metrics, Analysis, Differential Calculus, and Optimization",
    series = "Advances in Design and Control",
    publisher = SIAM,
    address = SIAM:adr,
    edition = 2,
    year = 2011,
    doi = "10.1137/1.9780898719826",
    isbn = "978-0-898719-36-9"
}

@article{Dziuk_2008a,
    author = "G. Dziuk and C.M. Elliott",
    title = "{E}ulerian finite element method for parabolic {PDE}s on implicit surfaces",
    journal = IFB,
    volume = 10,
    pages = "119--138",
    year = 2008,
    doi = "10.4171/IFB/182"
}

@article{Dziuk_2008b,
    author = "G. Dziuk and C.M. Elliott",
    title = "An {E}ulerian approach to transport and diffusion on evolving implicit surfaces",
    journal = "Comput. Vis. Sci.",
    volume = 13,
    pages = "17--28",
    year = 2008,
    doi = "10.1007/s00791-008-0122-0"
}

@article{Dziuk_2013a,
    author = "G. Dziuk and C.M. Elliott",
    title = "Finite element methods for surface {PDE}s",
    journal = AN,
    volume = 22,
    pages = "289--396",
    year = 2013,
    doi = "10.1017/S0962492913000056"
}

@article{Echter_2013a,
    author = "R. Echter and B. Oesterle and M. Bischoff",
    title = "A hierarchic family of isogeometric shell finite elements",
    journal = CMAME,
    volume = 254,
    pages = "170--180",
    year = 2013,
    doi = "10.1016/j.cma.2012.10.018"
}

@book{Federer_1969a,
    author = "H. Federer",
    title = "Geometric measure theory",
    publisher = Spg,
    address = "New York",
    year = 1969,
    isbn = "9780387045054"
}

@article{Fries_2018a,
    author = "T.P. Fries",
    title = "Higher-order surface {FEM} for incompressible {N}avier-{S}tokes flows on manifolds",
    journal = IJNMF,
    volume = 88,
    pages = "55--78",
    year = 2018,
    doi = "10.1002/fld.4510"
}

@article{Fries_2020a,
    author = "T.P. Fries and D. Schöllhammer",
    title = "A unified finite strain theory for membranes and ropes",
    doi = "10.1016/j.cma.2020.113031",
    journal = CMAME,
    year = 2020,
    volume = 365,
    pages = "113031"
}

@article{Fries_2023a,
    author = "T.P. Fries and M.W. Kaiser",
    title = "On the Simultaneous Solution of Structural Membranes on all Level Sets within a Bulk Domain",
    journal = CMAME,
    volume = 415,
    pages = "116223",
    year = 2023,
    doi = "10.1016/j.cma.2023.116223"
}

@inproceedings{Fries_2024c,
    author = "T.P. Fries and M.W. Kaiser",
    title = "Simultaneous, Dynamical Analysis of Structural Ropes and Membranes on all Level-sets",
    booktitle = "Proceedings of the 9th European Congress on Computational Methods in Applied Sciences and Engineering (ECCOMAS 2024)",
    address = "Lisbon, Portugal",
    year = 2024,
    doi = "10.23967/eccomas.2024.006"
}

@inproceedings{Fries_2024b,
    author = "T.P. Fries and J. Neumeyer and M.W. Kaiser",
    title = "A new concept for embedding sub-structures via level-sets",
    booktitle = "Proceedings of the 16th World Congress on Computational Mechanics (WCCM 2024)",
    address = "Vancouver, Canada",
    year = 2024,
    doi = "10.23967/wccm.2024.025"
}

@article{Gfrerer_2021a,
    author = "M.H. Gfrerer",
    title = "A $C^1$-continuous {T}race-{F}inite-{C}ell-{M}ethod for linear thin shell analysis on implicitly defined surfaces",
    journal = CM,
    volume = 67,
    pages = "679--697",
    year = 2021,
    doi = "10.1007/s00466-020-01956-5"
}

@article{Gimena_2008a,
    author = "L. Gimena and F.N. Gimena and P. Gonzaga",
    title = "Structural analysis of a curved beam element defined in global coordinates",
    journal = ES,
    volume = 30,
    pages = "3355--3364",
    year = 2008,
    doi = "10.1016/j.engstruct.2008.05.011"
}

@article{Hansbo_2014a,
    author = "P. Hansbo and M.G. Larson",
    title = "Finite element modeling of a linear membrane shell problem using tangential differential calculus",
    journal = CMAME,
    volume = 270,
    pages = "1--14",
    year = 2014,
    doi = "10.1016/j.cma.2013.11.016"
}

@article{Hansbo_2014b,
    author = "P. Hansbo and M.G. Larson and K. Larsson",
    title = "Variational formulation of curved beams in global coordinates",
    journal = CM,
    volume = 53,
    pages = "611--623",
    year = 2014,
    doi = "10.1007/s00466-013-0921-0"
}

@article{Jankuhn_2017a,
    author = "T. Jankuhn and M.A. Olshanskii and A. Reusken",
    title = "Incompressible fluid problems on embedded surfaces: {M}odeling and variational formulations",
    journal = IFB,
    volume = 20,
    pages = "353--377",
    year = 2018,
    doi = "10.4171/IFB/405"
}

@article{Johnson_1973a,
    author = "C. Johnson",
    title = "On the convergence of a mixed finite-element method for plate bending problems",
    journal = NumerMath,
    volume = 21,
    pages = "43--62",
    year = 1973,
    doi = "10.1007/BF01436186"
}

@article{Kabaria_2015a,
    author = "H. Kabaria and A.J. Lew and B. Cockburn",
    title = "A hybridizable discontinuous {G}alerkin formulation for non-linear elasticity",
    journal = CMAME,
    volume = 283,
    pages = "303--329",
    year = 2015,
    doi = "10.1016/j.cma.2014.08.012"
}

@article{Kaiser_2023a,
    author = "M.W. Kaiser and T.P. Fries",
    title = "Curved, linear {K}irchhoff beams formulated using tangential differential calculus and {L}agrange multipliers",
    journal = PAMM,
    volume = 22,
    pages = "e202200042",
    year = 2023,
    doi = "10.1002/pamm.202200042"
}

@article{Kaiser_2024a,
    author = "M.W. Kaiser and T.P. Fries",
    title = "Simultaneous analysis of continuously embedded {R}eissner-{M}indlin shells in 3{D} bulk domains",
    journal = IJNME,
    volume = 125,
    pages = "e7495",
    year = 2024,
    doi = "10.1002/nme.7495"
}

@article{Kaiser_2024b,
    author = "M.W. Kaiser and T.P. Fries",
    title = "Simultaneous solution of implicitly defined curved, linear {T}imoshenko beams in two-dimensional bulk domains",
    journal = PAMM,
    volume = 24,
	pages = "e202400019",
    year = 2024,
    doi = "10.1002/pamm.202400019"
}

@article{Kaiser_2025a,
    author = "M.W. Kaiser and T.P. Fries",
    title = "Simultaneous solution of incompressible {N}avier-{S}tokes flows on multiple surfaces",
    journal = AAM,
    volume = 95,
    eid = 233,
    year = 2025,
    doi = "10.1007/s00419-025-02935-z"
}

@article{Kiendl_2009a,
    author = "J. Kiendl and K.U. Bletzinger and J. Linhard and R. Wüchner",
    title = "Isogeometric shell analysis with {K}irchhoff-{L}ove elements",
    journal = CMAME,
    volume = 198,
    pages = "3902--3914",
    year = 2009,
    doi = "10.1016/j.cma.2009.08.013"
}

@article{Kirchhoff_1850a,
    author = "G. Kirchhoff",
    title = "Über das {G}leichgewicht und die {B}ewegung einer elastischen {S}cheibe",
    journal = JRAM,
    volume = 40,
    pages = "51--88",
    year = 1850,
    doi = "10.1515/crll.1850.40.51"
}

@article{Love_1888a,
    author = "A.E.H. Love",
    title = "{XVI}. {T}he small free vibrations and deformation of a thin elastic shell",
    journal = PTRSA,
    volume = 179,
    pages = "491--546",
    year = 1888,
    doi = "10.1098/rsta.1888.0016"
}

@article{Mindlin_1951a,
    author = "R.D. Mindlin",
    title = "Influence of Rotatory Inertia and Shear on Flexural Motions of Isotropic, Elastic Plates",
    journal = JAM,
    volume = 18,
    pages = "31--38",
    year = 1951,
    doi = "10.1115/1.4010217"
}

@book{Morgan_1988a,
    author = "F. Morgan",
    title = "Geometric measure theory: a beginner's guide",
    publisher = "Academic press",
    address = "San Diego",
    year = 1988,
    isbn = "0-12-506855-7"
}

@article{Neumeyer_2025a,
    author = "J. Neumeyer and M.W. Kaiser and T.P. Fries",
    title = "Higher-Order, Mixed-Hybrid Finite Elements for {K}irchhoff--{L}ove Shells",
    journal = IJNME,
    volume = 126,
    pages = "e70175",
    year = 2025,
    doi = "10.1002/nme.70175"
}

@article{Neunteufel_2019a,
    author = "M. Neunteufel and J. Schöberl",
    title = "The {H}ellan--{H}errmann--{J}ohnson method for nonlinear shells",
    journal = CS,
    volume = 225,
    pages = "106109",
    year = 2019,
    doi = "10.1016/j.compstruc.2019.106109"
}

@article{Neunteufel_2024a,
    author = "M. Neunteufel and J. Schöberl",
    title = "The {H}ellan--{H}errmann--{J}ohnson and {TDNNS} methods for linear and nonlinear shells",
    journal = CS,
    volume = 305,
    pages = "107543",
    year = 2024,
    doi = "10.1016/j.compstruc.2024.107543"
}

@article{Olshanskii_2009b,
    author = "M.A. Olshanskii and A. Reusken and J. Grande",
    title = "A finite element method for elliptic equations on surfaces",
    journal = SIAM_JNA,
    volume = 47,
    pages = "3339--3358",
    year = 2009,
    doi = "10.1137/080717602"
}

@article{Olshanskii_2014a,
    author = "M.A. Olshanskii and A. Reusken and X. Xu",
    title = "A stabilized finite element method for advection-diffusion equations on surfaces",
    journal = IMA_JNA,
    volume = 34,
    pages = "732--758",
    year = 2014,
    doi = "10.1093/imanum/drt016"
}

@incollection{Olshanskii_2017a,
    author = "M.A. Olshanskii and A. Reusken",
    title = "Trace finite element methods for {PDE}s on surfaces",
    booktitle = "Geometrically Unfitted Finite Element Methods and Applications",
    editor = "S.P.A. Bordas and E. Burman and M.G. Larson and M.A. Olshanskii",
    series = "Lecture notes in computational science and engineering",
    publisher = "Springer Nature",
    address = "Cham",
    volume = 121,
    pages = "211--258",
    year = 2017,
    doi = "10.1007/978-3-319-71431-8_7",
    isbn = "978-3-319-71430-1"
}

@article{Osher_2001a,
    author = "S. Osher and R.P. Fedkiw",
    title = "Level set methods: an overview and some recent results",
    journal = JCP,
    volume = 169,
    pages = "463--502",
    year = 2001,
    doi = "10.1006/jcph.2000.6636"
}

@book{Osher_2003a,
    author = "S. Osher and R.P. Fedkiw",
    title = "Level Set Methods and Dynamic Implicit Surfaces",
    publisher = Spg,
    address = Spg:adr,
    year = 2003,
    doi = "10.1007/b98879",
    isbn = "978-1-4684-9251-4",
    issn = "0066-5452"
}

@article{Rafetseder_2018a,
    author = "K. Rafetseder and W. Zulehner",
    title = "A decomposition result for {K}irchhoff plate bending problems and a new discretization approach",
    journal = SIAM_JNA,
    volume = 56,
    pages = "1961--1986",
    year = 2018,
    doi = "10.1137/17M1118427"
}

@article{Rafetseder_2019a,
    author = "K. Rafetseder and W. Zulehner",
    title = "A new mixed approach to {K}irchhoff--{L}ove shells",
    journal = CMAME,
    volume = 346,
    pages = "440--455",
    year = 2019,
    doi = "10.1016/j.cma.2018.11.033"
}

@article{Reissner_1945a,
    author = "E. Reissner",
    title = "The Effect of Transverse Shear Deformation on the Bending of Elastic Plates",
    journal = JAM,
    volume = 12,
    pages = "A69--A77",
    year = 1945,
    doi = "10.1115/1.4009435"
}

@article{Schoellhammer_2019a,
    author = "D. Schöllhammer and T.P. Fries",
    title = "{K}irchhoff-{L}ove shell theory based on tangential differential calculus",
    journal = CM,
    volume = 64,
    pages = "113--131",
    year = 2019,
    doi = "10.1007/s00466-018-1659-5"
}

@article{Schoellhammer_2019b,
    author = "D. Schöllhammer and T.P. Fries",
    title = "{R}eissner-{M}indlin shell theory based on tangential differential calculus",
    journal = CMAME,
    volume = 352,
    pages = "172--188",
    year = 2019,
    doi = "10.1016/j.cma.2019.04.018"
}

@article{Schoellhammer_2020a,
    author = "D. Schöllhammer and T.P. Fries",
    title = "A higher-order {T}race finite element method for shells",
    journal = IJNME,
    volume = 122,
    pages = "1217--1238",
    year = 2021,
    doi = "10.1002/nme.6558"
}

@book{Sethian_1999b,
    author = "J.A. Sethian",
    title = "Level Set Methods and Fast Marching Methods",
    publisher = CUP,
    address = CUP:adr,
    edition = 2,
    year = 1999,
    doi = "10.1017/S0263574799212404",
    isbn = "0-521-64557-3"
}

@article{Timoshenko_1921a,
    author = "S.P. Timoshenko",
    title = "{LXVI}. On the correction for shear of the differential equation for transverse vibrations of prismatic bars",
    journal = LEDPMJS,
    volume = 41,
    pages = "744--746",
    year = 1921,
    doi = "10.1080/14786442108636264"
}

@book{Zienkiewicz_2013a,
    author = "O.C. Zienkiewicz and R.L. Taylor and J.Z. Zhu",
    title = "The Finite Element Method: Its Basis and Fundamentals",
    publisher = BH,
    address = BH:adr,
    edition = 7,
    year = 2013,
    doi = "https://doi.org/10.1016/C2009-0-24909-9",
    isbn = "978-1-85617-633-0"
}

@book{Zingoni_2017a,
    author = "A. Zingoni",
    title = "Shell Structures in Civil and Mechanical Engineering",
    publisher = ICE,
    address = ICE:adr,
    edition = 2,
    year = 2017,
    doi = "10.1680/ssicame.60289",
    isbn = "978-0-7277-6028-9"
}
 
\end{document}